# Analysis of experimental data on neutron decay for the possibility of the existence of the right vector boson $W_R$

A.P. Serebrov, O.M. Zherebtsov, A.K. Fomin, R.M. Samoilov, N.S. Budanov

*NRC "Kurchatov Institute" - Petersburg Institute of Nuclear Physics, 188300, Gatchina, Russia*

**e-mail:* serebrov_ap@pnpi.nrcki.ru

The analysis of the latest most accurate experimental data on neutron decay for the possibility of the existence of the right vector boson $W_R$ is carried out. As a result of the analysis within the framework of the left-right symmetric model, it was found that there is an indication of the existence of the right vector boson $W_R$ with a mass of $M_{W_R} = 304^{+24}_{-20}$GeV, and a mixing angle with $W_L$: $\zeta = -0.039 \pm 0.014$. It is shown that this result does not contradict the experiments at colliders to search for a hypothetical vector boson. In addition, it is shown that it is possible to describe the effects of CP violation in decays of neutral $K$-mesons, $D$-mesons and $B^0 -$ mesons using the parameters of the extended left-right symmetric model obtained from neutron decay, i.e. squared masses of the left and right bosons ratio and the mixing angle. The formation of baryon-lepton asymmetry of the Universe is considered within the framework of the left-right model of weak interaction with CP violation. A new experimental design for measuring neutron decay asymmetries with increased accuracy is presented.

## 1. INTRODUCTION

There is a hypothesis that sterile neutrinos are actually righthanded neutrinos [1, 2]. This hypothesis is quite appropriate, for example, it is discussed in connection with the possibility of explaining dark matter by righthanded neutrinos. However, this idea requires experimental justification. Righthanded neutrinos can appear together with right vector bosons: $W_R^{\pm}, Z_R$. In this regard, the analysis of experimental data on neutron decay on the possibility of the existence of a right vector boson $W_R$ was carried out.

Theoretical models with the introduction of right vector bosons have been well known [3-6] since the late 1970s. The most detailed analysis of neutron decay is presented in [7], where aspects of the possible contribution of right currents are also considered. Before analyzing the current experimental situation in neutron decay for the possible presence of right currents, we present a review that illustrated the increase in measurement accuracy and shows trends in the measured values for the neutron lifetime and decay asymmetries.

## 2. A REVIEW ILLUSTRATING THE INCREASE IN MEASUREMENT ACCURACY

The results of the neutron lifetime measurements, electron and neutrino asymmetries of neutron decay are presented in Figs. 1, 2, 3 and 4. It can be seen that significant progress has been made in the accuracy of neutron lifetime measurements over the past 25 years. In 2005, a revolutionary refinement of the neutron lifetime occurred thanks to the work [9] - the measurement of the neutron lifetime with a gravitational trap of ultracold neutrons. This result was later confirmed by measurements with a magnetic UCN trap at PNPI [11, 12] in 2009 and finally, with even greater accuracy, by measurements with a magnetic UCN trap at LANL in 2018 [13] and in 2021 [14].

Finally, in December 2024, the result of a new beam experiment with electron detection instead of proton detection appeared, which confirmed the neutron lifetime measured with UCN [15].

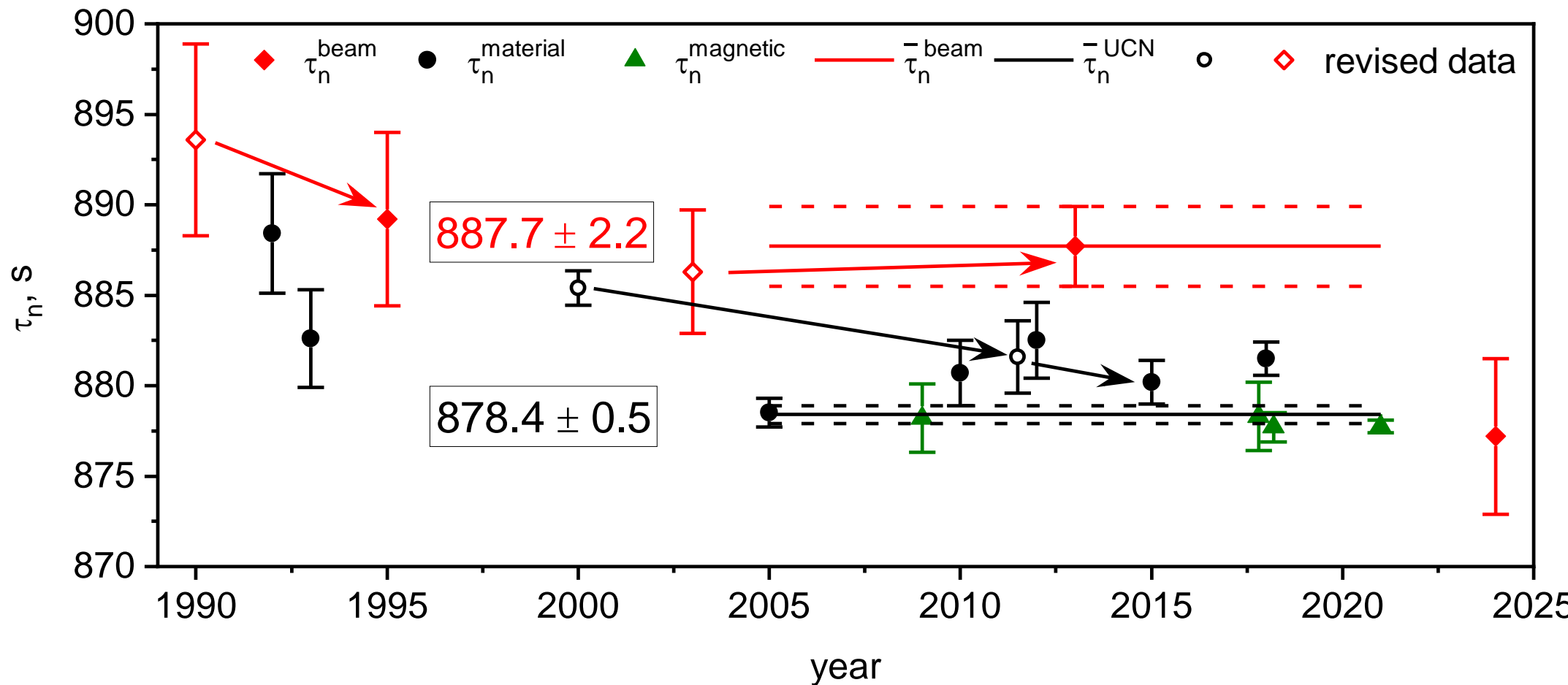

Fig. 1. Data of experimental results for the neutron lifetime, starting from 1990 from [8], discrepancy of data in 2005 [9] with data of 2000 [10], new results with magnetic trap (marked in green), which are decisive [11-14]. New beam experiment [15].

Trends in neutron decay electron asymmetry measurements are shown in Fig. 2. In decay electron asymmetry measurements, significant changes in accuracy occurred at the PERKEO II [16] and PERKEO III [17,18]. The accuracy of measurements of decay asymmetry first increased by 3 times and then by another 2.5 times, and ultimately amounted to 0.17%. At the same time, the absolute value of the electron decay asymmetry increased by 2%.

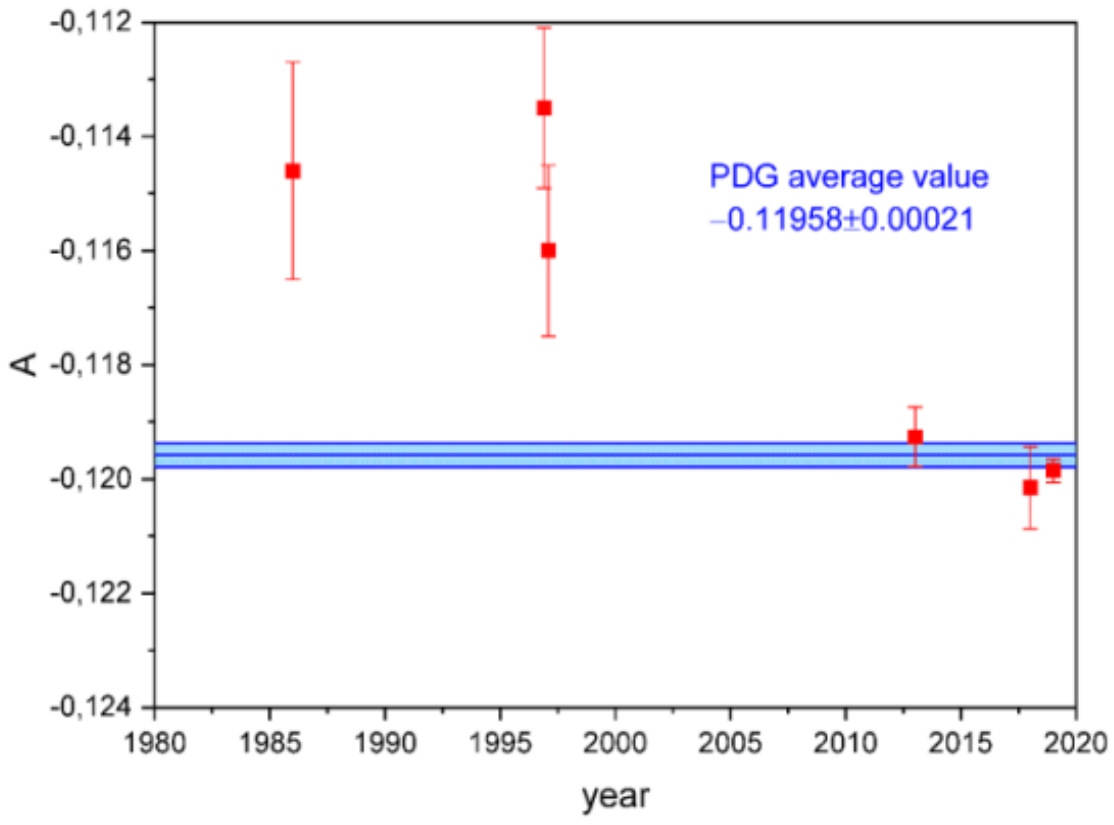

Fig. 2. Measurements of the electron asymmetry of neutron decay (A) and the averaged result from PDG [19].

Finally, and particularly important for our analysis of right currents, we must use the results of measurements of the neutrino asymmetry of neutron decay, where the measurement accuracy was significantly increased in 1998 by the experiment [20]. The result was later confirmed by the experiment [21] with the same accuracy. As a result, the value of the neutrino asymmetry presented in PDG [19] was determined.

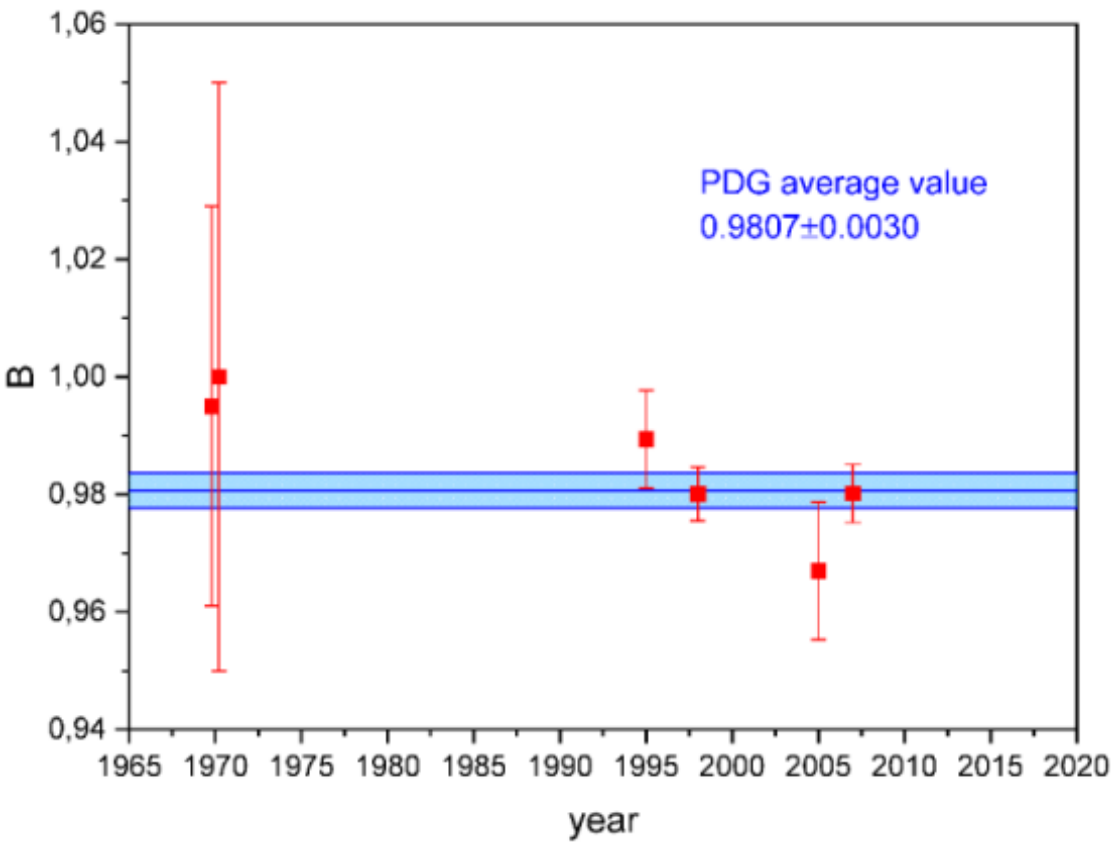

Fig. 3. Experimental results of neutrino asymmetry of neutron decay ($B$)and averaged result from PDG .

Significant progress in the measurements of electron-neutrino asymmetries of neutron decay (a) has been achieved in recent years in the experiment [22]. The experimental results of electron-neutrino asymmetries of neutron decay (a) and the averaged result from PDG are presented in Fig. 4.

In the following analysis, we will use the most precise result $a = -0.10402(82)$ from [24].

In addition, for further analysis, the unitarity condition of the CKM matrix [25] and the data of experiments with Fermi superallowed nuclear $0^+ - 0^+$ transitions [26]. A graphical analysis of the listed measurement results is presented in Fig. 5.

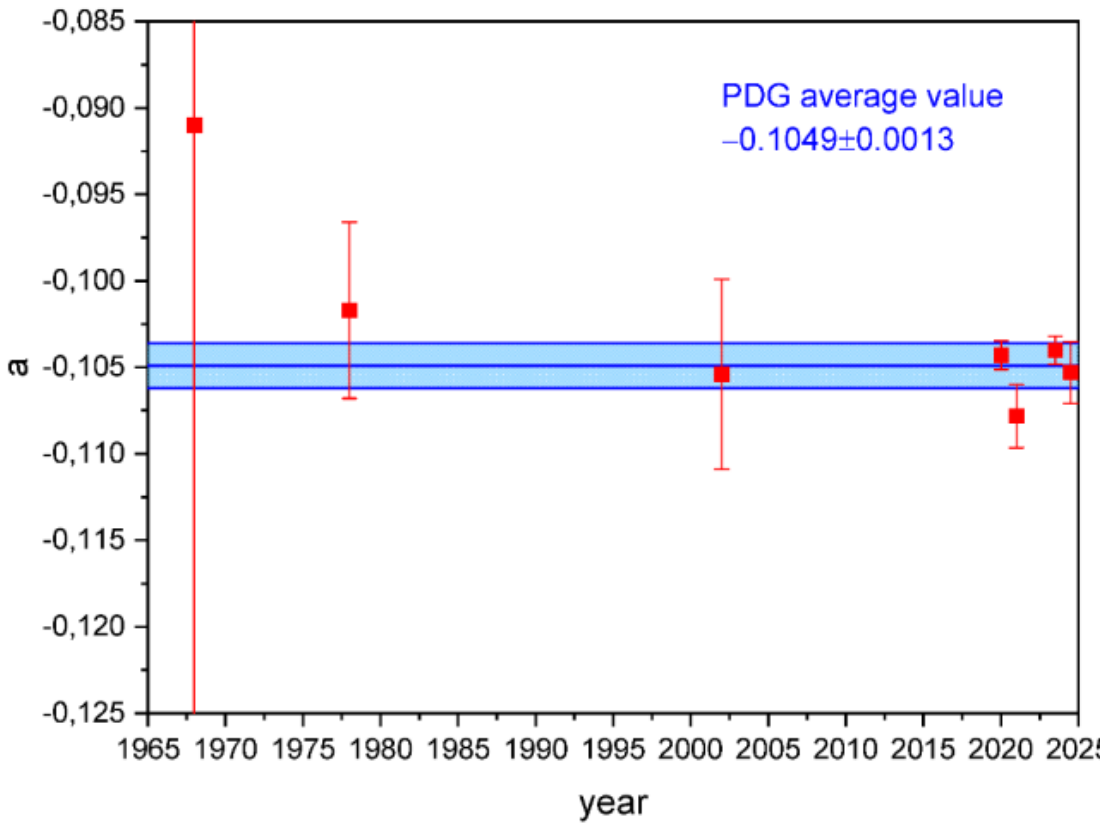

Fig. 4. The experimental results for the electron-neutrino asymmetry in neutron decay a, including the results from the aCORN [23] and aSPECT [24] experiments, as well as the averaged result from PDG [19].

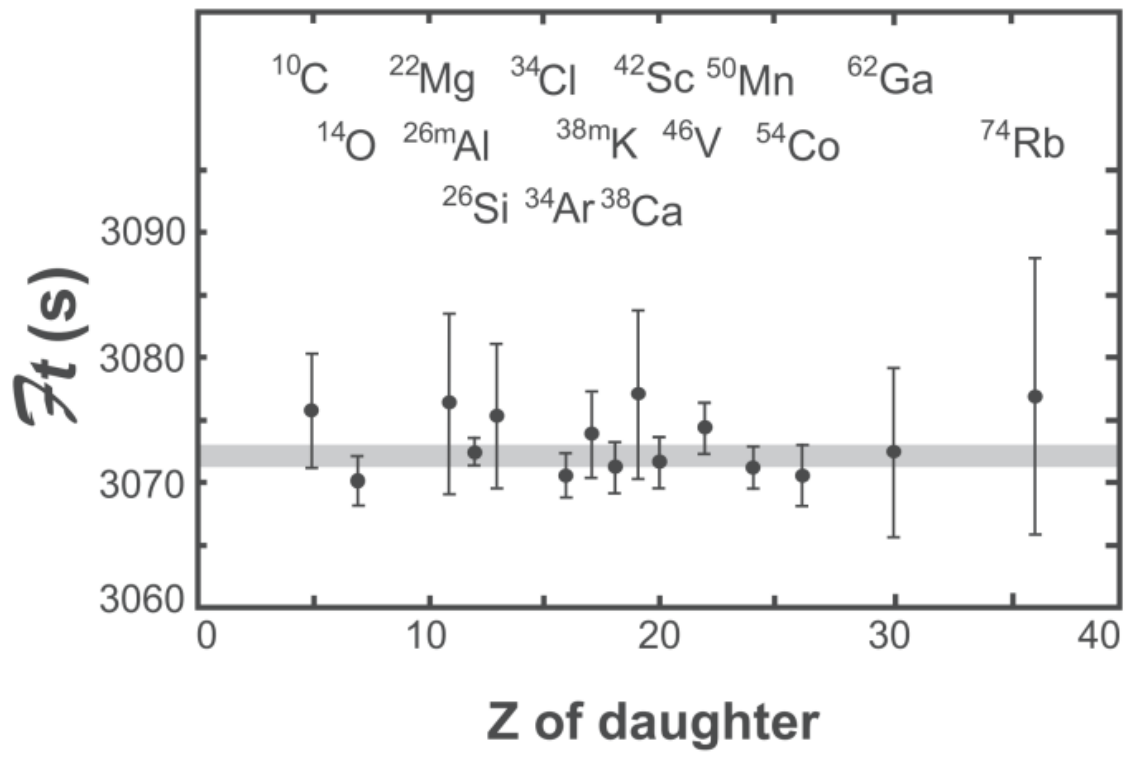

Fig. 5. Results of measuring the quantity $F\tau$ in experiments with Fermi superallowed nuclear $0^+ - 0^+$transitions from work [26].

Within the V - A theory of weak interaction, all three methods of $V_{ud}$ determination (from neutron decay, from experiments with Fermi superallowed nuclear $0^+ - 0^+$ transitions and from the unitarity of the CKM matrix) must coincide. The results of determination $V_{ud}$ from neutron decay are determined by the accuracy of measuring the neutron lifetime $877.75 \pm 0.35\ s$ [14] (blue area in Fig. 6) and the accuracy of measuring the ratio of the axial and vector constants from the electron asymmetry of neutron decay - $\lambda = -1.2757(5)$ [18] (green area in Fig. 6). The intersection of the data for the neutron lifetime and the value of the ratio of the axial and vector constants of the weak interaction $\lambda$ from the electron asymmetry of neutron decay $A$ gives the value $V_{ud}^n = 0.97477(37)$[27].

From the unitarity of the CKM matrix, using the value $V_{us} = 0.2243(8)$ [19] and $|V_{ub}|^2 = 1.7{\cdot}10^{-5}$ [19] can be calculated $V_{ud}^{unit} = \sqrt{1 - V_{us}^2 - V_{ub}^2} = 0.97452(18)$. This value agrees within the error limits with the value from neutron decay $V_{ud}^n$, however, the matrix element $V_{ud}^{00}$from $0^+ - 0^+$ transitions is noticeably different. $V_{ud}^{00} = 0.97373(32)$ [26] (Fig. 6). The difference $V_{ud}$ between the matching values $V_{ud}^n$ and the value $V_{ud}^{00}$ from $0^+ - 0^+$ the transitions is $2.6\sigma$. It is important to note that in [26] a violation of unitarity at 2.4 is indicated $\sigma$, which is also discussed in [28].

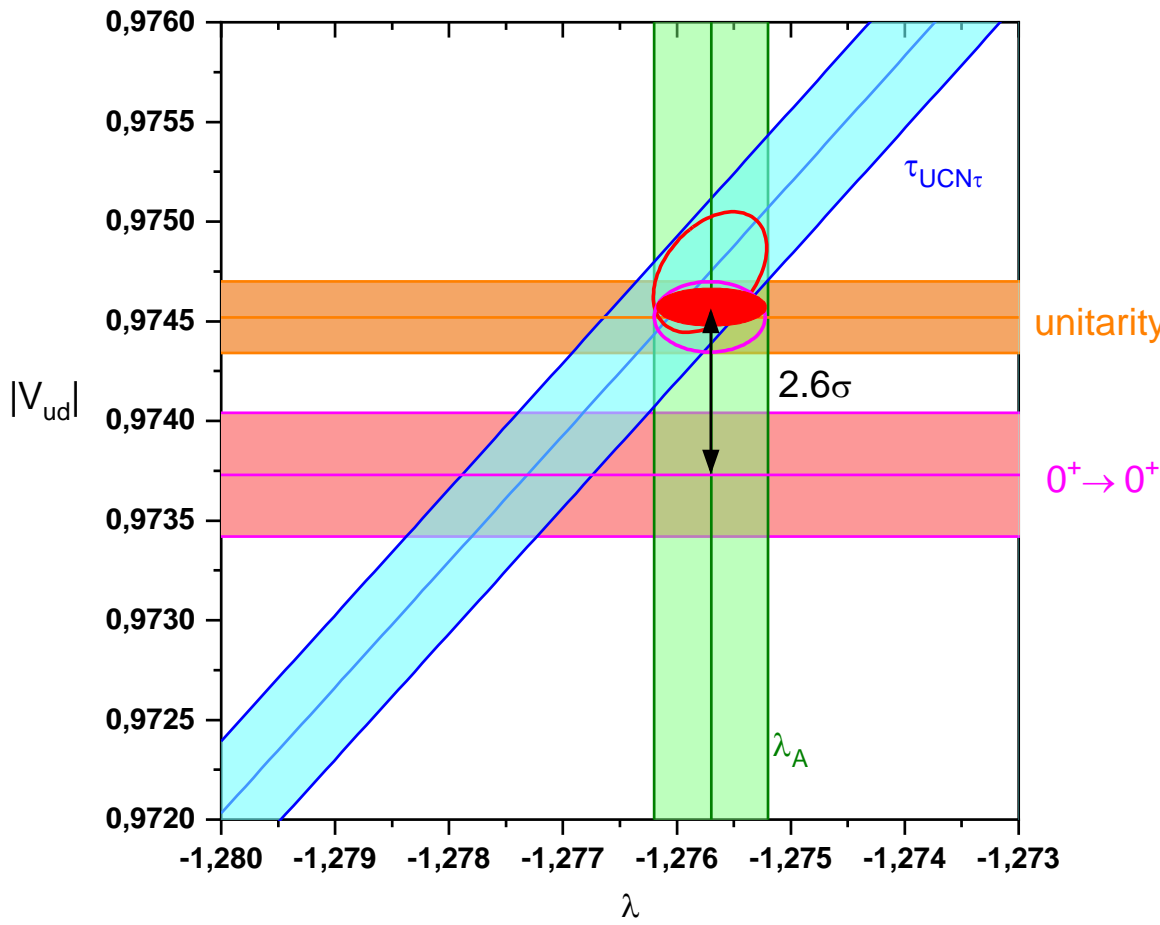

Fig. 6. Dependence of the matrix element of quark mixing $V_{ud}$ on $\lambda$, calculated using the SM formulas from neutron decay, from experiments with Fermi superallowed nuclear $0^+ - 0^+$ transitions and from the unitarity of the CKM matrix, using measurements $V_{us}$ [19].

Fig. 7 shows a comparison of the experimental neutrino decay asymmetry values $B = 0.9807(30)$ [19] and the calculated asymmetry within the SM framework depending on $\lambda$. In this case, there is also a discrepancy between the experimental value of the neutrino asymmetry and the SM prediction. The difference in the values of these quantities is $2.1\sigma$. The best experimental test is to measure the neutrino asymmetry of neutron decay. Note that there are practically no radiative corrections for the neutrino asymmetry, internal radiative corrections occur at a level of about $10^{-5}$ [29], therefore, measuring neutrino asymmetry is the purest test for right currents. It was for this purpose that the works [20, 21, 30] were carried out.

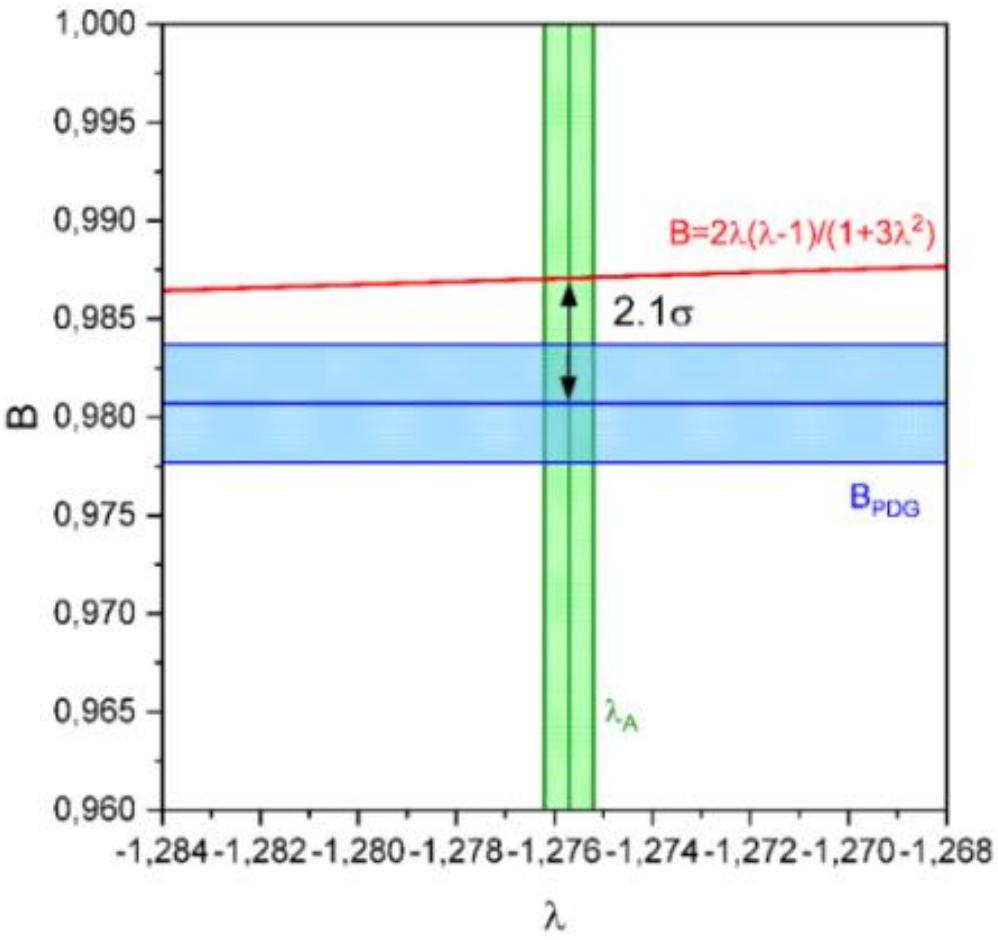

Fig. 7. Comparison of the experimental neutrino asymmetry of neutron decay and that calculated within the SM framework as a function of the ratio of the axial and vector constants of weak interaction $\lambda$.

At the same time, in [31] was shown that, consideration of the emission of real photons leads to a change in the proton recoil momentum, which is used to determine the antineutrino emission angle when processing experimental data. The magnitude of such a correction for the coefficient $B$ is on the order of 0.1%. If such a correction is introduced to the value of $B$ from [19], this will reduce the discrepancy between the experimentally measured value and the theoretical value from 2.1σ to 1.8σ, but will not eliminate it completely.

Finally, the most precise measurements of the electron-neutrino asymmetry $(a)$ allow for an independent calculation of the $\lambda_a$ value. However, it turns out that $\lambda_a = -1.2686 \pm 0.0025$, as reported in reference [24], differs from $\lambda_{A/\tau} = -1.2754 \pm 0.0013$ by $2.5\sigma$. This discrepancy is illustrated in Figure 8 from [23].

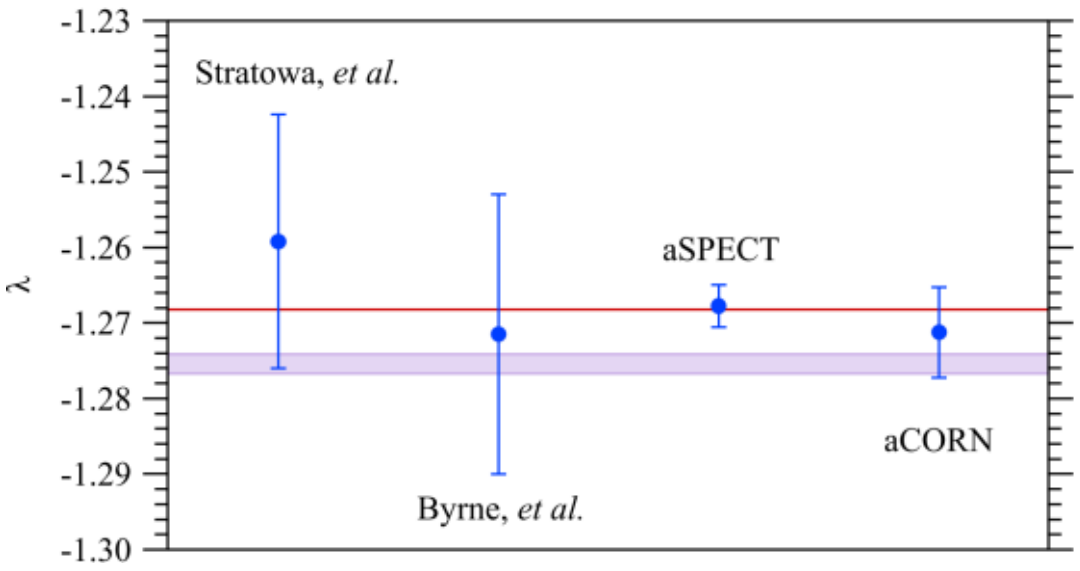

Fig. 8. The main results for the determination of λ from experiments measuring the coefficient a. The horizontal line represents the weighted average $\lambda = -1.2686 \pm 0.0025$. The shaded band shows the value $\lambda = -1.2754 \pm 0.0013$ from PDG 2022, which was obtained from measurements of the parameters $A$ and $\tau$.

From the presented picture of experimental data it follows that significant progress has been achieved in the accuracy of measurements and at the same time deviations in the interpretation of data within the framework of the V - A theory of weak interaction has been discovered. In this connection it is advisable to conduct an analysis taking into account the possible influence of right currents.

## 3. MANIFEST LEFT-RIGHT SYMMETRIC MODEL OF MIXING LEFT AND RIGHT VECTOR BOSONS

The analysis of the observed discrepancy can be done within the framework of the model taking into account the right currents. In the simplest left-right symmetric model [3-5], the mixing of left and right vector bosons is considered, and for flavor states $W_L, W_R$ and mass states $W_1, W_2$ we can write [5]:

$$W_L = W_1 \cos\zeta + W_2 \sin\zeta$$
$$W_R = e^{i\omega}(-W_1 \sin\zeta + W_2 \cos\zeta) \qquad (3.1)$$

where $\zeta$ is the angle of mixing of current states $W_L$ and $W_R$, and $\delta$ is the ratio of the squares of the masses of states $W_1$ and $W_2$.

$\omega$- CP- violating phase. Complete CP violation at $\omega = \pi/2$ and absence of CP- violation at $\omega = 0$.

However, it turned out that the use of the left-right manifest model does not resolve the unitarity problem for $0^+ \to 0^+$ transitions. Therefore, we propose considering an extended version of the left-right model of weak interaction with CP violation, which is introduced differently than in the manifest model. In this new version of the left-right model, we treat $W^-$ and $W^+$ as particle and antiparticle, respectively. Consequently, the mixing matrices for negatively and positively charged bosons are Hermitian conjugates of each other, which explains the sign changes for the sines. We introduce CP violation between particles and antiparticles through $W^-$ and $W^+$, which define the nature of the weak interaction and serve as particles and antiparticles. Thus, we should write:

$$\begin{pmatrix} W_L^{\pm} \\ W_R^{\pm} \end{pmatrix} = \begin{pmatrix} \cos\zeta & \mp\sin\zeta \\ \pm\sin\zeta & \cos\zeta \end{pmatrix} \begin{pmatrix} W_1^{\pm} \\ W_2^{\pm} \end{pmatrix} \quad (3.2)$$

In the scheme of mixing left and right charged vector bosons ($W^-$)the plus sign is chosen for particles, and ($W^+$)the minus sign is chosen for antiparticles at the sine in the top row.

It should be noted that in the scheme (3.2) we are considering there is an important difference compared to the commonly used scheme (3.1), which does not take into account the different sign of mixing for particles and antiparticles. In this scheme, we essentially introduce a difference in the coupling constants of ($\mathrm{W}^-$) and ($\mathrm{W}^+$), i.e. particles and antiparticles, which will lead to CP violation.

We have examined the unitarity problem for $0^+ \to 0^+$ transitions within the framework of the left-right manifest model and the extended left-right model. The results of the analysis are presented in Figure 9.

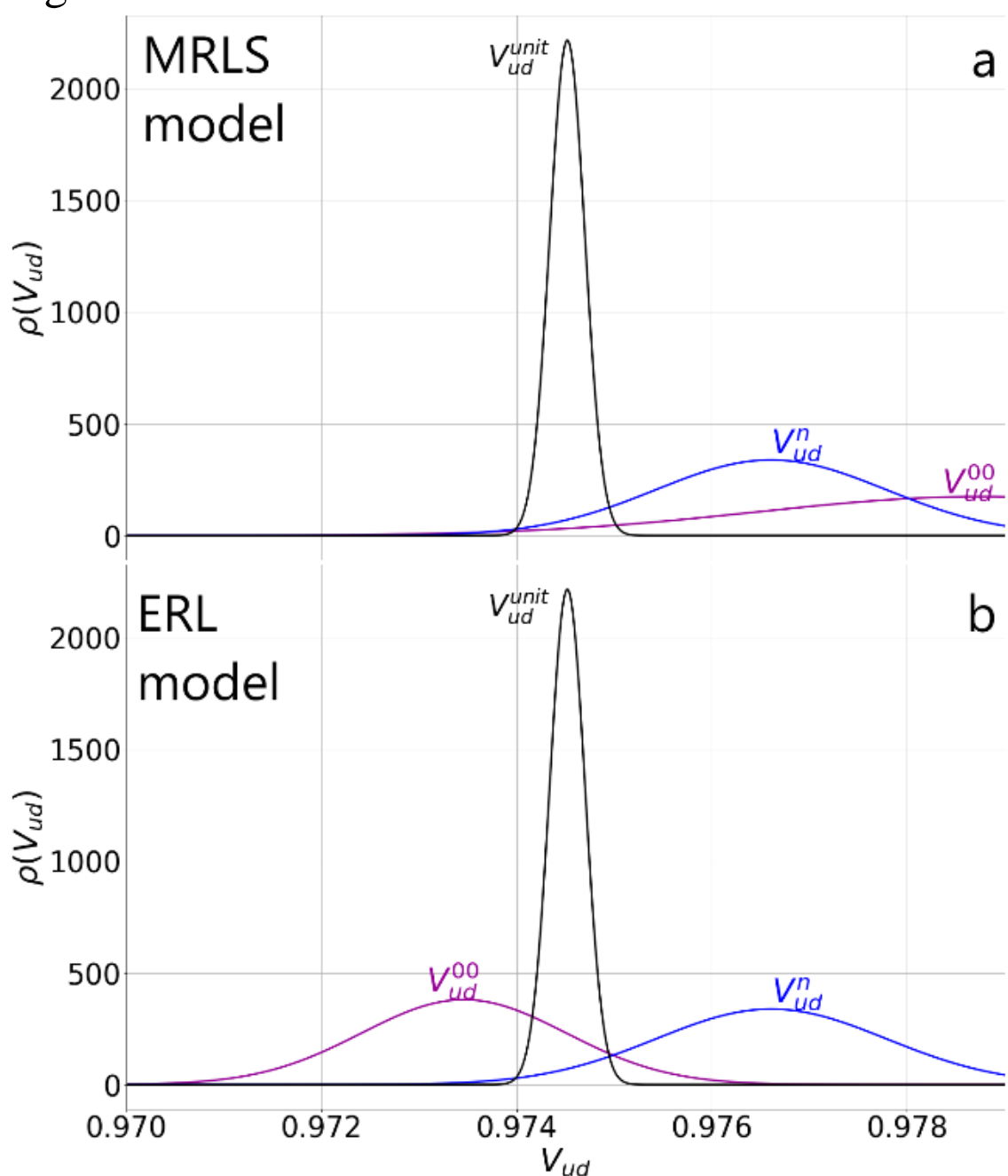

Fig. 9. Comparison of the values of $V_{ud}$ obtained from data on neutron decay, superallowed Fermi transitions, and the requirement of CKM matrix unitarity within the framework of two models. In all plots, $V_{ud}^{\mathrm{unit}} = 0.97452(18)$, derived from the unitarity condition and experimental values $V_{us} = 0.2243(8)$ and $|V_{ub}|^2 = 1.7 \times 10^{-5}$. (a) The left-right manifest model as presented in work [5]. (b) The extended left-right model, introduced in this work.

As can be seen, it is not possible to reconcile the values of $V_{ud}^{00}$ and $V_{ud}^{n}$ with $V_{ud}^{\mathrm{unit}}$ in either case. However, the case of the extended Standard Model is of interest due to the symmetric placement of $V_{ud}^{00}$ and $V_{ud}^{n}$ around $V_{ud}^{\mathrm{unit}}$. It is important to note that the value $V_{ud}^{00}$ is determined by the decay of $W^+$, while the value $V_{ud}^{n}$ is determined by the decay of $W^-$. This asymmetry between particles and antiparticles arises due to CP violation. In the extended left-right model, this asymmetry originates from the different signs of the mixing of the right-handed vector boson $W_R^{\pm}$ with the left-handed vector boson $W_L^{\pm}$.

In the presented scheme, transitions are allowed only between $W_1^-$ and $W_2^-$, as well as $W_1^+$ and $W_2^+$. All other transitions are prohibited due to the violation of the law of conservation of electric charge.

In this model, we consider left bosons. $W_1^-$ and $W_1^+$ as particles and antiparticles, respectively. Moreover, $W_1^-$ has a negative charge in accordance with the charge of the electron, appearing after the decay, and has a negative (left) parity in accordance with the negative (left) chirality of the antineutrino. For $W_1^+$ in accordance with the charge and spatial conjugation (CP) we have a positive charge in accordance with the charge of the positron, appearing after the decay, and have a positive (right) parity in accordance with the positive (right) chirality of the neutrino. Thus, we introduce into consideration the concept of P-parity for $W_1^-$ and $W_1^+$, identifying it with the P-parity of the neutrino during decay.

Then the right vector bosons $W_2^-$ and $W_2^+$ have opposite parity to the left vector bosons, but are antiparticles, so the charge parity changes. In accordance with charge and spatial conjugation (CP), we have:

$W_1^-$—left particle (C = -1, P =-1), CP=+1 (3.3)
$W_1^+$—left antiparticle (C =+1, P =+1), CP=+1
$W_2^-$—right particle (C =-1, P =-1), CP=+1
$W_2^+$—right antiparticle (C =+1, P =+1), CP=+1

C= -1, P= -1, CP= +1
$W_1^-$
s=1
$p_e$ $s_{\tilde{\nu}}$ $s_e$ $p_{\tilde{\nu}}$
C= +1, P= +1, CP= +1
$W_2^+$
s=1
$p_{e^+}$ $s_\nu$ $s_{e^+}$ $p_\nu$
VM
$W_1^+$
s=1
$p_\nu$ $s_\nu$ $s_{e^+}$ $p_{e^+}$
C= +1, P= +1, CP= +1
$W_2^-$
s=1
$p_{\tilde{\nu}}$ $s_{\tilde{\nu}}$ $s_e$ $p_e$
C= -1, P= -1, CP= +1
AM

Fig. 10. Mixing diagram between $W_1^-$and $W_2^-$, as well as $W_1^+$and $W_2^+$. VM is a mirror of vectors, AM is a mirror of axial vectors.

Note that in the final state, in the case of $W_1^-$ and $W_2^-$ the light antineutrino and heavy antineutrino are mixed, and in the case of $W_1^+$and , $W_2^+$the light neutrino and heavy neutrino are mixed.

$W_1^-$and $W_1^+$have positive CP parity.

$W_2^-$and $W_2^+$have positive CP parity.

Thus, CP parity is conserved for $W_1$and $W_2$, although these are particles of different masses. This is the most important feature of the left-right scheme, in which there is a mixing of a light antineutrino and a heavy antineutrino, as well as mixing of a light neutrino and a heavy neutrino, i.e. neutrinos of the same chirality.

The weak interaction Hamiltonian in the case where only vector and axial currents (V - A) are present can be represented in the well-known general form, for example - [7].

$$H_{V,A}^{N} = \bar{e}\gamma_{\mu}\left(C_V + C_V'\gamma_5\right)\nu \cdot \bar{p}\gamma_{\mu}n - \bar{e}\gamma_{\mu}\gamma_5\left(C_A + C_A'\gamma_5\right)\nu \cdot \bar{p}\gamma_{\mu}\gamma_5 n + h.c. \qquad (3.4)$$

In the presence of right currents with mixing parameters $\delta = m_1^2/m_2^2$, where $m_1$ is the mass of the light $W_1$ and $m_2$ is the mass of the heavy $W_2$.

The coefficients $C_V, C_V'$ and $C_A, C_A'$ within the framework of the simplest manifest left-right symmetric model: $g_R = g_L, V_{ud}^R = V_{ud}^L$, $\omega = 0$ are given by the following expressions [7].

$$C_V = g_V \frac{G_F V_{ud}}{\sqrt{2}}(1 - 2\zeta + \delta), \quad C_V' = g_V \frac{G_F V_{ud}}{\sqrt{2}}(1 - \delta) \qquad (3.5)$$

$$C_A = g_A \frac{G_F V_{ud}}{\sqrt{2}}(1 + 2\zeta + \delta), \qquad C_A' = g_A \frac{G_F V_{ud}}{\sqrt{2}}(1 - \delta)$$

According to the general formula from [7], in the absence of scalar and tensor contributions, the probability of transition (neutron decay) is proportional to the product of the phase space and the lifetime to the power of -1, i.e. a value $(f\tau)_n^{-1}$equal to

$$(f\tau)_n^{-1} = |M_F|^2(|C_V|^2 + |C_V'|^2) + |M_{GT}|^2(|C_A|^2 + |C_A'|^2) \qquad (3.6)$$

$$|C_V|^2 + |C_V'|^2 = |g_V G_F V_{ud}|^2(1 - \zeta)^2(1 + ((\delta - \zeta)/(1 - \zeta))^2).$$

$$|C_A|^2 + |C_A'|^2 = |g_A G_F V_{ud}|^2(1 + \zeta)^2(1 + ((\delta + \zeta)/(1 + \zeta))^2).$$

If we are only interested in quadratic contributions, we can write a simplified expression:

$$|C_V|^2 + |C_V'|^2 = |g_V G_F V_{ud}|^2(1 - \zeta)^2(1 + (\delta - \zeta)^2) \qquad (3.7)$$

$$|C_A|^2 + |C_A'|^2 = |g_A G_F V_{ud}|^2(1 + \zeta)^2(1 + (\delta + \zeta)^2) \qquad (3.8)$$

For $0 \leftrightarrow 0$ transitions that occur with the decay of the mixed state $W_1^+$ and $W_2^+$ we must choose the appropriate sign of the mixing angle. This means that in formula 3.7 for $0 - 0$ transitions we must change the sign in front of $\zeta$.

From the experiment we can extract only the matrix elements taking into account the mixing of vector bosons, i.e. $\tilde{V}_{ui}$, where $(i = d, s, b)$. This means that we must renormalize the matrix elements for the model-independent approach considered in the work of P. Herczeg [6]. In the linear approximation this means that for $0^+ \to 0^+$ transitions, i.e. for $W^+$

$$\tilde{V}_{ud}^{+} = V_{ud}^{+}(1+\zeta) = V_{ud}^{+(V)} \equiv V_{ud}^{00(V)},$$

and for the decay of a neutron, i.e. for $W^-$

$$\tilde{V}_{ud}^{-} = V_{ud}^{-}(1-\zeta) = V_{ud}^{-(V-A)} \equiv V_{ud}^{n(V-A)}.$$

$$(f\tau)_{00}^{-1} = |M_F|^2(|C_V|^2 + |C_V'|^2) = = |M_F|^2|g_V\ G_F V_{ud}|^2(1 + \zeta)^2(1 + (\delta + \zeta)^2) \qquad (3.9)$$

Thus, the matrix element within the left-right model, extracted from $0^+ - 0^+$ the transitions ($V_{ud}^{00LR}$) equals:

$$V_{ud}^{00LR} \simeq V_{ud}^{00(V)}\sqrt{[1+(\delta+\zeta)^2]} \qquad (3.10)$$

However, it is necessary to additionally take into account the effect of renormalization of the decay probability due to the additional decay channel through $W_2^+$. The total decay probability considering the mixing angle is proportional $1+\zeta^2$, so for renormalization it is necessary to introduce a factor $(1+\zeta^2)^{-1}$.

$$V_{ud}^{00LR} = V_{ud}^{00(V)}\sqrt{\frac{[1+(\delta+\zeta)^2]}{1+\zeta^2}} \qquad (3.11)$$

Note that $V_{ud}^{00LR}$ this is a matrix element extracted from the process of antiparticle decay $W_1^+$. $V_{ud}^{00LR} \equiv V_{ud}^{LR}(W^+)$

For the decay of a neutron

$$(f\tau)_n^{-1} = |M_F|^2(|C_V|^2 + |C_V'|^2) + |M_{GT}|^2(|C_A|^2 + |C_A'|^2)$$
$$= |M_F|^2|g_V G_F V_{ud}|^2(1 - \zeta)^2(1 + (\delta - \zeta)^2) +$$
$$+|M_{GT}|^2|g_A G_F V_{ud}|^2(1 + \zeta)^2(1 + (\delta + \zeta)^2), \qquad (3.12)$$

where $|M_F|^2 = 1$,$|M_{GT}|^2 = 3$

We take out $2|M_F|^2|g_V G_F V_{ud}|^2(1 - \zeta)^2$the common factor and for the neutron we get

$$(f\tau)_n^{-1} = |M_F|^2|g_V G_F V_{ud}|^2(1 - \zeta)^2 \times$$

$$\{(1 + (\delta - \zeta)^2) + \frac{|M_{GT}|^2 |g_A|^2}{|M_F|^2 |g_V|^2} \frac{(1 + \zeta)^2}{(1 - \zeta)^2} \times$$

$$(1 + (\delta + \zeta)^2)\}$$

(3.13)

where, as a first approximation $\widetilde{V_{ud}^{n}}^{2} = V_{ud}^2 (1 - \zeta)^2$,

$$\tilde{V}_{ud}^{-} = V_{ud}^{-}(1-\zeta) = V_{ud}^{-(V-A)}$$

$$\tilde{\lambda}_n^2 \equiv \frac{|g_A|^2}{|g_V|^2} \frac{(1+\zeta)^2}{(1-\zeta)^2},$$

which corresponds (taking into account the notations) to the definition $\lambda$ in formula (29) from the work of P. Herczeg [6]. $\frac{|M_{GT}|^2}{|M_F|^2} = 3$, $\tilde{\lambda}_n^2 \equiv \lambda_{n,V-A}^2$

Taking into account the quadratic terms, including the effect of renormalization of the decay probability due to the additional decay channel via $W_2^-$ :

$$\left(f\tau\right)_n^{-1} = G_F^2 \left|g_V\right|^2 (V_{ud}^{n(V-A)})^2 \left(1+3\lambda_{n,V-A}^2\right) \times$$
$$(1+\zeta^2)^{-1} \left\{1+(\delta^2+\zeta^2)+2\frac{\left(3\lambda_{n,V-A}^2-1\right)}{\left(3\lambda_{n,V-A}^2+1\right)}\delta\zeta\right\}$$

(3.14)

Let us introduce into consideration the matrix element $V_{ud}^{n,LR}$, defined according to the formula $G_F^2 \left|g_V\right|^2 (V_{ud}^{n,LR})^2 \left(1+3\lambda_{\exp,LR}^2\right) = \left(f\tau\right)_n^{-1}$ (3.15),

$\lambda_{\exp,LR}$– a value determined from the experiment using the formulas of the left-right model, taking into account the results of measurements: the neutron lifetime $\tau$ and decay asymmetries: $a, B$ and $A$. The relations for $\tau$ , $a, B$ and $A$ within the left-right model will be presented in the next section.

Thus, the matrix element within the left-right model extracted from the decay of a neutron $(V_{ud}^{nLR})$ is:

$$V_{ud}^{nLR} = V_{ud}^{n(V-A)} \times$$
$$\sqrt{\frac{1+3\lambda_{n,V-A}^2}{1+3\lambda_{\exp,LR}^2} \frac{[1+(\delta^2+\zeta^2)+2\frac{\left(3\lambda_{n,V-A}^2-1\right)}{\left(3\lambda_{n,V-A}^2+1\right)}\delta\zeta]}{(1+\zeta^2)}} \quad (3.16)$$

Note that $V_{ud}^{nLR}$ is a matrix element extracted from the decay process with particle $W_1^-$. $V_{ud}^{nLR} \equiv V_{ud}^{LR}(W^-)$.

Since the mixing angle has a different sign for particles and antiparticles, it is possible that the values of $V_{ud}^{LR}(W^-)$ and $V_{ud}^{LR}(W^+)$ may differ, in addition, these are vector and axial-vector transitions. The calculation carried out using formulas (3.11), (3.16) is presented in Fig. 11 and confirms this difference.

The calculation carried out using formulas (3.11) and (3.16) is presented in Figure 9 (b) and confirms this difference. Thus, due to CP violation, a splitting of the values of $V_{ud}$ for $W^-$ and $W^+$ occurs.

We assume that the unitarity of the CKM within the left-right model should be satisfied for the mean value of the sum of squares vector $V_{ud}^{LR}(W^+)$ and axial-vector $V_{ud}^{LR}(W^-)$ matrix elements. According to the left-right model, we use different signs in the mixing matrix for $W^-$ and $W^+$. As a result, we have got that the elements of the CKM matrix are split. However, the average value of $V_{ud}$ is conserved because it represents the average probability.

$$(V_{ud}^{LR})^2 = \frac{1}{2}[(V_{ud}^{LR} W^+)^2 + (V_{ud}^{LR} W^-)^2] \quad (3.17)$$

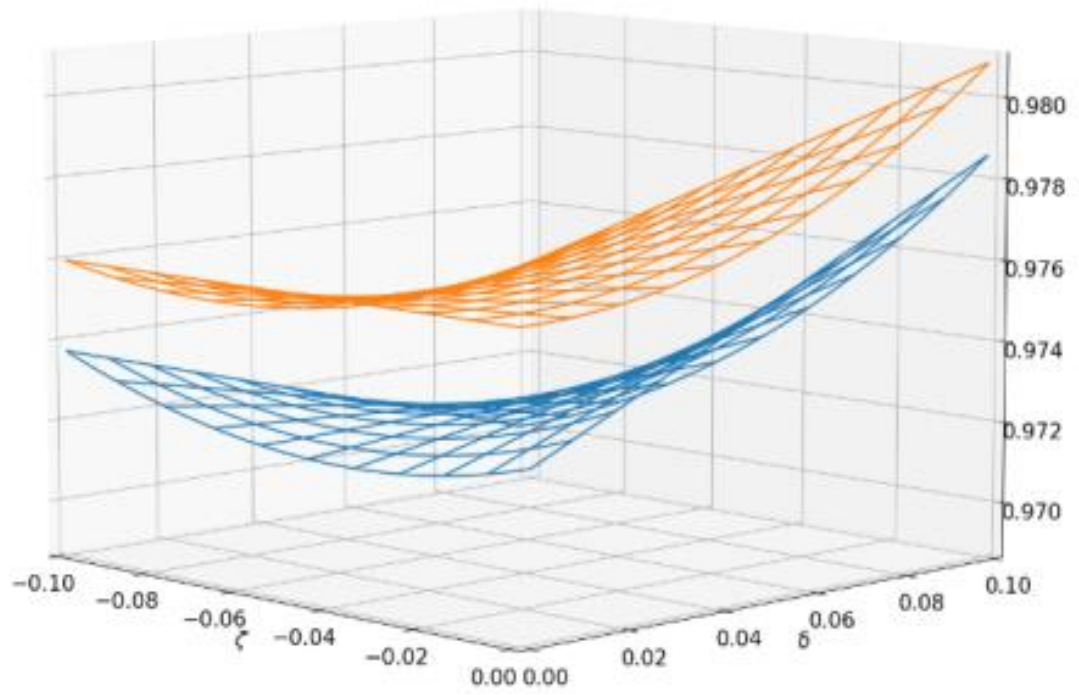

Fig. 11. Values of axial-vector $V_{ud}^{LR}(W^-)$ and vector $V_{ud}^{LR}(W^+)$ on the plane $\delta$ and $\zeta$calculated using formulas (3.11), (3.16).

Above, the notations $\tilde{V}_{ud}$ from the work of P. Herczeg [6] were used, which are identical to the notations presented below, and the values of these quantities have already been determined from the experiment.

$$\tilde{V}_{ud}^{n} \equiv V_{ud}^{n(V-A)} = 0.97477(37) \; [27]. \quad (3.18)$$

$$\tilde{V}_{ud}^{00} \equiv V_{ud}^{00(V)} = 0.97367(32) \; [26] \quad (3.19)$$

## 4. ANALYSIS OF NEUTRON DECAY WITHIN THE LEFT-RIGHT MODEL

Experimental studies of neutron decay provide extremely important information for testing the Standard Model of particle physics (SM). Within the SM, neutron decay is described by the V - A version of the weak interaction theory. The probability of decay is determined by a number of parameters extracted from the experiment. The general formula for describing neutron decay within the V - A version of the weak interaction theory can be represented by the following expression.

$$\frac{d^3\Gamma}{dE_e d\Omega_e d\Omega_\nu} = \frac{1}{2(2\pi)^5} G_F^2 {V_{ud}}^2 \left(1+3\lambda^2\right) p_e E_e (E_0 - E_e)^2 \times$$
$$\times\left[1 + a\frac{\vec{p}_e \cdot \vec{p}_\nu}{E_e E_\nu} + b\frac{m_e}{E_e} + \frac{\langle\vec{\sigma}_n\rangle}{\vec{\sigma}_n}\left(A\frac{\vec{p}_e}{E_e} + B\frac{\vec{p}_\nu}{E_\nu} + D\frac{\vec{p}_e \times \vec{p}_\nu}{E_e E_\nu}\right)\right] \quad (4.1)$$

The decay probability is proportional to the square of the coupling constant $G_F$ with a fairly good accuracy obtained from the muon decay and is also proportional to the square of the matrix element $V_{ud}$ of the CKM matrix. The matrix element is determined fairly accurately from the unitarity of the CKM matrix, given that the matrix elements $V_{us}$ and $V_{ub}$ are determined from the decay of strange and charmed mesons. The remaining parameters in this formula $a, A, B$ and $D$, as well as the neutron lifetime $\tau$ must be determined from experiment. The parameter $\lambda$, which is the ratio of the axial and vector weak coupling constants $G_A/G_V$, must be determined using the experimental values of $a, A, B$ and $\tau$, based on the V - A version of the weak interaction theory. The ratio $G_A/G_V$ is renormalized by the strong interaction of quarks, and therefore differs from unity. For the V - A version of the theory, the Fierz term $b$ is zero. In the following we will assume that there is no T violation, since $D = -1.2(2.0)10^{-4}$[19].

For experimental results $a, A, B$ and $\tau$ within the framework of the V - A version of the theory, the following formulas can be written below.

$$\tau_{\text{exp}} = \frac{4905.7}{V_{ud}^2(1+3\lambda^2)}$$
$$a_{\text{exp}} = \frac{(1-\lambda^2)}{(1+3\lambda^2)}$$
$$A_{\text{exp}} = -\frac{2\lambda(\lambda+1)}{1+3\lambda^2} \quad (4.2)$$
$$B_{\text{exp}} = \frac{2\lambda(\lambda-1)}{1+3\lambda^2}$$

And the experimental results:

$$\tau_{\exp} = 877.75(35)$$
$$a_{\exp} = -0.10402(82)$$
$$A_{\exp} = -0.11958(21) \quad (4.3)$$
$$B_{\exp} = 0.9807(30)$$
$$V_{ud}^{unit} = 0.97452(18)$$

From the formulas for $A_{\exp}$ and $B_{\exp}$ it follows that

$$A_{\exp}/B_{\exp} = \frac{1+\lambda}{1-\lambda}.$$

For $\tau$, we use the most precise value of the neutron lifetime from the PDG, although the latest refined result, $\tau = 877.82 \pm 0.22^{+0.20}_{-0.17}$ s, is presented in work [32]. The refined result does not affect the results of our analysis within the current level of precision.

Using these relations and experimental results, taking into account internal and external radiation corrections, it is possible to calculate the corresponding values of the parameter $\lambda$. The results of these calculations are presented in Fig. 12.

When calculating the value of $\lambda$ (Fig. 12), the ratio of the coefficients $A/B$ is used, so in this case the relative change in $\lambda$ is an order of magnitude smaller than the relative change in the coefficient B. Thus, a relative change in $B$ by $\mathcal{O}(10^{-3})$ leads to a relative change in $\lambda$ of the order of $10^{-4}$. This means that the value of $\lambda_{opt} = -1.2738 \pm 0.001$ used in the analysis for right-hand currents will not shift beyond its error limits when the value of $B$ changes by a value of the order of $10^{-3}$.

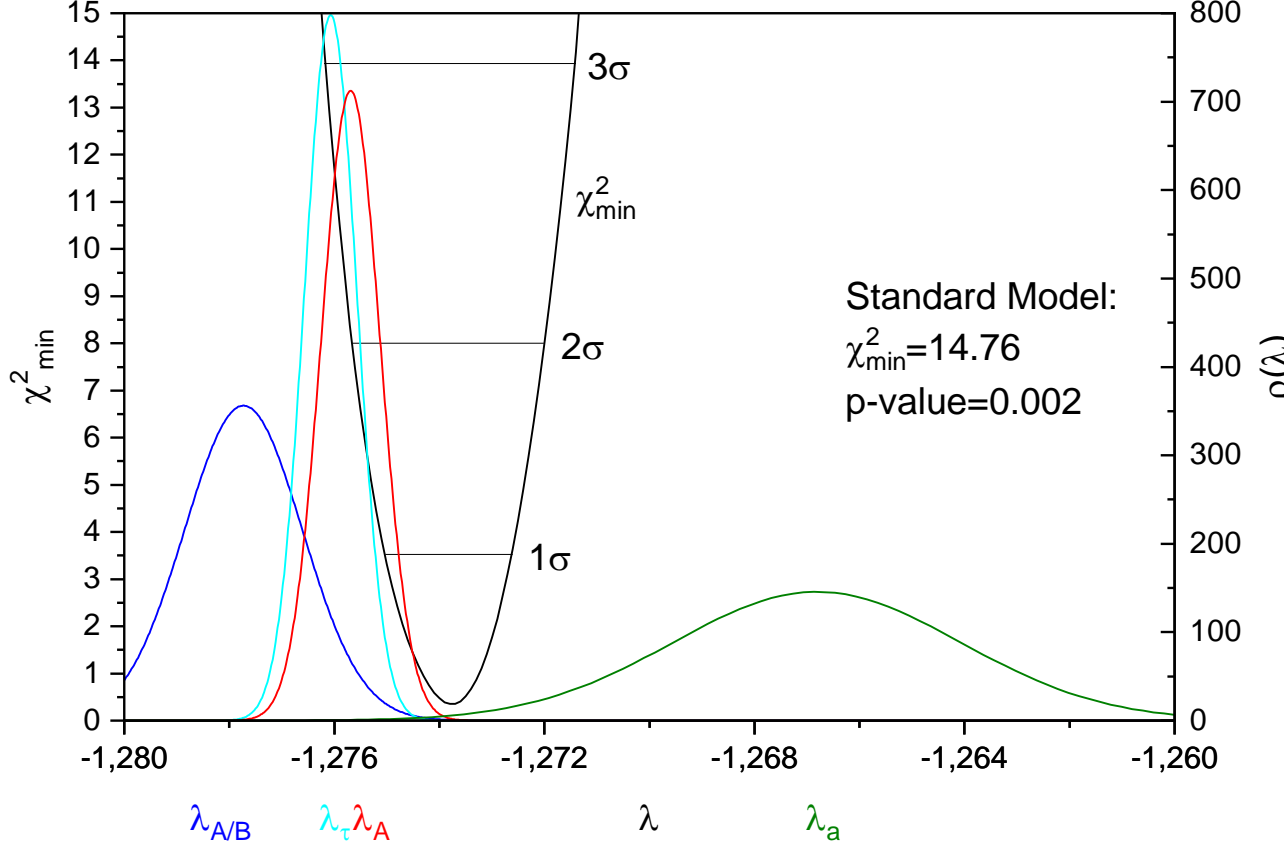

Fig.12. Results of calculating the $\lambda$ parameter value within the framework of the V-A theory of weak interaction.

It can be seen that the proposed description within the V-A of theory turns out to be unsatisfactory. For the best fit using only the Standard Model $\chi^2$=14.76, p-value=0.002. Thus, there is a deviation from the SM by more than 3σ. In connection with the current situation in determining the parameter, $\lambda$ we decided to conduct an analysis of the possible contribution of right currents in weak interaction, i.e. for the presence of a right vector boson $W_R$. The formulas describing the possible mixing $W_L$ and $W_R$ are presented in detail in the review [7].

The task of the planned analysis of the possible contribution of right currents in the weak interaction is to find the best agreement between the experimental quantities $\tau_0$, $a_0$, $A_0$, $B_0$with a single set of parameters $\lambda$, $\delta$and $\zeta$. In this analysis, the value of $V_{ud}^n = 0.97477(37)$ [27], obtained from neutron decay, is used. This value differs from $V_{ud}^{\text{unit}} = \sqrt{1 - V_{us}^2 - V_{ub}^2} = 0.97452(18)$ by an amount that is negligible within the limits of experimental uncertainty.

$$\tau_{\exp} \pm \Delta\tau_{\exp} = \frac{4905.7}{V_{ud}^2[1+(\delta-\zeta)^2+3\lambda^2(1+(\delta+\zeta)^2)]}$$

$$a_{\exp} \pm \Delta a_{\exp} = \frac{(1-\lambda^2)[1+(\delta+\zeta)^2]-4\delta\zeta}{(1+3\lambda^2)[1+(\delta+\zeta)^2]-4\delta\zeta}$$

$$A_{\exp} \pm \Delta A_{\exp} = -\frac{2\lambda[\lambda(1-(\delta+\zeta)^2)+(1-\delta^2+\zeta^2)]}{1+(\delta-\zeta)^2+3\lambda^2(1+(\delta+\zeta)^2)}$$

$$B_{\exp} \pm \Delta B_{\exp} = \frac{2\lambda[\lambda(1-(\delta+\zeta)^2)-(1-\delta^2+\zeta^2)]}{1+(\delta-\zeta)^2+3\lambda^2(1+(\delta+\zeta)^2)}$$

(4.4)

Within the framework of the manifest left-right symmetric model, the ratio of the difference between the experimental values and the values obtained by the standard $V-A$ model to the value of the standard model can be represented (in an expansion in $\delta$ and $\zeta$ not higher than the second order) by the following expressions. In essence, this is a transition to a representation in relative deviations from the V-A theory.

Each of the equations gives three lines on the plane $\delta$, $\zeta$- for positive, negative and zero values of the measurement error. To obtain the optimal set of parameters $\lambda$, $\delta$ and $\zeta$ calculations were carried out on the plane $\delta$, $\zeta$ according to the formulas given above for different values of $\lambda$. As expected, for extreme values of $\lambda = -1.2677$ and $\lambda = -1.2784$ it is impossible to find a point on the plane $\delta$, $\zeta$, satisfying the above equations. This is the upper and lower graph in Fig. 13. However, it turned out to be possible to obtain the optimal set of parameters $\lambda$, $\delta$ and $\zeta$ for the value of $\lambda_{opt} = -1.2738 \pm 0.0011$. This is the middle graph in Fig. 13.

$$\frac{\tau_{\exp} \pm \Delta\tau_{\exp} - \tau_{V-A}}{\tau_{V-A}} \simeq -\left[\delta^2+\zeta^2+2\frac{(3\lambda^2-1)}{(3\lambda^2+1)}\delta\zeta\right]$$

(4.5)

$$\frac{a_{\exp} \pm \Delta a_{\exp} - a_{V-A}}{a_{V-A}} \simeq -\frac{16}{(1-\lambda^2)(1+3\lambda^2)}\delta\zeta$$

(4.6)

$$\frac{A_{\exp} \pm \Delta A_{\exp} - A_{V-A}}{A_{V-A}} \simeq -2\delta^2 - 2\delta\zeta\frac{[6\lambda^3+3\lambda^2-1]}{(\lambda+1)(1+3\lambda^2)} - 2\frac{\lambda}{\lambda+1}\zeta^2$$

(4.7)

$$\frac{B_{\exp} \pm \Delta B_{\exp} - B_{V-A}}{B_{V-A}} \simeq -2\delta^2 - 2\delta\zeta\frac{[6\lambda^3-3\lambda^2+1]}{(\lambda-1)(1+3\lambda^2)} - 2\frac{\lambda}{\lambda-1}\zeta^2$$

(4.8)

The calculations were carried out using formulas (4.4) and formulas (4.5-4.8), no significant difference was found.

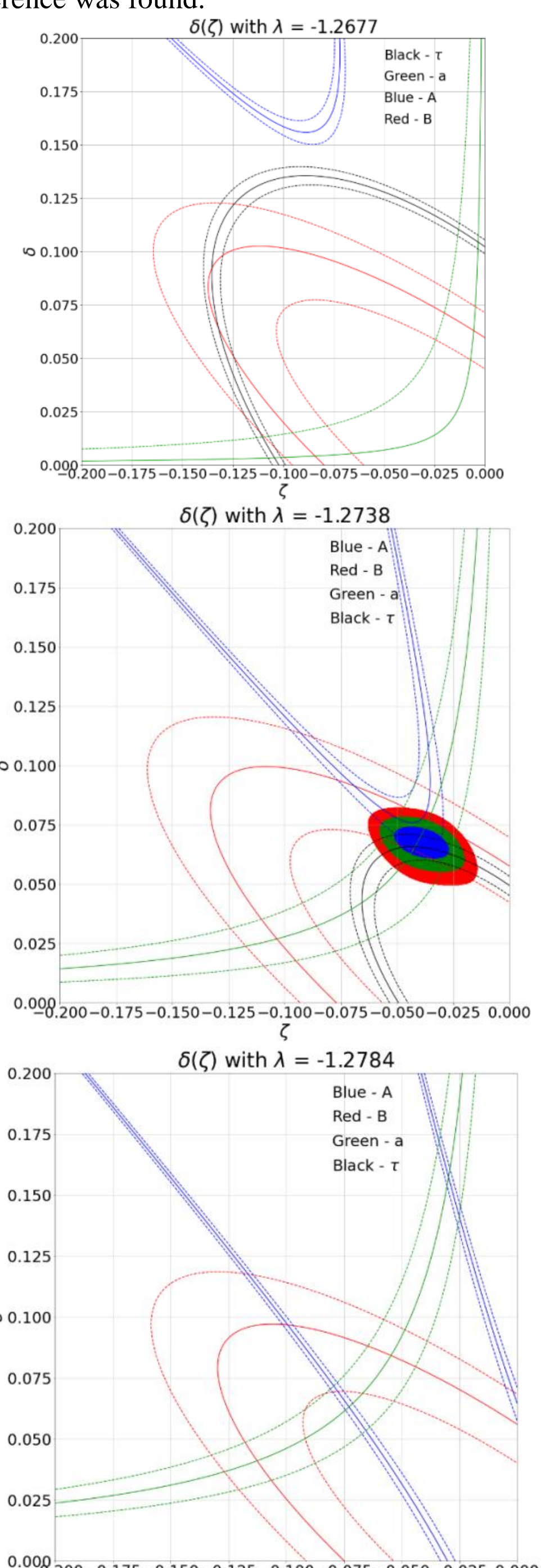

Fig. 13. Dependence of parameter $\delta$ on parameter $\zeta$ from equations (4.4) for the results of measuring quantities $a, A, B$ and $\tau$ for different values of parameter $\lambda$. Upper and lower graphs for extreme values of $\lambda = -1.2677$ and $\lambda = -1.2784$, where it is impossible to find a point on the plane $\delta$, $\zeta$, satisfying the above equations. Middle graph for the optimal set of parameters $\lambda$, $\delta$ and $\zeta$ for the value $\lambda_{opt} = -1.2738$.

A more accurate search for optimal values was made by the $\chi^2$ method. The result of the analysis for $\lambda_{opt}$ is shown in Fig. 12, and for $\delta_{opt}$ and $\zeta_{opt}$ the result is shown in Fig. 14. Due to this analysis, optimal values and the accuracy of their determination were determined. The analysis was made using the most accurate data from PDG 24 [19]. These are the neutron lifetime $\tau_n$, electron asymmetry $A$, electron-neutrino correlation $a$.

Using the most accurate experimental data for $a, \tau$ and PDG data for $A, B$ we obtained:

$$\lambda_{\text{opt}} = -1.2738 \pm 0.0012.$$

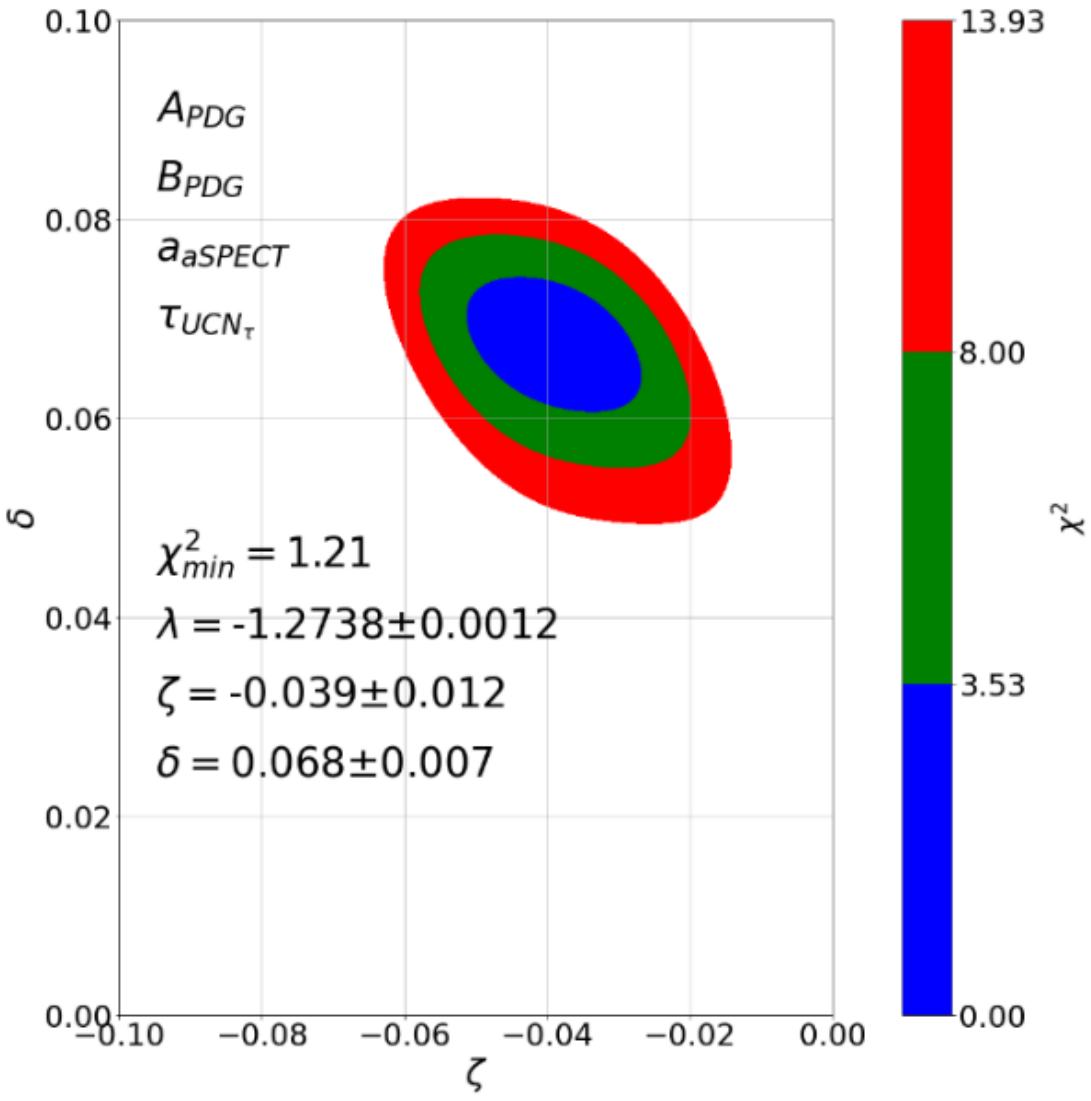

Fig. 14. Optimal values of the parameters $\lambda$, $\delta$ and $\zeta$, obtained by the method $\chi^2$, using experimental data on neutron decay for $a, A, B$ and $\tau$. $\chi^2_{min} = 1.21$.

As a result of the analysis within the framework of the manifest left-right symmetric model, using experimental data on neutron decay, an estimation of the mass of the right vector boson and the mixing angle was made: $M_{W_R} = 308^{+17}_{-15}\text{GeV}$, $\zeta_{opt} = -0.039 \pm 0.012$.

In conclusion of this section, we present a correlation matrix for the parameters used in this analysis.

$$\begin{array}{cccc} & \zeta & \delta & \lambda \\ \zeta & 1 & -0.33 & -0.19 \\ \delta & -0.33 & 1 & 0.90 \\ \lambda & -0.19 & 0.90 & 1 \end{array} \qquad (4.9)$$

## 5. CONSIDERATION OF THE ACCURACY OF CALCULATION OF RADIATION CORRECTIONS

Taking into account radiative corrections [33,34,35] plays an important role in precision experiments to measure the neutron lifetime - $\tau_n$ and the correlation coefficients of the neutron decay asymmetry – $A$ and $a$ to search for possible deviations from the predictions of the standard model [27,36,37]. In addition to radiative corrections, corrections for weak magnetism and the final mass of the nucleon (non-zero proton momentum) are also taken into account [38]. The neutron lifetime $\lambda$ and the square of the modulus of the matrix element $V_{ud}$ of the CKM matrix are related by the following expression [36].

$$\frac{1}{\tau_n} = \frac{G_F^2 {V_{ud}}^2}{2\pi^3} m_e^5 (1 + 3\lambda_n^2)(1 + \Delta_{\text{R}}) f \qquad (5.1),$$

where $f$ is a factor related to the phase space [38] and takes into account the Fermi function, as well as the finite size and mass of the nucleus $f = 1.6887(1)$, $\Delta_{\text{R}}$ are the full radiative corrections. The expression for $f$ can be found in [29]. The leading order of the full radiative corrections $\Delta_{\text{R}}$ is of the order of the fine structure constant $\alpha/\pi$ [34,35]. The so-called master formula is given in [27]:

$${V_{ud}}^2 = \frac{4905.7(1.7)s}{\tau_n(1+3\lambda_n^2)} \qquad (5.2),$$

which is obtained from (5.1) with the value of radiative corrections $\Delta_{\text{R}}$ equal to $0.03947(32)$. There is also the new calculation result equal to $0.03983(27)$ from [39]. But it does not significantly change the results of our calculations, so we will use the old data for formula (5.2).

Radiative corrections in the form of a factor $(1 + \Delta_{\text{R}})$ can be represented as a product $(1 + \Delta_{\text{R}}) = (1 + \delta_R)(1 + \Delta_R^{\text{V}})$, where the contribution of the “outer” radiative correction $\delta_R$ was obtained by computing QED one-loop and bremsstrahlung diagrams, assuming that nucleons are point-like, and the contribution of the “inner” radiative correction $\Delta_R^{\text{V}}$ that is sensitive to the details of SM electroweak theory and probes the hadronic structure of the nucleon [29]. In the [27] the master formula (5.2) was obtained from (5.1) with $\Delta_R^V = 0.02426(32)$. The new calculations of $\Delta_R^V$ based on data-driven dispersion relation can be found in [40,41,42] and based on lattice QCD in [43]. The average result is $\Delta_R^V = 0.02467(27)$ [39].

When measuring the correlation coefficient $A$ in work [17], when fitting the experimental value of asymmetry, the necessary corrections are introduced for the final mass of nucleons (recoil protons) and weak magnetism, $g_V - g_A$ interference. In addition, the experimental value of asymmetry $A$ takes into account radiation corrections, the value of which is of the order of $10^{-3}$ and the error is $10^{-4}$.

When measuring the correlation coefficient $a$ in [22], the proton energy spectrum was fitted using formulas from [44], which take into account radiation corrections of the order of the fine structure constant. Corrections for weak magnetism and Coulomb interaction were also taken into account.

In [29], expressions are given for the dependence of the coefficients $a$, $A$ and $B$ on the electron energy with an accuracy of up to contributions of the order of $10^{-4}$

It turns out that the influence of radiative corrections on the value of the coefficient $B$ is smaller [45, 27] than for the neutron lifetime.

In the analysis conducted above, the precision of the radiative corrections was already taken into account through the error of 1.7 s in formula (5.2). However, to demonstrate their role, calculations can be carried out with level of precision that is higher than real, effectively taking these corrections into account a second time.

Table 1

| Size | Experimental error % | Correction % | Correction error % | Work |
|---|---|---|---|---|
| $\tau_n$ | 0.040 | 3.947 | 0.032 | [27] |
| $A$ | 0.176 | -0.100 | 0.01 | [17] |
| $a$ | 0.788 | 0.005 | 0.005 | [46] |
| $B$ | 0.306 | <0.1 | <0.1 | [31] |

Below in Fig. 15 the results obtained for the PDG data are presented using experimental data on neutron decay for $a, A, B$ and $\tau$ taking into account accuracy of calculation of radiative corrections: $\lambda_{opt} = -1.2738 \pm 0.0012$, $\delta_{opt} = 0.070 \pm 0.010$, $M_{W_R} = 304^{+24}_{-20} GeV$, $\zeta_{opt} = -0.039 \pm 0.014$

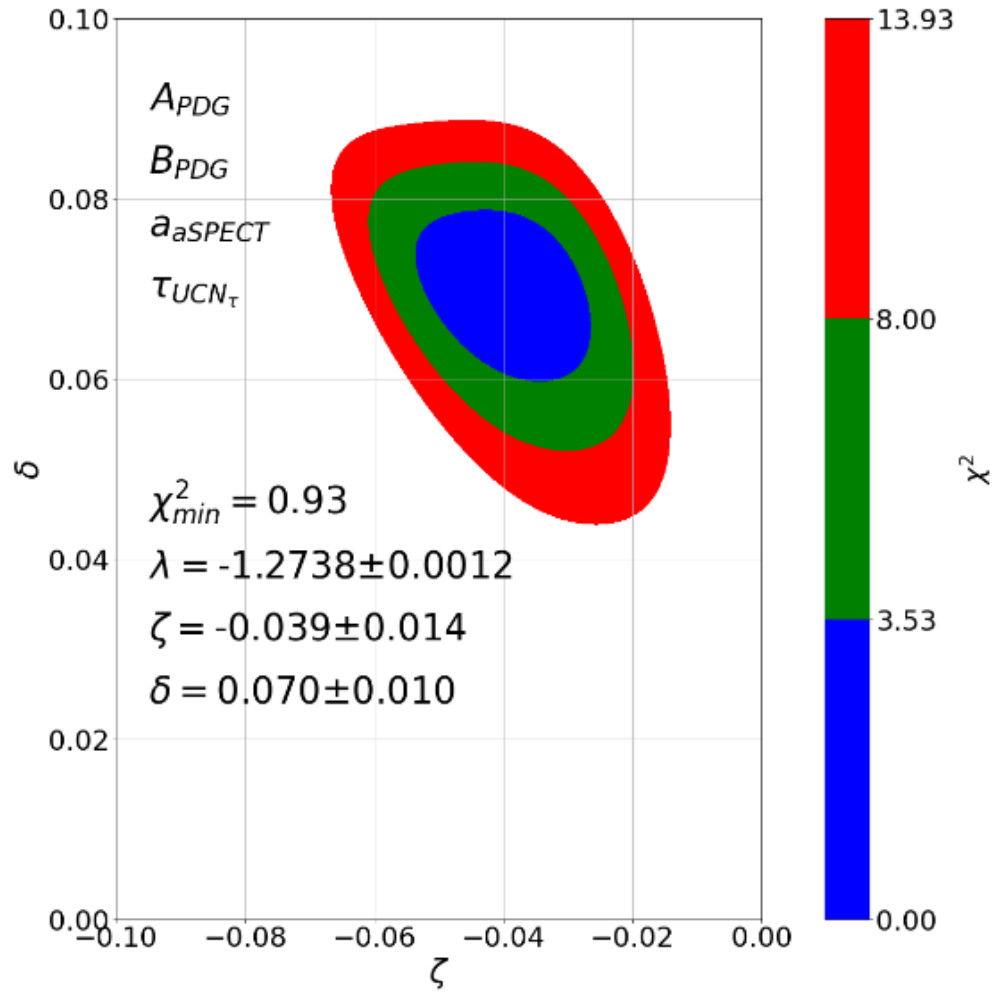

Fig.15. Optimal values of the parameters δ and ζ obtained by the $\chi^2$ method, using experimental data from neutron decay for $a, A, B$, and τ, taking into account an exaggerated level of precision for radiative corrections. The 1σ, 2σ, and 3σ regions are indicated. $\chi^2_{min} = 0.93$.

As can be seen, the precision of determining the main parameters of the left-right model has remained practically unchanged, while the value of the minimal $\chi^2$ has decreased because the errors were overstated.

Here we present a refined result for the parameters $\delta, \zeta$, which was previously presented in our work [47] with the following values $M_{W_R} \approx 319^{+26}_{-20}$ GeV and the mixing angle with $W_L$: $\zeta = -0.034 \pm 0.013$.

In conclusion, it is important to note that the analysis of the neutron decay was made within the framework of the conventional manifest left-right symmetric model considered in [3-5, 7]. The extended version of the model with CP violation was not used, since we have so far considered the process involving only $W^-$. The extended version of the left-right model will be required when we consider transitions where the process $0^+ - 0^+$ occurs with $W^+$. Then it will be possible to compare the direct transition from the d-quark to the u-quark and the reverse transition from the u-quark to the d-quark and study the CP-violating asymmetry of the processes.

## 6. CONSTRAINTS ON THE $W^{\prime}$ MASS OBTAINED IN COLLIDER EXPERIMENTS

PDG has significantly stronger (>6 TeV) constraints on the hypothetical vector boson $W'$ obtained in accelerator experiments. It is necessary to distinguish between these constraints and it should be noted that they are in different subsections in PDG not by chance. The point is that constraints on the mass $W_R$ can be obtained in the decay of neutrons and nuclei, these studies are a method of precision measurements alternative to research at colliders. This is a whole line of experiments that were discussed earlier, as well as a large number of theoretical works devoted to the analysis of these experimental data for possible deviations from the SM [3–6, 28, 48–54]. Our work complements this list and is focused mainly on the right currents, i.e. on the search for an admixture of the right vector boson.

The constraints on the hypothetical vector boson $W'$obtained in accelerator experiments mean that no resonances have been detected in the various decay modes up to energy of 6 TeV. at a cross-section statistically available at the present time. Of course, the signals $e\nu$ from $e\mu$ the decay of the right vector boson are no different from the signals from the decay of the left vector boson, so the search is carried out with the background of similar events and with the background of other decay modes associated with the interaction of quarks and gluons.

Let us calculate the resonance production cross section of $W_L$ and $W_R$ bosons in high-energy proton collisions. We use formula (4) from [55] and the coupling constants to quarks and leptons from [5] to calculate the cross section. It is important to note that the left resonance is a process occurring with a small mixing angle $\zeta$ of the left vector boson $W_L$ with the right vector boson $W_R$, and the right resonance is a process with a small mixing angle $\zeta$ of the right vector boson $W_R$ with the left vector boson $W_L$. For this reason, the coupling constants in Table 2 for the left and right resonances are different.

Below is a modified formula for calculating the cross section from [55] and a table of coupling constants for the left and right resonance from [5].

$$\sigma(s)=\frac{\pi\alpha_W^2}{6}\,V_{ud}^2\times\Bigg[\frac{a_{ud}^{L^2}a_{lv}^{L^2}+a_{ud}^{R^2}a_{lv}^{R^2}+a_{ud}^{R^2}a_{lv}^{L^2}+a_{ud}^{L^2}a_{lv}^{R^2}}{\left(s-m_{W_L}^2\right)^2+\gamma_{W_L}^2m_{W_L}^2}+2a_{ud}^{L}a_{lv}^{L}\frac{\left(s-m_{W_L}^2\right)\left(s-M_{W_R}^2\right)+\gamma_{W_L}^2\Gamma_{W_R}^2}{\left(\left(s-m_{W_L}^2\right)^2+\gamma_{W_L}^2m_{W_L}^2\right)\left(\left(s-M_{W_R}^2\right)^2+\Gamma_{W_R}^2M_{W_R}^2\right)}+\frac{a_{ud}^{L^2}a_{lv}^{L^2}+a_{ud}^{R^2}a_{lv}^{R^2}+a_{ud}^{R^2}a_{lv}^{L^2}+a_{ud}^{L^2}a_{lv}^{R^2}}{\left(s-M_{W_R}^2\right)^2+\Gamma_{W_R}^2M_{W_R}^2}\Bigg] \tag{6.1}$$

Table 2. Coupling constants to quarks and leptons for left and right resonances.

| For the left resonance | For the right resonance |
|---|---|
| $a_{ud}^{L^2}a_{lv}^{L^2}=(\cos^2\zeta\ +\delta\sin^2\zeta)^2$ | $a_{ud}^{L^2}a_{lv}^{L^2}=(\sin^2\zeta+\delta\cos^2\zeta)^2$ |
| $a_{ud}^{R^2}a_{lv}^{R^2}=(sin^2\zeta+\delta\, cos^2\zeta)^2$ | $a_{ud}^{R^2}a_{lv}^{R^2}=(sin^2\zeta+\delta\, cos^2\zeta)^2$ |
| $a_{ud}^{R^2}a_{lv}^{L^2}=(\delta-1)^2\, sin^2\zeta\, cos^2\zeta\, e^{2i\omega}$ | $a_{ud}^{R^2}a_{lv}^{L^2}=(\delta-1)^2\, sin^2\zeta\, cos^2\zeta\, e^{2i\omega}$ |
| $a_{ud}^{L^2}a_{lv}^{R^2}=(\delta-1)^2\sin^2\zeta\cos^2\zeta\, e^{-2i\omega}$ | $a_{ud}^{L^2}a_{lv}^{R^2}=(\delta-1)^2\sin^2\zeta\cos^2\zeta\, e^{-2i\omega}$ |

$$\sigma(s)=\frac{\pi\alpha_W^2}{6}\,V_{ud}^2\times\Bigg[\frac{(cos^2\zeta+\delta\sin^2\zeta\,)^2+(\sin^2\zeta+\delta\cos^2\zeta\,)^2+(\delta-1)^2\sin^2\zeta\cos^2\zeta\, e^{2i\omega}+(\delta-1)^2\sin^2\zeta\cos^2\zeta\, e^{-2i\omega}}{\left(s-m_{W_L}^2\right)^2+\gamma_{W_L}^2m_{W_L}^2}+2(\sin^2\zeta+\delta\cos^2\zeta\,)\,\frac{\left(s-m_{W_L}^2\right)\left(s-M_{W_R}^2\right)+\gamma_{W_L}^2\Gamma_{W_R}^2}{\left(\left(s-m_{W_L}^2\right)^2+\gamma_{W_L}^2m_{W_L}^2\right)\left(\left(s-M_{W_R}^2\right)^2+\Gamma_{W_R}^2M_{W_R}^2\right)}+\frac{(\sin^2\zeta+\delta\cos^2\zeta\,)^2+(\sin^2\zeta+\delta\cos^2\zeta\,)^2+(\delta-1)^2\sin^2\zeta\cos^2\zeta\, e^{2i\omega}+(\delta-1)^2\sin^2\zeta\cos^2\zeta\, e^{-2i\omega}}{\left(s-M_{W_R}^2\right)^2+\Gamma_{W_R}^2M_{W_R}^2}\Bigg]$$

$$=\frac{\pi\alpha_W^2}{6}\,V_{ud}^2\times\Bigg[\frac{(1+\delta^2)+4\delta\sin^2\zeta\cos^2\zeta+(\delta-1)^2\sin^2\zeta\cos^2\zeta\,(e^{2i\omega}+e^{-2i\omega})}{\left(s-m_{W_L}^2\right)^2+\gamma_{W_L}^2m_{W_L}^2}+2(\sin^2\zeta+\delta\cos^2\zeta)\,\frac{\left(s-m_{W_L}^2\right)\left(s-M_{W_R}^2\right)+\gamma_{W_L}^2\Gamma_{W_R}^2}{\left(\left(s-m_{W_L}^2\right)^2+\gamma_{W_L}^2m_{W_L}^2\right)\left(\left(s-M_{W_R}^2\right)^2+\Gamma_{W_R}^2M_{W_R}^2\right)}+\sin^2\zeta\,\frac{2\sin^2\zeta+4\delta\cos^2\zeta+2\dfrac{\cos^4\zeta}{\sin^2\zeta}\delta^2+(\delta-1)^2\cos^2\zeta\,(e^{2i\omega}+e^{-2i\omega})}{\left(s-M_{W_R}^2\right)^2+\Gamma_{W_R}^2M_{W_R}^2}\Bigg]$$

Neglecting the small terms containing $\zeta$ and $\delta$, we obtain a simplified expression without taking into account the interference effect. As can be seen, the resonance term for the right $W_R$ contains the order suppression factor $\zeta^2\approx 1.4\cdot 10^{-3}$. This circumstance is the main reason why the resonance of the right $W_R$ has not been detected in collider experiments.

$$\sigma(s)=\frac{\pi\alpha_W^2}{6}\,V_{ud}^2\times\Bigg[\frac{1}{\left(s-m_{W_L}^2\right)^2+\gamma_{W_L}^2m_{W_L}^2}+\sin^2\zeta\,\frac{2\sin^2\zeta+4\delta\cos^2\zeta+2\dfrac{\cos^4\zeta}{\sin^2\zeta}\delta^2+(\delta-1)^2\cos^2\zeta\,(e^{2i\omega}+e^{-2i\omega})}{\left(s-M_{W_R}^2\right)^2+\Gamma_{W_R}^2M_{W_R}^2}\Bigg] \tag{6.2}$$

Let us calculate the cross-section of resonance processes using the following values of the quantities included in this expression: $\alpha_W=\frac{1}{137}, V_{ud}=0.97477,\ m_{W_L}=80.385$ GeV, $\gamma_{W_L}=2.141$ GeV, $M_{W_R}=303.8$ GeV, $\Gamma_{W_R}=11$ GeV, $\zeta=-0.039, \delta=0.07$ . Below in Fig. 16 we present the results of the calculation of the production cross-section in relative units depending on the transverse mass at theoretical resonance widths: $\gamma_{W_L}=2.141$ GeV, $\Gamma_{W_R}=11$ GeV. The upper figure is without CP violation at $\omega=0$, the lower figure is with CP violation at $\omega=\pi/2$. In Fig. 17 we present the results of the calculation taking into account the instrumental broadening in the experiment.

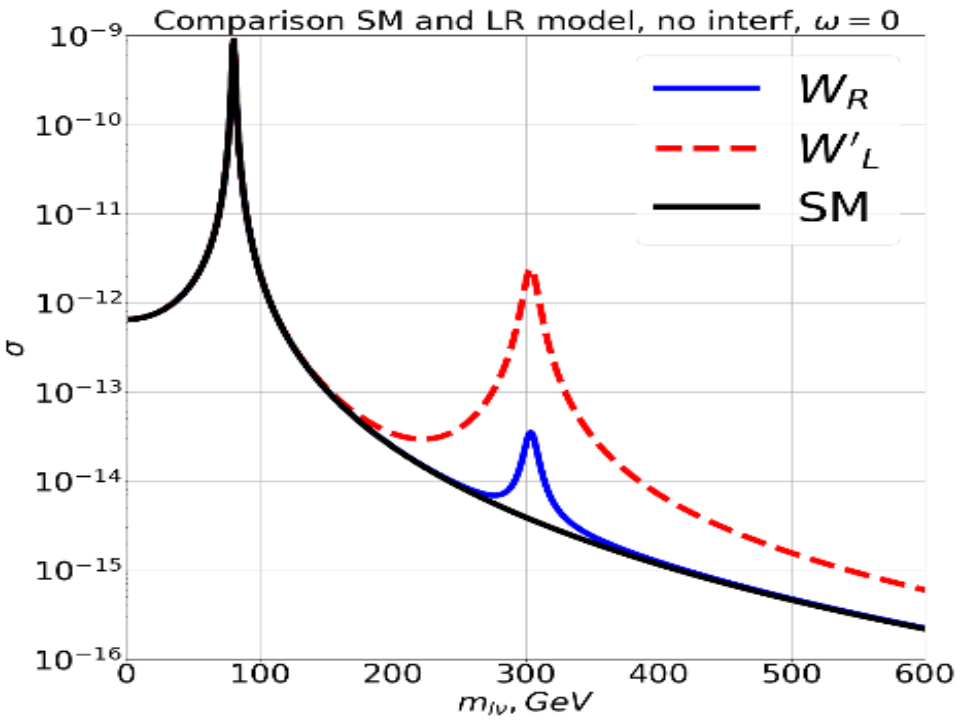

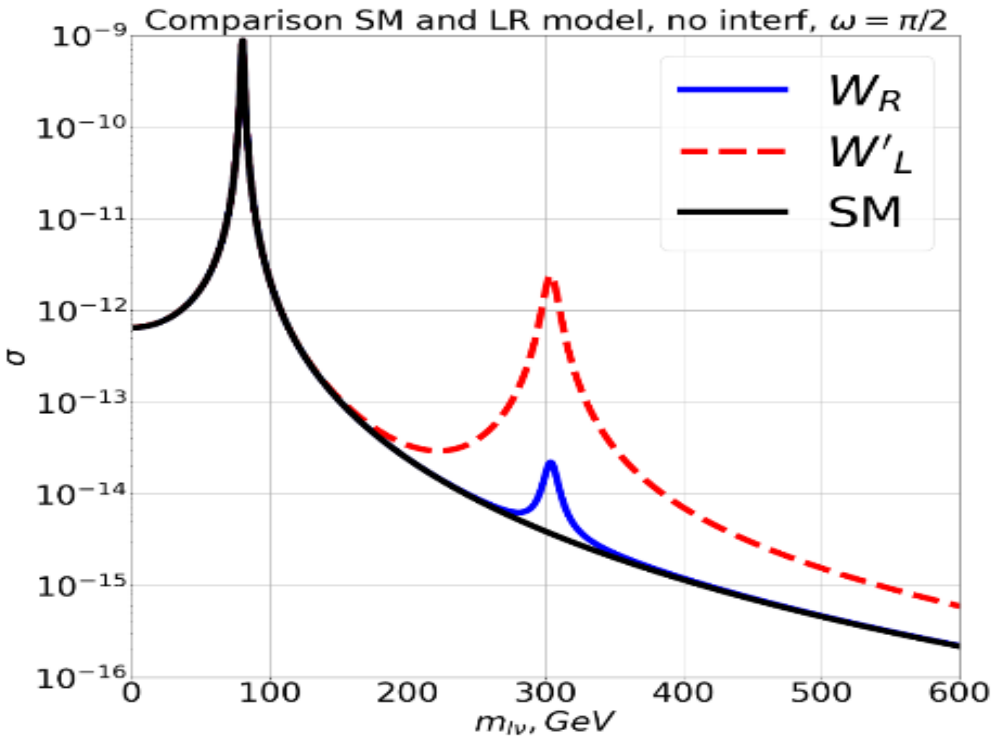

Fig. 16. Relative cross-section of the $W_R$ and $W_L'$ production process at theoretical resonance widths: $\gamma_{W_L} = 2.141$ GeV, $\Gamma_{W_R} = 11$ GeV. The upper figure is for $\omega = 0$, the lower figure is for $\omega = \pi/2$.

Fig. 17 presents the resonance widths are increased by 5 times to take into account the instrumental broadening in the experiment.

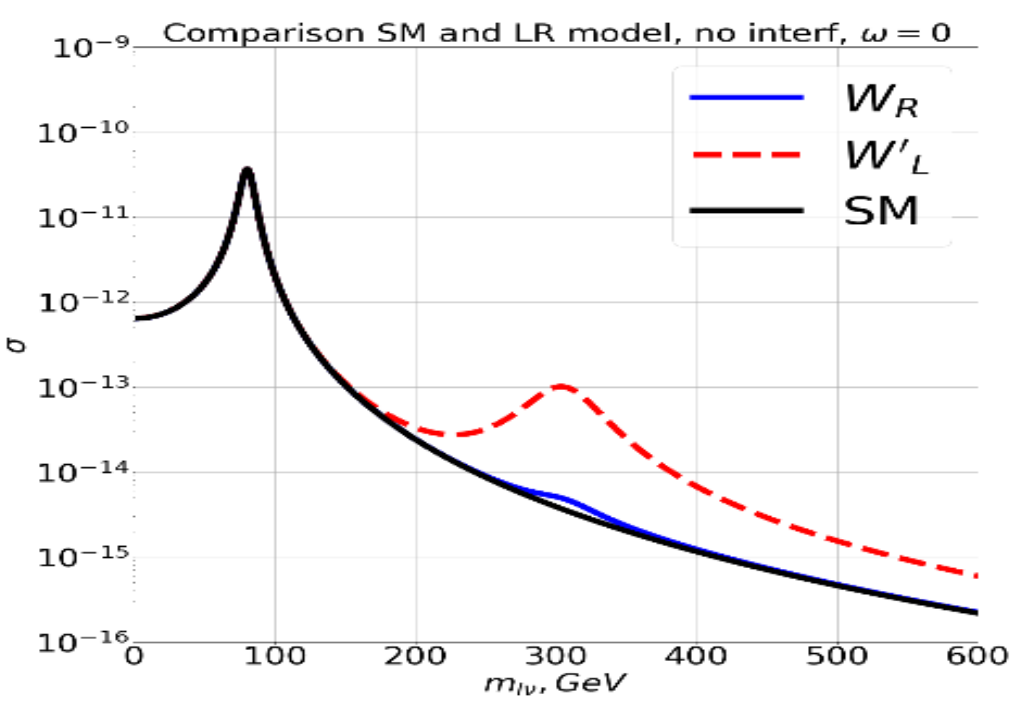

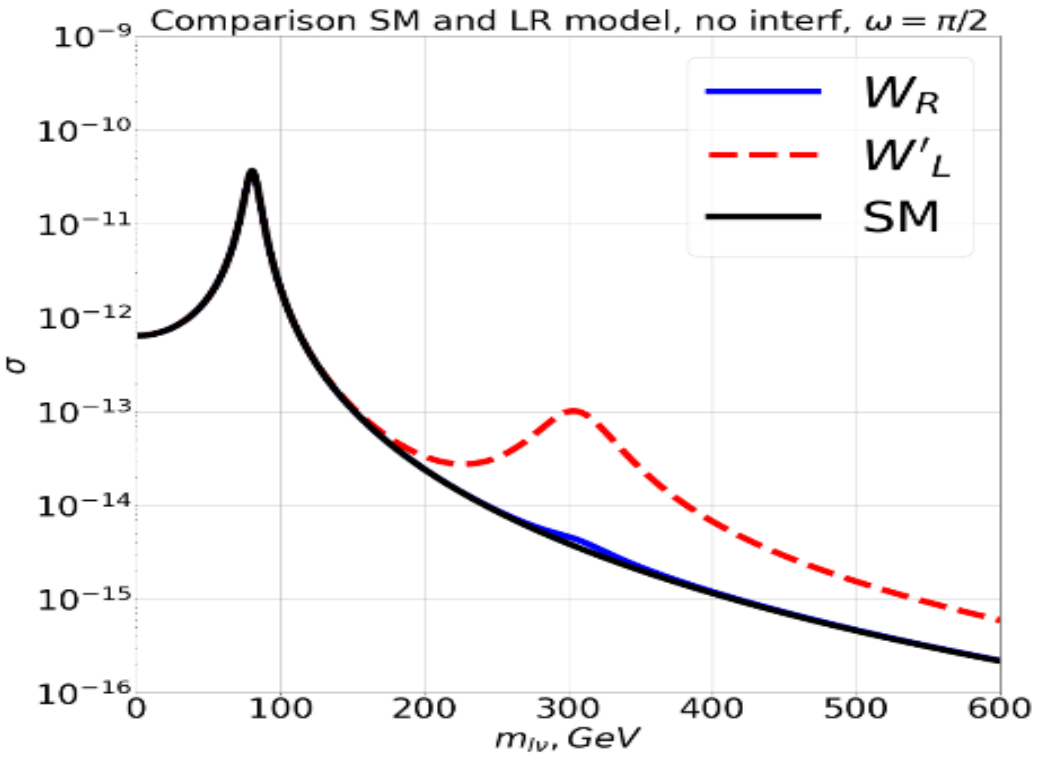

Fig. 17. Relative cross-section of the $W_R$ and $W_L'$ production process at theoretical resonance widths: $\gamma_{W_L} = 10.7$ GeV, $\Gamma_{W_R} = 55$ GeV. The upper figure is for $\omega = 0$, the lower figure is for $\omega = \pi/2$.

As can be seen, the reason why the right resonance $W_R$ was not detected in the collider experiments is that the resonance term for the right $W_R$ contains an order suppression factor $\zeta^2 \approx 1.4 \cdot 10^{-3}$ and that the instrumental broadening of the resonances significantly reduces the sensitivity of the experiment.

Below, in Figure 18, we present a comparison of the calculation results with experimental data for the experiment at the Tevatron at Fermilab from publication [56] and for the ATLAS experiment [57] at CERN.

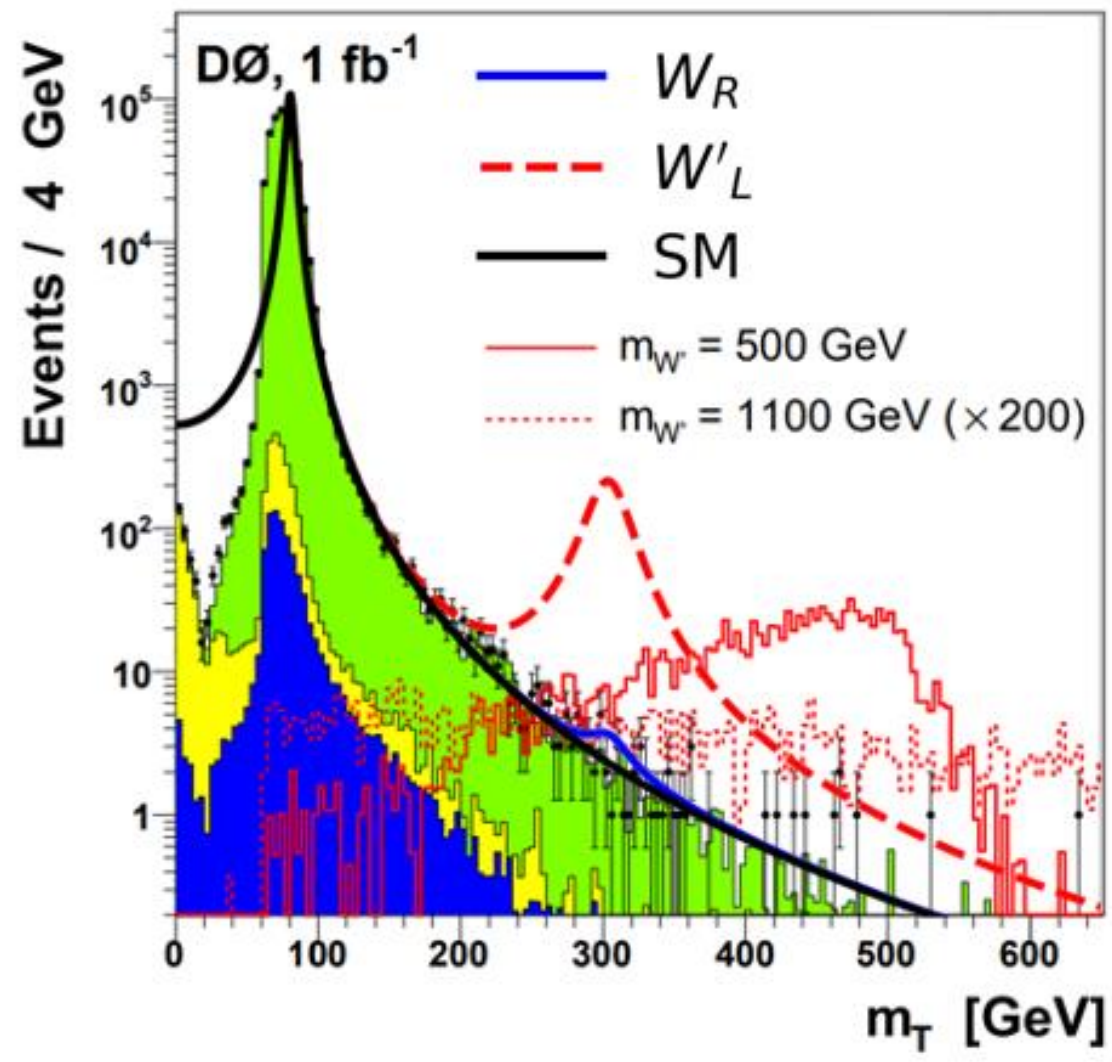

Fig. 18. The process of registration of the left vector boson at the Tevatron in Fermilab from the [56]. The resonance from the left vector boson is shown in green on the left. An example of the calculated signals for the resonance $W_L'$ with masses 500 GeV and 1100 GeV are shown on the right. The results of our calculations are also shown here: 1) the black bold line shows the calculation according to the SM, 2) the dotted red line shows the resonance $W_L'$ with a mass 304 GeV, 3) the bold blue line shows the expected resonance $W_R$ with a mass $M_{W_R} = 304 \pm 24$ GeV.

From this comparison, it follows that the expected effect from $W_R$ with a mass of $M_{W_R} = 304$ GeV could not be detected in the experiment at the Tevatron at Fermilab, as this effect lies below the statistical significance of the experimental data.

Let us now consider the results of the ATLAS experiment [57], where the statistical precision is expected to be higher. The results of the ATLAS experiment [57] are shown in Figure 19. Unfortunately, the region corresponding to the left resonance is not displayed on this plot, so the normalization was carried out only based on its tail. The total count of events above an energy of 100 GeV is indicated by the black stepped line with experimental data points in Figure 19. The same plot also shows a black normalization line for the Standard Model (SM) based on the tail of the left resonance. The white area below the black stepped line corresponds to the count of $e\nu$ events from the tail of the left resonance. Below that is a red area representing events involving the production of a top quark, followed by additional possible event types.

Our calculation results are presented on this plot for a resonance $W'$ with a mass of 304 GeV as a red dashed line. Below it is a barely noticeable blue line, corresponding to the expected resonance with a mass of $M_{W_R} = 304$ GeV. It should be noted that in the resonance region, in addition to statistical errors from the left resonance, there are systematic errors associated with event identification, which are an order of magnitude larger. In this case, the statistics would have been sufficient to identify the resonance if it were not for the systematic errors of the experiment, which are an order of magnitude larger than the statistical errors.

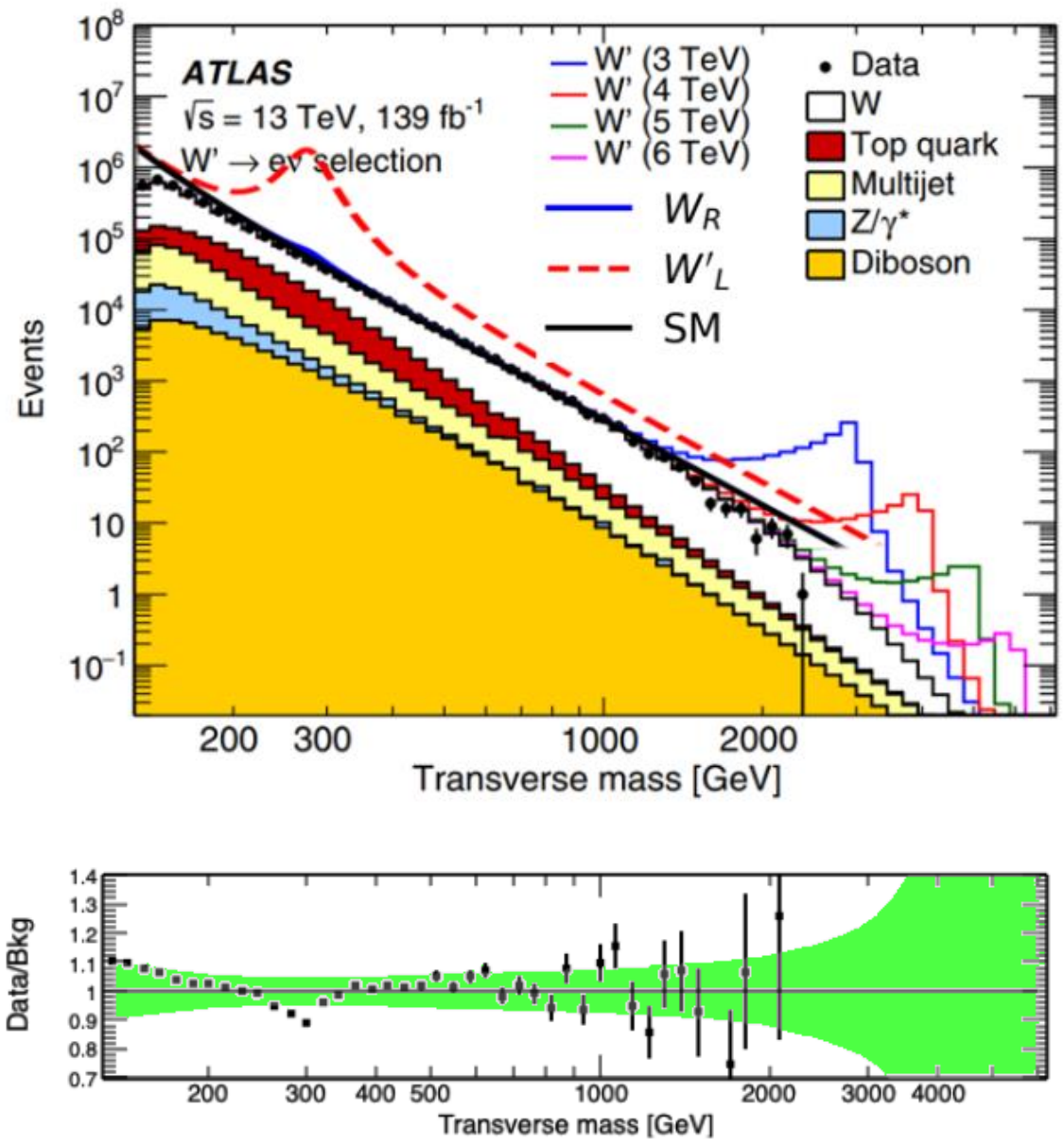

Fig. 19. Results of the ATLAS experiment [57]. The published results of the experiment are presented. The results of our calculations are also shown here: 1) the calculation according to the SM is shown by the black bold line, 2) the resonance $W_L'$ with the mass 304 GeV is shown by the dotted red line, 3) the expected resonance $W_R$ with the mass $M_{W_R} = 304 \pm 24$ GeV is shown by the bold, barely noticeable blue line. Systematic errors are indicated in green on the bottom panel.

The possibility of detecting right-handed currents in accelerator experiments was also discussed in work [58]. In it, the authors introduce the parameter ξ, which corresponds to the mixing angle ζ from the extended left-right model. According to formula (29) from [58], $|\xi_{ud}| < 0.04$ with 90% confidence, which can be compared to our result for the mixing angle $|\zeta| = 0.039 \pm 0.014$. The results are consistent with each other within $1.64\sigma$.

Thus, in this section it was shown that the collider experiments do not contradict the results of our analysis about the possible existence of a right vector boson $M_{W_R} = 304^{+24}_{-20}$ GeV and a mixing angle with $W_L$: $\zeta = -0.039 \pm 0.014$.

## 7. RESULTS OF DETERMINING THE MATRIX ELEMENT $V_{ud}$ WITHIN THE LEFT-RIGHT MODEL OF $0^+ - 0^+$ TRANSITIONS AND NEUTRON DECAY

Now we should move on to the analysis of the situation with the superallowed Fermi $0^+ - 0^+$ transitions. These transitions occur due to the decay of $W^+$. In this process, the spin of the nucleus, equal to zero, and the positive parity of the nucleus are preserved, so this is a vector transition. The axial part is absent, and the parameter $\lambda$ is not included in the definition of the quantity $(f\tau)^{00}$. It is this value that was carefully determined in the works J. C. Hardy and I. S. Towner [26]. They obtained average values $f\tau$ for each of the 21 transitions that have a complete data set and then took into account the radiative and isospin corrections that break the symmetry. Fifteen of these values $f\tau$ have an accuracy of 0.3% or better, and all take the same value within the statistics, as expected from the conservation of vector current. Their average value, $f\tau$, combined with the muon lifetime, gives $V_{ud}$ a quark mixing element of the CKM matrix of 0.97373 ± 0.00031. Thus, the value of the matrix element obtained within the V - A theory $V^{00}_{ud,SM} = 0.97373 \pm 0.00031$ and for neutron $V^n_{ud} = 0.97477(37)$[27].

Using optimal parameter values $\lambda_{opt} = -1.2738 \pm 0.0012$, $\delta_{opt} = 0.070 \pm 0.010$, $\zeta_{opt} = -0.039 \pm 0.014$ and formulas 3.11, 3.16 we obtain values $V_{ud}$ for $0^+ \rightarrow 0^+$transitions and for neutron decay: $V^{00LR}_{ud} = 0.97346 \pm 0.00105$ $V^{nLR}_{ud} = 0.97661 \pm 0.00118$. These results are shown in Fig. 20.

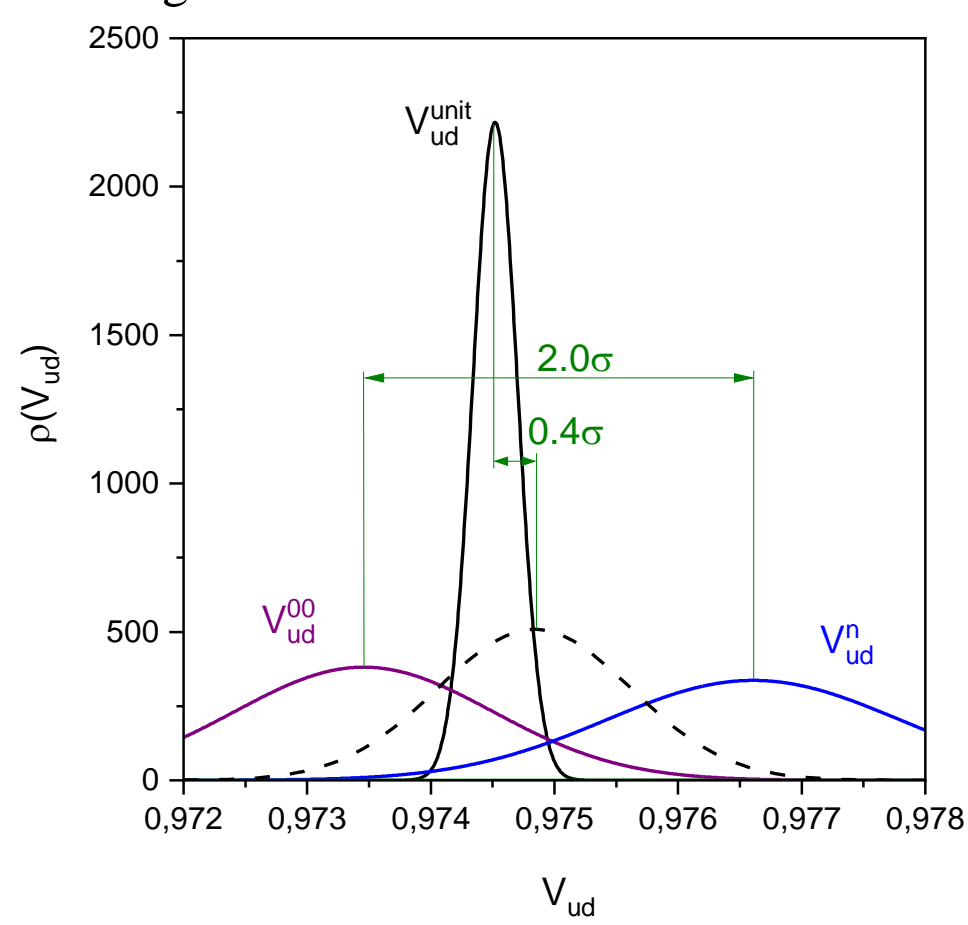

Fig. 20. Distribution for quantities $V_{ud}$ for $0^+ \rightarrow 0^+$ transitions from neutron decay within the left-right model, as well as $V_{ud}$ from the unitarity condition - $V^{unit}_{ud} = 0.97452 \pm 0.00018$.

Discrepancy between values $V^{00LR}_{ud}$ and $V^{nLR}_{ud}$within the left-right model, is $2.0\sigma$. A the deviation of their mean value from unitarity is $0.4\sigma$. After considering the parameters of the left-right symmetric model, the difference between $V_{ud}^{nLR}$ and $V_{ud}^{00LR}$ increased by 3 times, but the error also increased by 3 times due to the errors $\delta, \zeta$ and $\lambda_{opt}$. Therefore, the difference between $V_{ud}^{nLR}$ and $V_{ud}^{00LR}$ is $2.0\sigma$.

It should be noted that the discrepancy between the values of $V_{ud}^{00LR}$ and $V_{ud}^{nLR}$ can be interpreted as a violation of CP invariance. since $V_{ud}^{nLR}$ corresponds to the transition of the d quark to the u quark, and $V_{ud}^{00LR}$ corresponds to the transition of the u quark to the d quark (Fig. 21). Note that the difference is that the neutron decay occurs via a negative vector boson in the mixed state $W_1^-(W_2^-)$, and the proton decays in the nucleus via a positive vector boson in the mixed state $W_1^+(W_2^+)$, and the sign of the mixing angle is opposite.

The CP violation parameter can be determined according to the Standard Model scheme, where to calculate the asymmetry, the probability of the process occurring via a positive vector boson is subtracted from the probability of the process occurring via a negative vector boson:

$$A_{p-n} = \frac{(V_{ud}^{00LR})^2 - (V_{ud}^{nLR})^2}{(V_{ud}^{00LR})^2 + (V_{ud}^{nLR})^2} = (-3.2 \pm 1.6) \cdot 10^{-3} (2.0\sigma) \quad (7.1)$$

Unfortunately, the accuracy of determining this parameter is still insufficient, but it is surprising that the value is in order of magnitude the values of the CP violation parameters in $K$-meson decays. However, the sign of asymmetry is opposite. Note that this is a baryon asymmetry that has not been observed experimentally before.

In this circumstance, it is useful to analyze the CP violation processes in $K$-meson decays within the framework of the extended left-right model, using the parameters $\delta$ and $\zeta$.

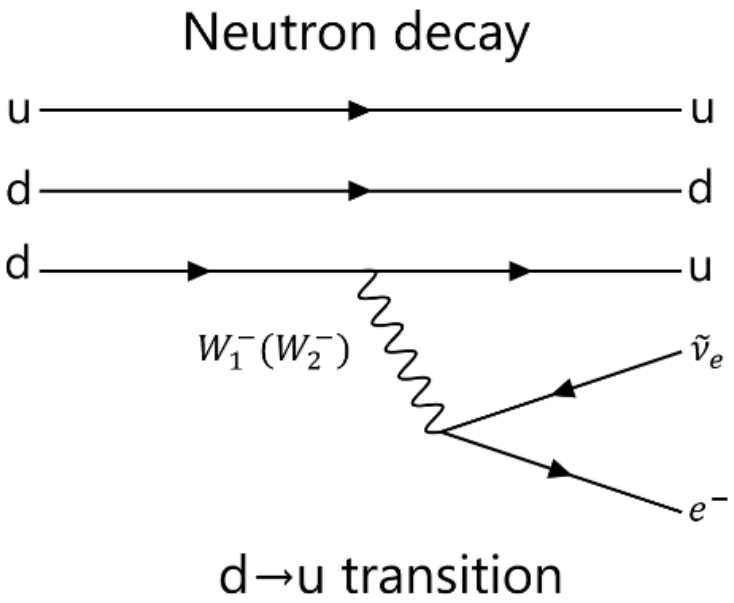

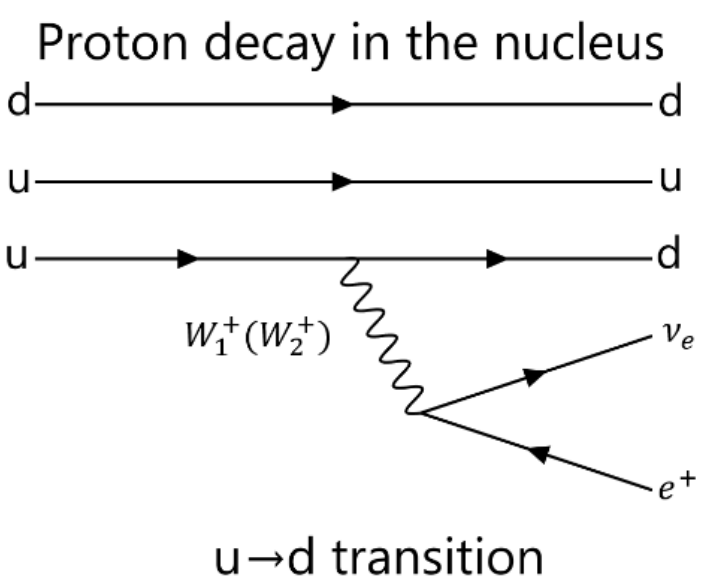

Fig. 21. The process of neutron decay and the process of proton decay in the nucleus. Direct transition from d-quark to u-quark and reverse transition from u-quark to d-quark.

## 8. ANALYSIS OF CP VIOLATION PROCESSES IN $K$-MESON DECAYS WITHIN THE EXTENDED LEFT-RIGHT MODEL USING THE PARAMETERS $\delta$ AND $\zeta$

In the process of oscillations of the $K^0\bar{K}^0$ system, a decay into the state $e^-\pi^+\bar{\nu}$ or into the state $e^+\pi^-\nu$ can occur.

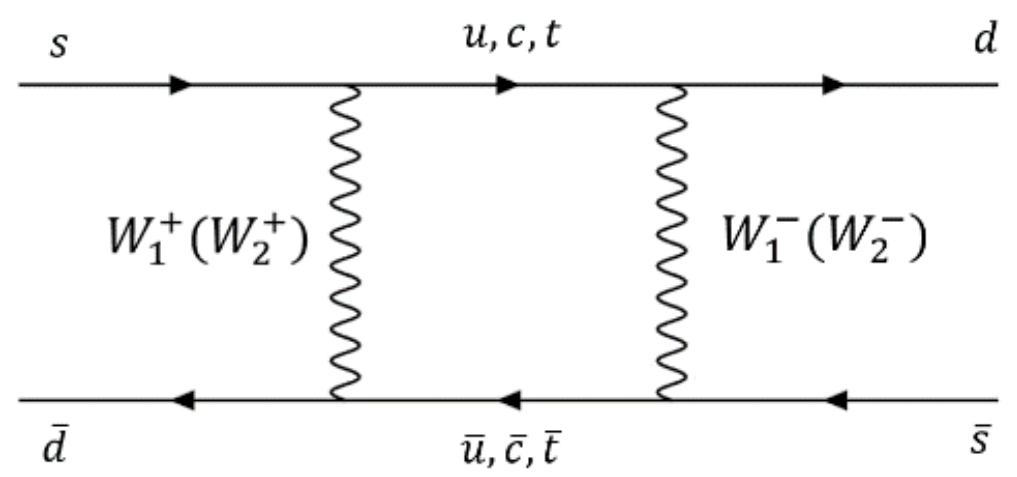

The weak interaction Hamiltonian with only vector currents can be represented in the same general form as for $0^+ \leftrightarrow 0^+$ transitions. However, K-mesons are pseudoscalar particles with zero spin and negative parity, so there are transitions $0^- \leftrightarrow 0^-$. Therefore, there is a $\zeta$ sign change compared to $0^+ \leftrightarrow 0^+$ transitions. For $0^- \leftrightarrow 0^-$ kaon-antikaon transitions we must write the Lagrangian with axial current:

$$H_V^N = \bar{e}\gamma_\mu (C_V + C_V'\gamma_5)\nu \cdot \bar{\pi}\gamma_\mu K^0 \quad (8.1),$$

where the decay of $W_1^-(W_2^-)$ is associated with the relation

$$\left|C_V\right|^2 + \left|C_V'\right|^2 = G_F^2 \left|V_{us}\right|^2 \left(1 + \left(\delta - \zeta\right)^2\right) \quad (8.2),$$

And decay of $W_1^+(W_2^+)$ associated with the relation

$$\left|C_V\right|^2 + \left|C_V'\right|^2 = G_F^2 \left|V_{us}\right|^2 \left(1 + \left(\delta + \zeta\right)^2\right) \quad (8.3)$$

The amplitude (8.1) can be written as [58]

$$M \simeq f_+ \left(\bar{u}(3)\gamma_\mu \left(C_V + C_V'\gamma_5\right)\upsilon(2)\right) p^\mu \quad (8.4),$$

where $f_+$ – form factor, $p$ - total momentum of pion and kaon, $u$ – positron (electron), $\upsilon$ – neutrino (antineutrino)

$$M^* M \simeq \left|f_+\right|^2 \left[u(3)\bar{u}(3)\gamma_\mu \left(C_V + C_V'\gamma_5\right) \times \upsilon(2)\bar{\upsilon}(2)\gamma^\sigma \left(C_V^* + C_V'^*\gamma_5\right)\right] p^\mu p_\sigma \quad (8.5)$$

The probability of kaon decay is proportional to

$$\Gamma \propto \left|f_+\right|^2 \left[\left|C_V\right|^2 + \left|C_V'\right|^2\right] \quad (8.6)$$

When decaying to the final state $e^+\pi^-\nu$ we have, up to quadratic terms,

$$\Gamma^{W^+} \propto \left|V_{us}\right|^2 \left|f_+\right|^2 \left[1 + \left(\delta - \zeta\right)^2\right] \quad (8.7)$$

When decaying into the final state $e^-\pi^+\bar{\nu}$, we have, up to quadratic terms,

$$\Gamma^{W^-} \propto \left|V_{us}\right|^2 \left|f_+\right|^2 \left[1+\left(\delta+\zeta\right)^2\right] \qquad (8.8)$$

Thus, taking into account (8.7) and (8.8), we obtain the formula for T-violating asymmetry:

$$A_T = \frac{\Gamma\left(\bar{K}^0 \to e^+\pi^-\nu\right) - \Gamma\left(K^0 \to e^-\pi^+\bar{\nu}\right)}{\Gamma\left(\bar{K}^0 \to e^+\pi^-\nu\right) + \Gamma\left(K^0 \to e^-\pi^+\bar{\nu}\right)} \quad (8.9)$$

which is equal to:

$$A_T = \frac{1+\left(\delta-\zeta\right)^2 - \left(1+\left(\delta+\zeta\right)^2\right)}{2\left(1+\delta^2+\zeta^2\right)} \approx -2\delta\zeta \quad (8.10)$$

Using the previously obtained values $\delta = 0.070(10)$ and $\zeta = -0.039(14)$, we obtain the value for the asymmetry

$$A_T^{LR} = (5.5 \pm 2.1) \times 10^{-3} (2.6\sigma) \qquad (8.11)$$

This value is in absolute agreement with the experimentally measured asymmetry with available accuracy [59].

$$A_T^{\exp} = \left(6.6 \pm 1.3 \pm 1.0\right) \times 10^{-3} (4\sigma) \qquad (8.12)$$

Thus, the predictions of the extended left-right symmetric model are confirmed by the experiment with confidence level $2.6\sigma$.

It is interesting to note that the experimental value of CP-violating asymmetry in the decay of neutral $D$ mesons has a value quite close to the value of CP-violating asymmetry in the decay of neutral $K$ mesons [60].

$$A_{D_0} = \frac{\Gamma\left(D_0 \to K^-K^+, \pi^-\pi^+\right) - \Gamma\left(\tilde{D}_0 \to K^-K^+, \pi^-\pi^+\right)}{2\Gamma}$$

$$A_{D_0}^{\exp} = (6.30 \pm 0.33) \times 10^{-3}$$

As can be seen, the experimental values of CP-violating asymmetries are in good agreement. These are the so-called direct CP-violating effects, which arise during the oscillations of $K_0\bar{K}_0$ and $D_0\bar{D}_0$. The description of this effect within the framework of the left-right CP-violating model has already been presented for $K_0\bar{K}_0$. The procedure can be repeated in exactly the same way for $D_0\bar{D}_0$. In both cases, the effect of CP violation with mixing is described by the formula: $A_T^{LR} = -2\delta\zeta = (5.5 \pm 2.1) \times 10^{-3} (2.6\sigma)$ and gives the same result. Table 3 presents a comparison of the experimental results of direct CP-violating asymmetries and estimates within the framework of the left-right CP-violating model.

Table 3. Comparison of experimental results for CP-violating asymmetries with mixing for $K_0\bar{K}_0$ and $D_0\bar{D}_0$ and estimations within the framework of the left-right model in units of $10^{-3}$.

| | $K^0$ | $D^0$ |
|---|---|---|
| $A^{exp}$ | $6.6 \pm 1.6$ | $6.30 \pm 0.33$ |
| $A^{LR}$ | $5.5 \pm 2.1$ | $5.5 \pm 2.1$ |

Besides, CP-violating lepton asymmetry $A_L$ is measured more accurately in decays of neutral $K$-mesons with decay product detection in final state [61]:

$$A_L = \frac{\Gamma\left(K_L \to e^+\pi^-\nu\right) - \Gamma\left(K_L \to e^-\pi^+\bar{\nu}\right)}{\Gamma\left(K_L \to e^+\pi^-\nu\right) + \Gamma\left(K_L \to e^-\pi^+\bar{\nu}\right)} \qquad (8.13)$$

$$A_L^{\exp} = \left(3.32 \pm 0.06\right) \times 10^{-3} \qquad (8.14)$$

This asymmetry $A_L^{\exp}$ is two times smaller than $A_T^{\exp}$. The fact is that the effect of CP violation with mixing is measured during the progress of the $K_0\bar{K}_0$ oscillation process over 10 periods of the lifetime of the $K_S$-state, which is $0.86 \cdot 10^{-10}$ s. And the effect of CP violation in the final state is measured at the lifetimes of the $K_L$-state, which is $5.4 \cdot 10^{-8}$ s. By this time, the effect associated with the $K_S$-state reduces. Therefore, $A_T/A_L = 2$ as shown in [60].

Of course, in our desire to understand the nature of CP violation, we must focus on the effect of CP violation with mixing. It determines the essence of the initial process. Its further development depends on a specific chain of decays. Therefore, within the framework of the model, we must explain only the main starting effect. It is surprising that this can be done using one simple relation $A_T^{LR} = -2\delta\zeta = (5.5 \pm 2.1) \times 10^{-3} (2.6\sigma)$. Of course, the confidence level does not give complete confidence, but it is at least higher than 99%.

Now let us consider to the CP-violating baryon asymmetry. The CP-violating baryon asymmetry from our comparison of $V_{ud}^{nLR}$ and $V_{ud}^{00LR}$ has an opposite sign.

$$A_{p-n} = \frac{\left(V_{ud}^{00LR}\right)^2 - \left(V_{ud}^{nLR}\right)^2}{(V_{ud}^{00LR})^2 + (V_{ud}^{nLR})^2} = (-3.2 \pm 1.6) \cdot 10^{-3} (2.0\sigma)$$

The opposite sign of the baryon asymmetry indicates that the B-L conservation condition, which was given in the famous work of A.D. Sakharov [62,63], is apparently satisfied.

The results for CP-violating asymmetries in the final state can be summarized in the following Table 4

Table 4. Experimental results for CP-violating asymmetries in the final state in units of $10^{-3}$.

| | $p-n$ | $K_L^0$ |
|---|---|---|
| $A^{exp}$ | $-3.2 \pm 1.6$ | $3.32 \pm 0.06$ |

Thus, we can summarize:

1. there is agreement within the available accuracy between the experimental results of CP-violating asymmetries with mixing for the $K_0\overline{K}_0$ system and $D_0\overline{D}_0$ system and the estimations within the left-right model with the parameters obtained from neutron decay. The relation $A^{LR} = -2\delta\zeta$ is satisfied.

2. there is agreement for absolute values within the available accuracy between the experimental results for CP-violating asymmetries in the final state for neutral K-mesons and CP-violating baryon asymmetry from our comparison $V_{ud}^{nLR}$ and $V_{ud}^{00LR}$ within the extended left-right model of the weak interaction with CP violation.

3. Finally, it is very important to note that the signs of the baryon and lepton CP-violating asymmetries are different, which may be the reason for the conservation of the difference in the number of baryons and leptons during the formation of the Universe, $B - L =$ const, $\Delta B = \Delta L$ [62,63].

## 9. BARYON-LEPTON ASYMMETRY OF THE UNIVERSE AND THE LEFT-RIGHT MODEL OF WEAK INTERACTION WITH CP VIOLATION

Now let us return to the question of the baryon-lepton asymmetry of the Universe and to the famous article by A.D. Sakharov [62,63]. This is a review report at a conference dedicated to the 100th anniversary of A.A. Friedman. Leningrad, June 22-26, 1988. This article formulates: (quote)
“Three basic conditions for cosmological formation of baryonic asymmetry
I. Absence of baryonic charge conservation
II. Difference between particles and antiparticles, manifesting itself in the violation of CP-invariance.
III. Nonstationarity. Formation of BA is only possible under nonstationary conditions in the absence of local thermodynamic equilibrium.”

In addition, the violation of both baryon and lepton asymmetry is discussed, and the conservation of the B-L difference is noted. (quote)
"If the baryon-lepton asymmetry with $B \neq L$ arises at a temperature above the range $T = 10^2$— $10^4$ GeV … then a state will be established (with high precision) in the low temperature region which corresponds to the entropy maximum at given constant value of B — $L =$ const (and under the condition of electric neutrality)."

Thus, all the main predictions of the baryon asymmetry of the Universe were made by A.D. Sakharov more than half a century ago, soon after the discovery of CP violation in $K$-mesons. Since then, the effects of CP violation have been thoroughly studied at accelerators, but the nature of CP violation has not been established. Although a method for describing CP violation through a complex phase in the CKM matrix has been proposed.

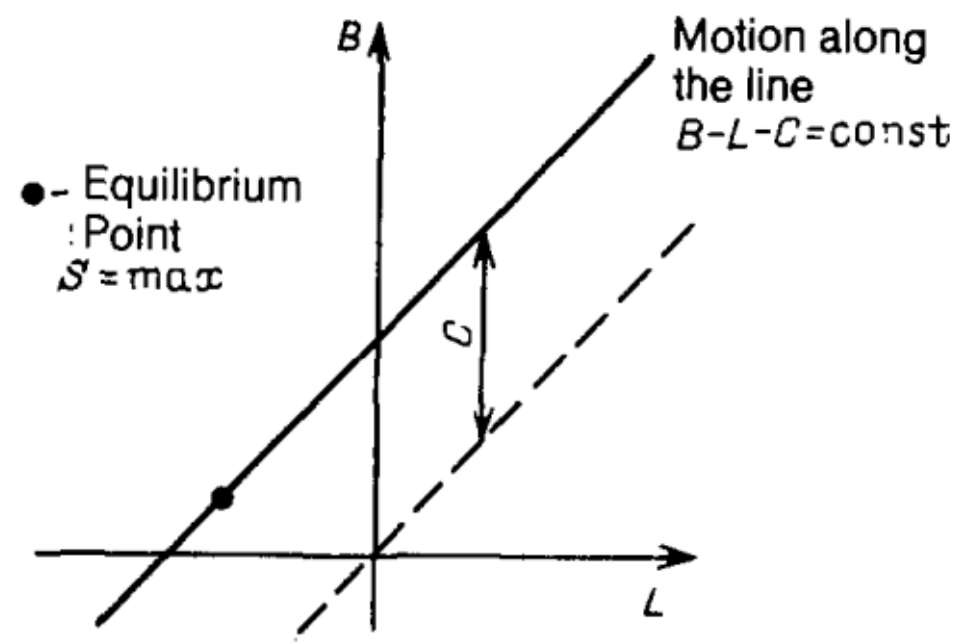

Fig. 22. A plot from A.D. Sakharov's paper, where the baryon and lepton asymmetries of the universe are shown.

In our work, a left-right model of weak interaction with CP violation is proposed at mixing $W_L$ and $W_R$, which indicates the nature of CP violation. In this regard, it is important to note that A.D. Sakharov's work speaks of the simultaneous formation of baryon and lepton asymmetry.

Now, based on the fact that the nature of CP violation is associated with the difference between left and right weak interactions, we can propose the following scenario for the formation of baryon and lepton asymmetry in the Universe.

At temperatures of the order of $10^4$ GeV, the degree of symmetry between left and right processes (processes with opposite CP parity) was quite high. But when the temperature decreased, an advantage arose in preserving neutrons and protons in relation to antineutrons and antiprotons and simultaneously in preserving antineutrinos in relation to neutrinos.

Another important factor contributing to the generation of matter in the universe may be the asymmetry of neutron-antineutron oscillations due to CP violation. This violation can lead to a violation of the law of conservation of baryon number, which is necessary for Sakharov's first condition. Also, because neutrons and antineutrons decay by emitting different $W$ bosons, the lifetime of the antineutron will differ from the neutron lifetime, which may also contribute to the generation of baryon asymmetry in the universe.

Therefore, proton accumulation occurs from neutron decay relative to antiprotons, leading to the formation of a positive baryon asymmetry of the universe. Simultaneously, there is an accumulation of antineutrinos, resulting in the formation of a negative lepton asymmetry of the universe.

Indeed, in the modern Universe we have protons and neutrons in nuclei, it is obvious that the baryon number is positive. Charged leptons are electrons in atoms, which indicates a positive asymmetry in the charged leptons sector, which is compensated excess of negative asymmetry due to a significant number of antineutrinos, therefore $L < 0$. If protons and neutrons formed galaxies, then neutrinos and antineutrinos formed dark matter located on the periphery of galaxies.

It is important to note that the existence of the $W_R$ suggests the presence of right (so-called sterile) antineutrinos, which have a significantly greater mass than active neutrinos. They provide a mass of dark matter approximately 5 times greater than the mass of baryonic matter. The requirement for the stability of dark matter [1] and astrophysical observations [67] limit the mass of sterile neutrinos to below a few keV.

## 10. CP and CPT invariance in the extended left-right model

CP violation with mixing of right- and left-handed W bosons is possible for neutral systems, such as $K^0-\bar{K}^0, D^0-\bar{D}^0, B^0-\bar{B}^0, B_s^0-\bar{B}_s^0$ mesons, as well as for $n-\bar{n}$ oscillations. Since mixing of right- and left-handed W bosons occurs with different signs of the mixing angle $\zeta$ for $W^+$ and $W^-$, and $W^+$ is an antiparticle, while $W^-$ is a particle, one might expect that in this left-right CP-violating model, CPT will be violated, but locally, since constant particle-antiparticle transformations occur. CPT is also violated locally in neutrino - sterile antineutrino (antineutrino - sterile neutrino) oscillations.

There are experimental constraints on the violation of CPT invariance. These can be divided into direct constraints from experiments with $\pi^+$and $\pi^-$ mesons and indirect constraints from experiments with $K^0-\bar{K}^0$oscillations. In experiments with $\pi^+$ and $\pi^-$ mesons, the lifetimes of the particle and antiparticle can be measured directly, and these values can be compared.

CPT- violation in decays $\pi^+$ and $\pi^-$ [65] has already been experimentally limited at the accuracy level of $10^{-3}$.

$$\frac{\left(\tau_{\pi^+}-\tau_{\pi^-}\right)}{\tau_{average}}=(7.1\pm5.5)\times10^{-4} \quad (10.1)$$

The same restrictions, proving CPT-conservation, exist from decays $K^0, \bar{K}^0$ [66] where

$$\frac{\left(\Gamma_{K^0}-\Gamma_{\bar{K}^0}\right)}{\Gamma_{\text{average}}}=\left(5.4\pm5.4\right)\times10^{-4} \quad (10.2)$$

However, this is an indirect limitation, since it is not possible to measure the lifetime separately for $K^0$ and separately for $\bar{K}^0$ , they are in the mode of mutual transformations, and the lifetime of each of them cannot be determined, as demonstrated by the experimental results (see Fig. 28 in Section 11).

The estimate (10.2) was obtained in the work [66], where the mixing matrix was considered with $\pm\delta\Gamma$ (different decay probabilities for $K^0$ and $\bar{K}^0$) on the diagonal of the matrix to account for CPT violation, but CP violation was taken into account in the off-diagonal matrix elements. This result (3.2) should be interpreted as follows: after taking into account CP violation in the off-diagonal matrix elements, nothing remained on CPT violation in the diagonal matrix elements with the specified accuracy. Therefore, one should rely on this experimental result and adopt the position, based on the results of the experiment [10], that CPT invariance is preserved on average, although it is violated locally. As for the asymmetry $A_T^{LR}=-2\delta\zeta=(5.5\pm2.1)\times10^{-3}$ $(2.6\sigma)$, it is completely determined by the effect of CP violation. At present, there is no reason to raise the question of CPT violation.

Thus, in the extended left-right model, CP invariance is violated and CPT invariance is preserved.

The constraints on the model parameters $\delta$ and $\zeta$ for charged and neutral particles do not overlap [1], as shown in Fig. 23. The result of the TWIST experiment [64] does not contradict the result of our analysis of neutron decay, since the TWIST experiment was performed with charged particles.

And these CP-violating effects for $K^0, D^0, B^0, B_s^0$ can be described by the same parameters $\delta$ and $\zeta$, extracted from neutron decay, they will be discussed in section 11.

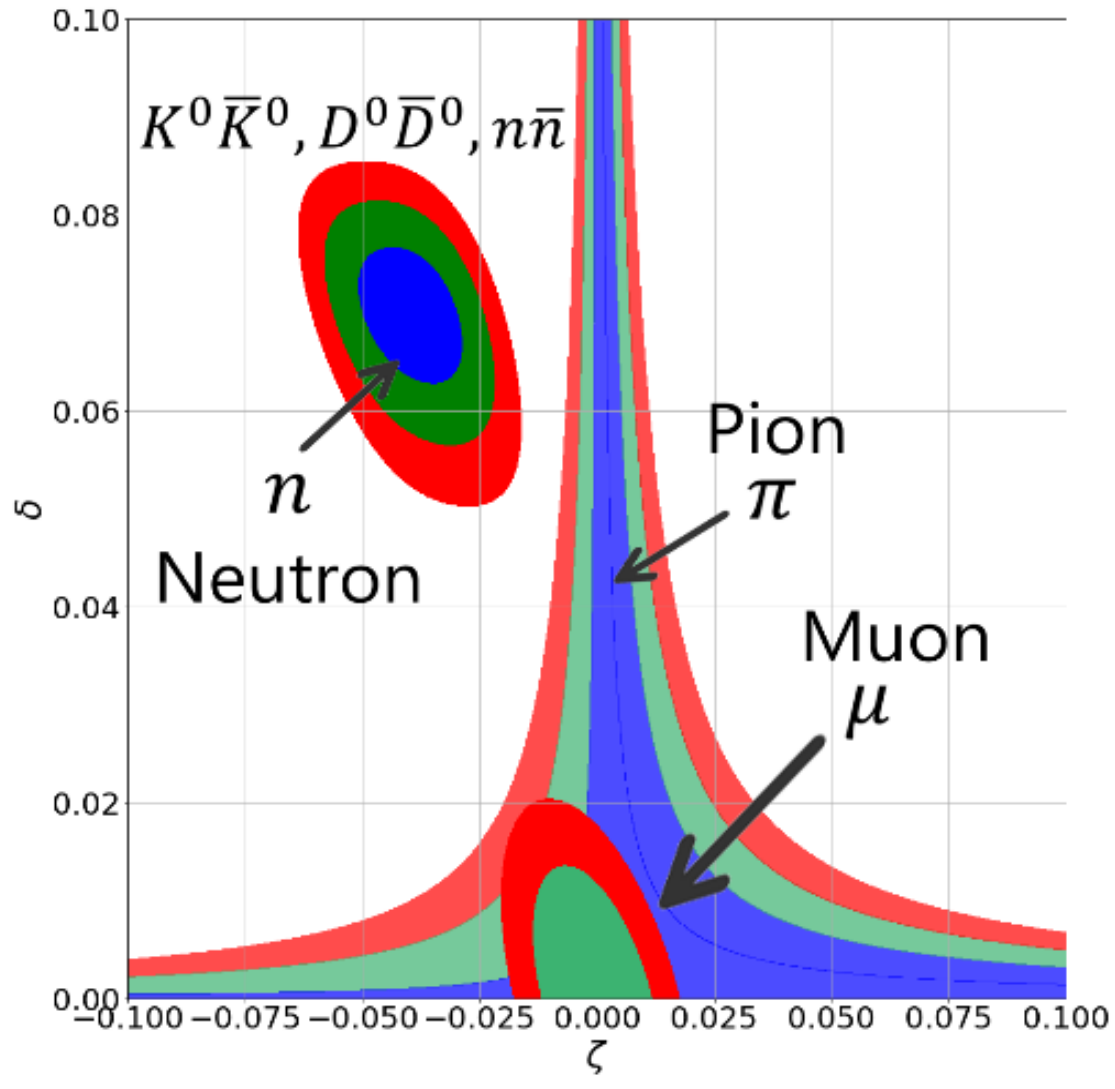

Fig. 23. Comparison of the results of the neutron decay analysis, which are also applicable to the asymmetries of the decays $K^0\bar{K}^0$ and $D^0\bar{D}^0$, as well as the constraints from CPT invariance for the decays $\pi^+\pi^-$ and $\mu^+\mu^-$. The asymmetry of the decays of charged particles such as $\mu^\pm$ and $\pi^\pm$ is impossible, as shown in the graph.

## 11. Description of neutral meson oscillations in the left-right CP-violating model

In accelerator studies, a mixing matrix of the form (11.2) shown in Fig. 4 is used to describe the oscillation processes of neutral mesons. The lifetimes of neutral mesons and the mixing processes are $10^{-8}$ – $10^{-12}$ s. The lifetime of $W_L$ and $W_R$ is $3 \cdot 10^{-25}$ s (Fig. 24). The mixing process of $W_L$ and $W_R$ can be represented by a diagonal matrix of the form (11.1) shown in Fig. 25. These two representations can be combined by a matrix of the form (11.3) in Fig. 25. And thus, we introduce an asymmetry in the decay probability for particles and antiparticles through the parameter $\delta\Gamma$ during the oscillation process.

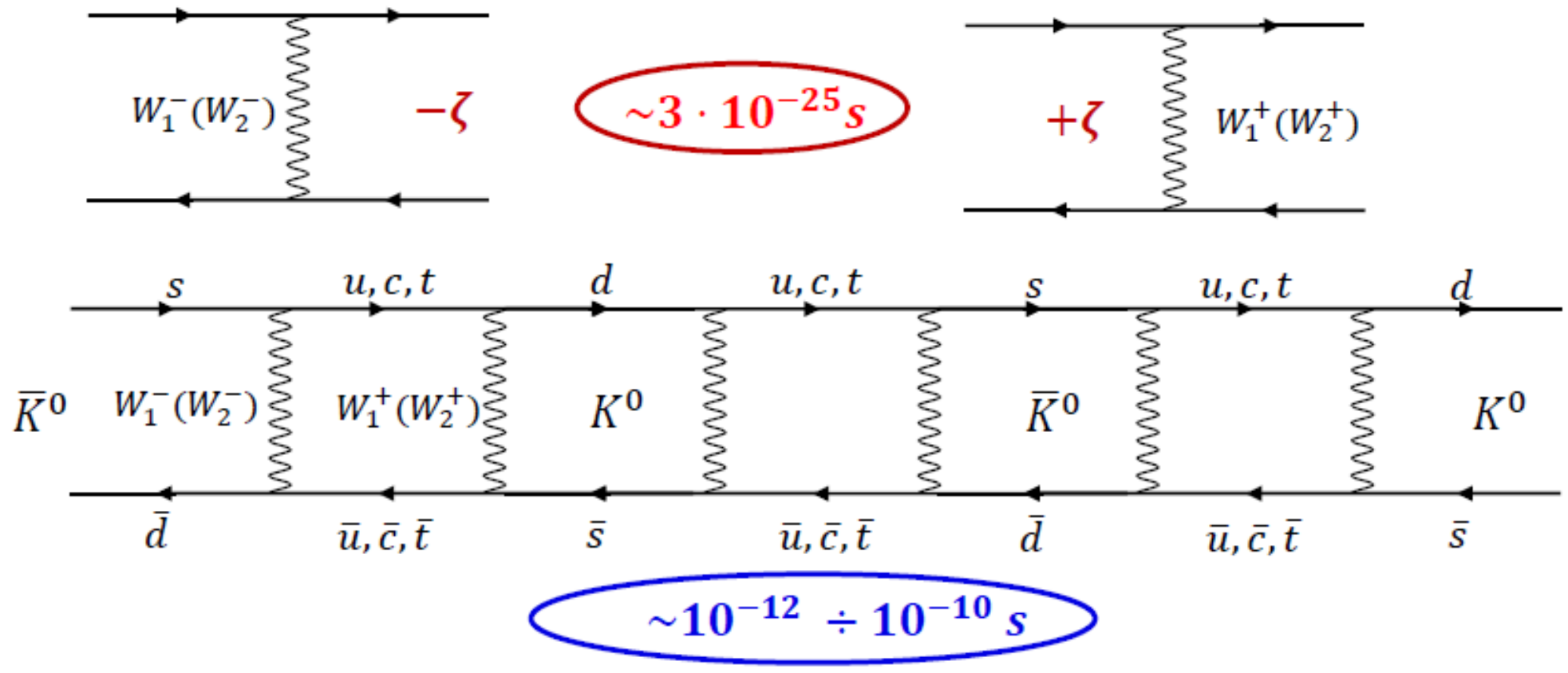

Fig. 24. The process of mixing left and right vector bosons $W_1^-(W_2^-)$ and $W_1^+(W_2^+)$ occurs at times $\sim 3 \cdot 10^{-25}$ s, and the mixing process $K^0\bar{K}^0$ at times $\sim 10^{-8} - 10^{-10}$ s.

$$\begin{pmatrix} M - i\dfrac{\Gamma_0 - \delta\Gamma}{2} & 0 \\ 0 & M - i\dfrac{\Gamma_0 + \delta\Gamma}{2} \end{pmatrix} (4.1) \quad \begin{pmatrix} M - i\dfrac{\Gamma_0}{2} & \Delta m - i\dfrac{\Delta\Gamma}{2} \\ \Delta m - i\dfrac{\Delta\Gamma}{2} & M - i\dfrac{\Gamma_0}{2} \end{pmatrix} (4.2) \quad \begin{pmatrix} M - i\dfrac{\Gamma_0 - \delta\Gamma}{2} & \Delta m - i\dfrac{\Delta\Gamma}{2} \\ \Delta m - i\dfrac{\Delta\Gamma}{2} & M - i\dfrac{\Gamma_0 + \delta\Gamma}{2} \end{pmatrix} (4.3)$$

Fig. 25. The mixing matrix of left and right vector bosons $W_1^-(W_2^-)$ and $W_1^+(W_2^+)$- (11.1), the mixing matrix of neutral mesons in SM - (11.2) and the combined mixing matrix - (11.3).

Mixing matrix (11.2) for neutral mesons in the Standard Model takes into account the effects of flavor oscillations and CP violation in off-diagonal matrix elements, with CP violation effects being accounted for through the complex phase in the CKM matrix as a phenomenological parameter. The physical causes of CP violation are unknown. With such a SM matrix, flavor oscillations and CP violation effects in off-diagonal matrix elements interfere, making their separation significantly difficult.

The mixing matrix (11.3) for neutral mesons in the extended left-right model is structured differently. It takes into account the effects of flavor oscillations in the off-diagonal matrix elements and the effects of CP violation in the diagonal matrix elements through the parameter $\delta\Gamma$. We believe that the physical reason for $\delta\Gamma$ is related to the different signs of the mixing angle of the right-handed vector boson with the positive and negative left-handed vector bosons, since $A_T^{LR} = -2\delta\zeta = (5.5 \pm 2.1) \cdot 10^{-3}$, based on the results of the neutron experiment for left-right asymmetry. It is believed that in such a representation of the matrix (with different signs $\delta\Gamma$ on the diagonal), CPT invariance is violated. Here it is necessary to repeat what we already mentioned in the previous section about the possibility of only local CPT violation. But since the experiment [66] showed the absence of CPT violation with an accuracy of $10^{-3}$, it can be considered that the effect exceeding this value is associated mainly with CP violation.

Let us analyze the results of experiments with neutral mesons using the mixing matrix representation (11.3), in which the parameters responsible for flavor oscillations ( $\Delta\Gamma$ and $\Delta$m) and the parameters responsible for CP violation effects ( $\delta\Gamma$) are separated. Moreover, the parameters $\Delta\Gamma$ and $\Delta$m are taken from the experimental data, and the parameter $\delta\Gamma = -2\delta\zeta \times \Gamma = 5.5 \times 10^{-3} \times \Gamma$.

First, let us recall the classical consideration of the problem, but without taking into account CP-violation.

We represent the effective Hamiltonian as $H = \begin{pmatrix} m_0 - i\frac{\Gamma_0}{2} & \Delta m - i\frac{\Delta\Gamma}{2} \\ \Delta m - i\frac{\Delta\Gamma}{2} & m_0 - i\frac{\Gamma_0}{2} \end{pmatrix}$ (4.4)

Solution of the Schrödinger equation for the Hamiltonian $H\Psi(t) = i\hbar\frac{\partial\Psi(t)}{\partial t}$ we will search for it in the form of a two-dimensional vector $\Psi(t) = \binom{a}{b} e^{-i\omega t}$. The solution to the Schrödinger equation is reduced to solving the eigenvalue problem – $\omega$ and the eigenvectors - $\binom{a}{b}$. The equation for the eigenfrequencies $\omega$ is determined from the following condition.

$$\begin{pmatrix} m_0 - i\frac{\Gamma_0}{2} & \Delta m - i\frac{\Delta\Gamma}{2} \\ \Delta m - i\frac{\Delta\Gamma}{2} & m_0 - i\frac{\Gamma_0}{2} \end{pmatrix}\binom{a}{b} = \omega\binom{a}{b} \quad (4.5),$$

$$\det\begin{Vmatrix} m_0 - i\frac{\Gamma_0}{2} - \omega & \Delta m - i\frac{\Delta\Gamma}{2} \\ \Delta m - i\frac{\Delta\Gamma}{2} & m_0 - i\frac{\Gamma_0}{2} - \omega \end{Vmatrix} = 0 \quad (4.6),$$

Natural frequencies $\omega_\pm = m_0 \pm \Delta m - i\frac{\Gamma_0 \pm \Delta\Gamma}{2}$. Frequency difference $\omega_+ - \omega_- = \Delta\omega = 2\Delta m - i\Delta\Gamma$.
For $\omega_+ = m_0 + \Delta m - i\frac{\Gamma_0 + \Delta\Gamma}{2}$ the normalized eigenvector is equal to: $\Psi_L(t) = \frac{1}{\sqrt{2}}\binom{1}{1} e^{-i\,\omega_+ t}$
For $\omega_- = m_0 - \Delta m - i\frac{\Gamma_0 - \Delta\Gamma}{2}$ the normalized eigenvector is equal to: $\Psi_S(t) = \frac{1}{\sqrt{2}}\binom{1}{-1} e^{-i\omega_- t}$
By the construction of the Hamiltonian, the state of a particle $|\psi\rangle$ is described as a projection onto the vector $\binom{1}{0}$, and the state of an antiparticle $|\bar{\psi}\rangle$ is described as a projection onto the vector $\binom{0}{1}$. Thus, it turns out that $\Psi_L(t) = \frac{e^{-i\,\omega_+ t}}{\sqrt{2}}\left(|\psi\rangle + |\bar{\psi}\rangle\right)$, $\Psi_S(t) = \frac{e^{-i\omega_- t}}{\sqrt{2}}\left(|\psi\rangle - |\bar{\psi}\rangle\right)$ (4.7),

If we use the generally accepted phase rules for mesons in the $CP$- transformation $CP|\psi\rangle = -|\bar{\psi}\rangle$, $CP|\bar{\psi}\rangle = -|\psi\rangle$ then we get that $\Psi_L(t)$- is a $CP$ odd state, and $\Psi_S(t)$- is a $CP$ even state.

If at a moment in time $t = 0$ we have a state of a particle, then $\psi(t) = \frac{1}{\sqrt{2}}\left(\Psi_L(t) + \Psi_S(t)\right)$ (4.8).

If at the initial moment of time the state of the antiparticle is, then $\bar{\psi}(t) = \frac{1}{\sqrt{2}}\left(\Psi_L(t) - \Psi_S(t)\right)$ (4.9).

When the initial state is a particle, the probability of detecting the particle will be (4.10), and the probability of detecting the antiparticle will be (4.11).

$\psi^*(t)\psi(t) = \frac{1}{4}e^{-\Gamma_0 t}\left[e^{-\Delta\Gamma t} + e^{\Delta\Gamma t} + 2\cos(2\Delta m t)\right]$ (4.10), $\bar{\psi}^*(t)\bar{\psi}(t) = \frac{1}{4}e^{-\Gamma_0 t}\left[e^{-\Delta\Gamma t} + e^{\Delta\Gamma t} - 2\cos(2\Delta m t)\right]$ (4.11),

When the initial state is an antiparticle, the probability of detecting the antiparticle will be (4.12), and the probability of detecting the particle will be (4.13).

$\bar{\psi}^*(t)\bar{\psi}(t) = \frac{1}{4}e^{-\Gamma_0 t}\left[e^{-\Delta\Gamma t} + e^{\Delta\Gamma t} + 2\cos(2\Delta m t)\right]$ (4.12), $\psi^*(t)\psi(t) = \frac{1}{4}e^{-\Gamma_0 t}\left[e^{-\Delta\Gamma t} + e^{\Delta\Gamma t} - 2\cos(2\Delta m t)\right]$ (4.13)

It's clear that particles and antiparticles behave symmetrically: when the initial conditions change, it's as if the particle and antiparticle swap places. We'll use this fact to estimate the magnitude of the asymmetry between the decay probabilities of particles and antiparticles, taking mixing into account.

The next step is to introduce CP violation. In the Standard Model, CP violation is introduced through the complex phase in the CKM matrix. However, we propose introducing CP violation into the diagonal elements of the mixing matrix. Regarding the CKM matrix, we expect a splitting of the matrix elements for interactions via $W^+$ and $W^-$, as shown in Figure 1.
Thus, if we introduce different lifetimes for the particle and antiparticle in the diagonal matrix elements of the Hamiltonian, we can write:

$$H = \begin{pmatrix} m_0 - i\frac{\Gamma_0 - \delta\Gamma}{2} & \Delta m - i\frac{\Delta\Gamma}{2} \\ \Delta m - i\frac{\Delta\Gamma}{2} & m_0 - i\frac{\Gamma_0 + \delta\Gamma}{2} \end{pmatrix} \quad (4.14).$$

Such a modification of the Hamiltonian will lead to a change in both the eigenvalues and eigenvectors compared to the Hamiltonian without corrections $\delta\Gamma$

$$\omega_+ = m_0 - i\frac{\Gamma_0}{2} + \frac{\Delta\omega}{2},\ \omega_- = m_0 - i\frac{\Gamma_0}{2} - \frac{\Delta\omega}{2},$$

where $\Delta\omega = \sqrt{(2\Delta m)^2 - [(\Delta\Gamma)^2 + (\delta\Gamma)^2] - 2i(\Delta\Gamma)(2\Delta m)}$ (4.15)

The presence of corrections from $\delta\Gamma$ leads to the fact that in the first order in $\delta\Gamma$ the probability of detecting a particle, calculated in the absence of $\delta\Gamma$ in the Hamiltonian, receives an addition $\varepsilon_{pp}(t)$

$$\psi^*(t)\psi(t) \approx \frac{1}{4}e^{-\Gamma_0 t}\left[e^{\Delta\Gamma t} + e^{-\Delta\Gamma t} + 2cos(2\Delta mt)\right] + \varepsilon_{pp}(t) \quad (4.16)$$

For an antiparticle, the probability taking into account $\delta\Gamma$ will have a contribution from $\varepsilon_{pp}(t)$ with a different sign.

$$\bar{\psi}^*(t)\bar{\psi}(t) \approx \frac{1}{4}e^{-\Gamma_0 t}\left[e^{-\Delta\Gamma t} + e^{\Delta\Gamma t} + 2\cos(2\Delta mt)\right] - \varepsilon_{pp}(t) \quad (4.17)$$

For approximate calculations of asymmetry in the expansion by the smallness parameter $x = \delta\Gamma/((2\Delta m)^2 + (\Delta\Gamma)^2)^{1/2}$ we use the formula:

$$\varepsilon_{pp}(t) = \frac{\delta\Gamma}{\sqrt{(2\Delta m)^2+(\Delta\Gamma)^2}} \times \frac{e^{-\Gamma_0 t}\left((\Delta\Gamma)sh(\Delta\Gamma t)+(2\Delta m)sin(2\Delta mt)\right)}{\sqrt{(2\Delta m)^2+(\Delta\Gamma)^2}} \quad (4.18)$$

The calculation results for the probability of detecting a particle or antiparticle for the two schemes are presented in Fig. 26. As can be seen, the calculation results with $\delta\Gamma = 0$ and $\delta\Gamma/\Gamma = -2\delta\zeta = 5.5 \times 10^{-3}$ do not differ within the graphical accuracy for all mesons. The fact is that the flavor oscillation process is successfully described by the off-diagonal parameters of the mixing matrices, and the diagonal parameter $\delta\Gamma$, responsible for CP violation, has little effect on the oscillation process. However, as will be shown below, the diagonal parameter $\delta\Gamma$, plays a decisive role in calculating the CP-violating decay asymmetry. The lack of sensitivity to the parameter $\delta\Gamma/\Gamma$ is shown in Fig. 27 for $B_s^0$ meson, where the ordinate axis shows the ratio of the integrals $\int_0^\infty N^{B_0}(t)d\,t/\int_0^\infty N^{\tilde{B}_0}(t)dt = R$ for the probability of being in the state $N^{B_0}$ over the entire decay time to the probability of being in the state $N^{\tilde{B}_0}$ over the entire decay time. The R -ratio characterizes the integral ratio of the probability of decaying to the particle state to the probability of decaying to the antiparticle state, if the process began from the particle state. The R -ratio differs from unity because the oscillation process begins at unity for the particle and at zero for the antiparticle (the same as when the process begins with the antiparticle). When changing $\delta\Gamma/\Gamma$ from zero to 1%, the R -ratio changes by 0.5%.

Thus, in the mixing matrix (11.3) in the extended left-right model, flavor oscillations and CP-violation effects are separated to a certain extent.

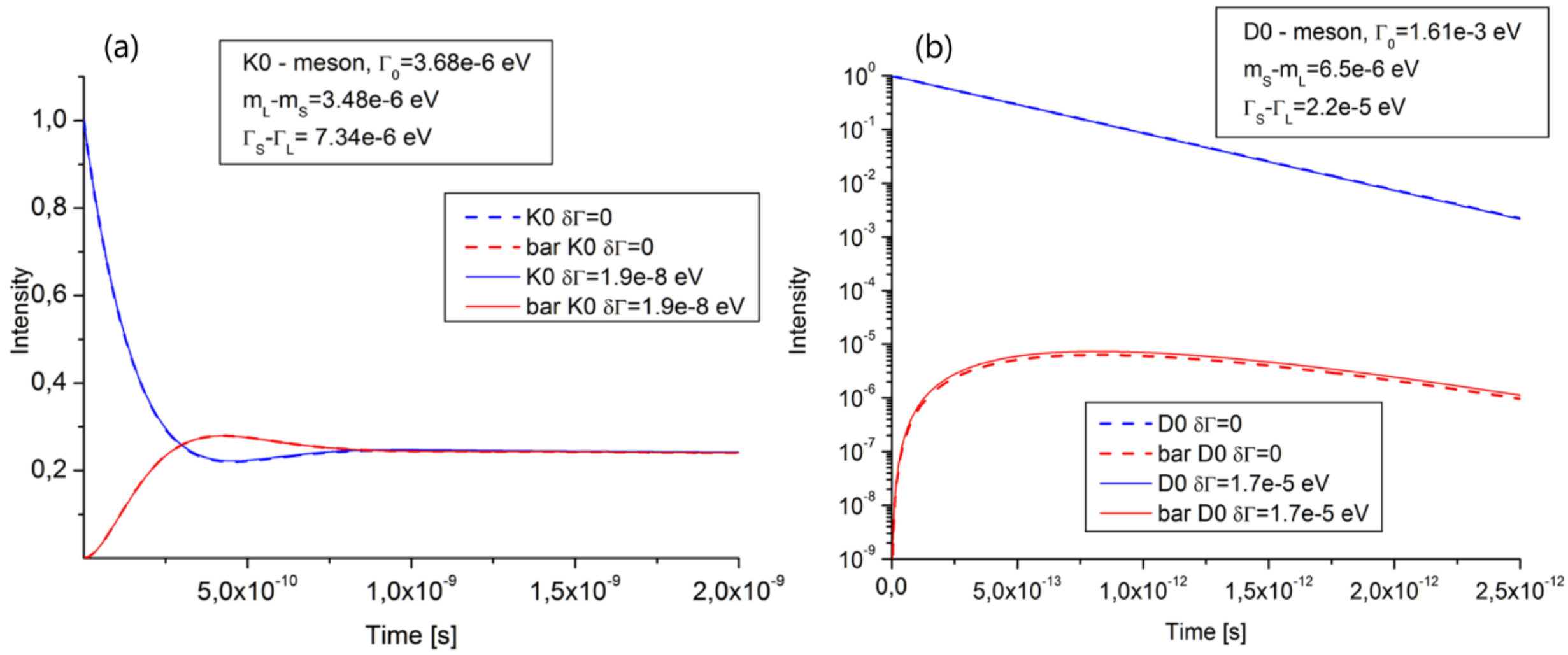

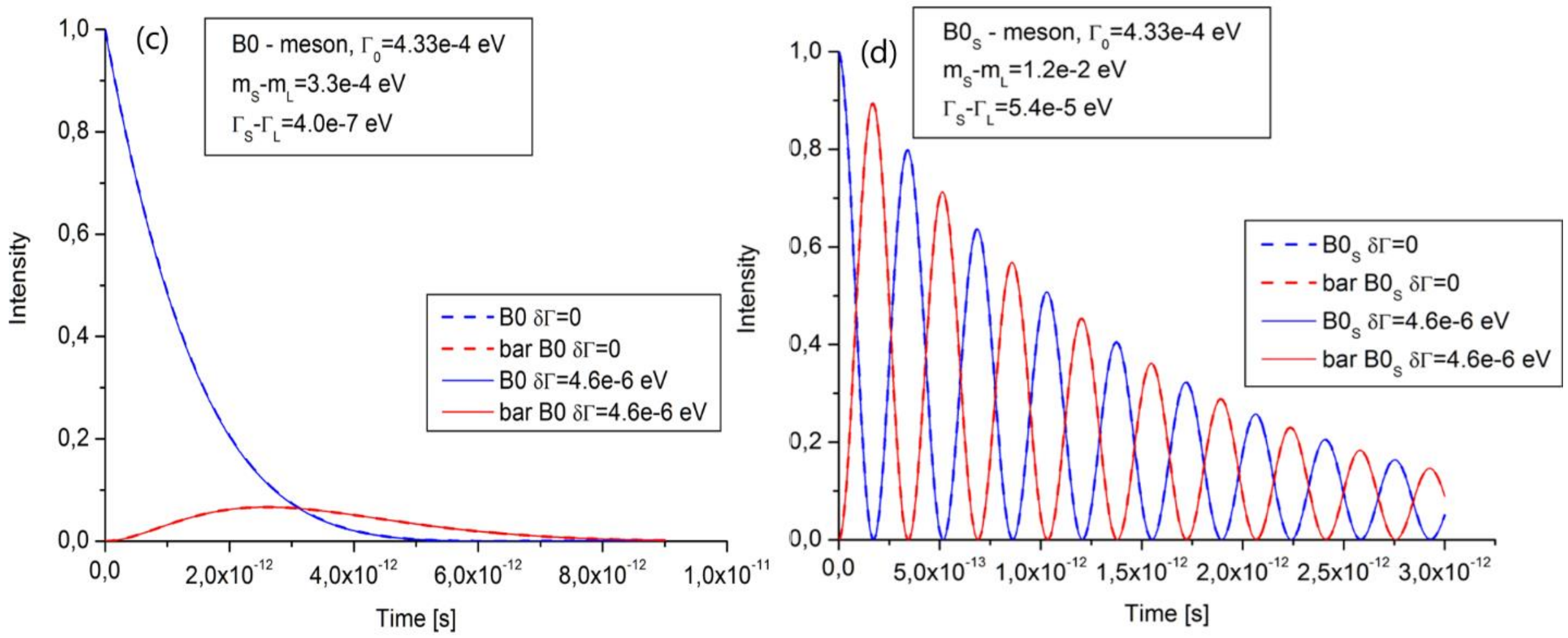

Fig. 26. The process of oscillations of neutral mesons. The blue curves correspond to the state of the particle, the red ones to the state of the antiparticle, the solid curves correspond to the calculations carried out with $\delta\Gamma = 0$, the dashed curves correspond to the calculations carried out with $\delta\Gamma = -2\delta\zeta \times \Gamma = 5.5 \times 10^{-3} \times \Gamma$, in the calculations the off-diagonal matrix elements are determined with $2\Delta m = (m_S - m_L)$, $2\,\Delta\Gamma = (\Gamma_S - \Gamma_L)$, a) - results of calculations for $K^0$ meson, b) - for $D^0$ meson, c) - for $B^0$ meson, d) for $B_S^0$ meson.

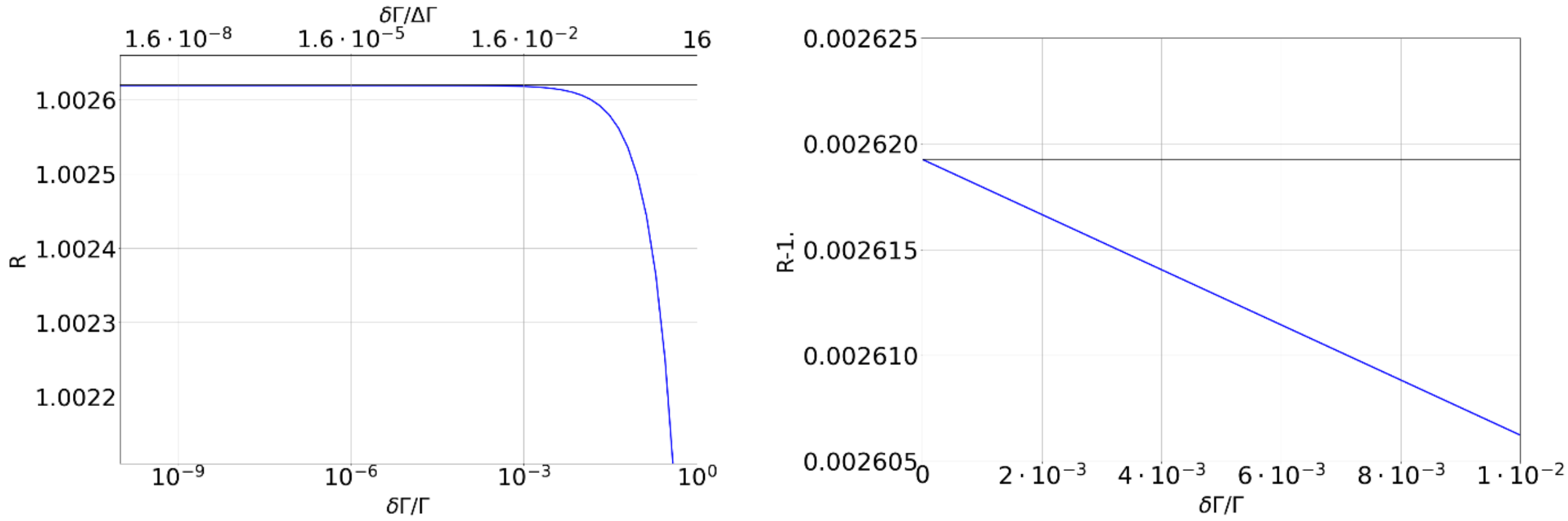

Fig. 27 . Dependence of R -ratio on $\delta\Gamma/\Gamma$ on a logarithmic and linear scale.

Now let's move on to calculating the integral asymmetry. For this, we take the ratio of the integral $\varepsilon_{pp}$ to the integral $\rho_{pp}$.

$$A_{pp\ \overline{pp}} = \frac{\left|\psi_{up,p}\right|^2 - \left|\psi_{up,\bar{p}}\right|^2}{\left|\psi_{up,p}\right|^2 + \left|\psi_{up,\bar{p}}\right|^2}\ ,\ \tilde{A}_{p\bar{p}} = \frac{\varepsilon_{pp}}{\rho_{pp}},\ \text{where}\ \varepsilon_{pp} = \int_0^\infty \varepsilon_{pp}(t)dt, \quad \rho_{pp} = \int_0^\infty |\psi(t)|^2 dt \qquad (4.19)$$

Table 5 presents the calculations of the smallness parameters $x$ for the approximate formulas. Table 6 presents the integral asymmetries $\tilde{A}_{p\bar{p}}$ for $K^0, B^0, B_S^0$ and $D^0$ mesons, calculated using the wave functions (Exact) and the approximate formulas (Approx). As can be seen, the approximate formulas agree reasonably well with the exact computer calculations. However, it is important to note that the formulas allow for an analytical analysis of various dependencies and relationships and are extremely useful.

Table 5. Coefficients at $sh(\Delta\Gamma t)$ and $sin(2\Delta m\ t)$ ($x_{sh}$ and $x_{sin}$) in the function $\varepsilon_{pp}(t)$ for different neutral mesons.

| meson | $x_{sh}$ | $x_{sin}$ | $x_{sh}/x_{sin}$ |
|---|---|---|---|
| $K^0$ | $2.8 \times 10^{-3}$ | $2.7 \times 10^{-3}$ | 1.1 |
| $D^0$ | 0.57 | 0.34 | 1.7 |
| $B^0$ | $4 \times 10^{-6}$ | $7 \times 10^{-3}$ | $6 \times 10^{-3}$ |
| $B_S^0$ | $4 \times 10^{-7}$ | $1.9 \times 10^{-4}$ | $2.2 \times 10^{-3}$ |

Table 6. ($2\Delta m = m_S - m_L$, $2\Delta\Gamma = \Gamma_S - \Gamma_L$)

| Meson | $2\Delta m[eV]$ | $2\Delta\Gamma[eV]$ | $\Gamma$ , $\delta\Gamma[eV]$ | Exact $\tilde{A}_{p\bar{p}}$ | Approx $\tilde{A}_{p\bar{p}}$ | $x$ |
|---|---|---|---|---|---|---|
| $K^0$ | $3.48 \times 10^{-6}$ | $7.3 \times 10^{-6}$ | $3.68 \times 10^{-6}$ $(1.9 \times 10^{-8})$ | $5.6 \times 10^{-3}$ | $5.6 \times 10^{-3}$ | $4 \times 10^{-3}$ |
| $B^0$ | $3.3 \times 10^{-4}$ | $4.0 \times 10^{-7}$ | $4.33 \times 10^{-4}$ $(2.3 \times 10^{-6})$ | $4.1 \times 10^{-3}$ | $4.1 \times 10^{-3}$ | $7 \times 10^{-3}$ |
| $B_S^0$ | $1.2 \times 10^{-2}$ | $5.4 \times 10^{-5}$ | $4.33 \times 10^{-4}$ $(2.3 \times 10^{-6})$ | $1.4 \times 10^{-5}$ | $1.4 \times 10^{-5}$ | $2 \times 10^{-4}$ |
| $D^0$ | $6.5 \times 10^{-6}$ | $2.2 \times 10^{-5}$ | $1.61 \times 10^{-3}$ $(8.5 \times 10^{-6})$ | $5.3 \times 10^{-3}$ | $5.3 \times 10^{-3}$ | 0.7 |

Let us return to the discussion of Table 6. It can be seen that for $K^0$ meson, $B^0$ meson and $D^0$ meson we have integral CP asymmetries that are practically determined by the value of the asymmetry: $A_T^{LR} = -2\delta\zeta = 5.5 \times 10^{-3}$. But for $B_S^0$ meson the CP violation effect is equal to $1.4 \times 10^{-5}$ , since in the process of oscillations of $B_s^0$ meson the CP asymmetry is averaged out. However, it should be noted that the CP-violating process existed throughout the entire oscillation process. This is because the sign of the CP-violating effect reversed during each oscillation period, according to formula $\varepsilon_{p\bar{p}} \approx e^{-\Gamma_0 t} \sin(2\Delta m t) \left(\frac{\delta\Gamma}{2\Delta m}\right)$, and was therefore compensated. Experimental observations with larger statistical data in 2021 confirmed the absence of asymmetry [12].

The above analysis of the calculation of integral asymmetries is somewhat simplified, while in experiment, the CP violation effect can be represented as a mixture of two effects: mixing and the final-state effect. For example, the case of oscillations $K^0\bar{K}^0$ is also a superposition of mixing and final-state decay. There is an experimental result for mixing: $A_T^{exp} = (6.6 \pm 1.3 \pm 1.0) \times 10^{-3}$ and there is an experimental result for final-state decay: $A_T^{exp} = (3.32 \pm 0.06) \times 10^{-3}$. The analysis of oscillations $K^0\bar{K}^0$ is complicated by the presence of two states $K_L$ and $K_S$ with significantly different lifetimes.

For example, for $B_s^0$ meson, the integral effect has already averaged out due to the large number of oscillations. Meanwhile, for $D^0$ meson, the ratio of the decay rate to the oscillation frequency is $\Gamma_{D^0}/2\Delta m_{D^0} = 2.5 \times 10^2$, so the oscillation process has time to fade before it even begins. For $B^0$ meson, the ratio of the decay rate to the oscillation frequency is $\Gamma_{B^0}/2\Delta m_{B^0} = 1.3$, so the oscillation process is on the verge of suppression. Finally, for $B_s^0$ meson, the ratio of the decay rate to the oscillation frequency is $\Gamma_{B_s^0}/2\Delta m_{B_s^0} = 3.6 \times 10^{-2}$, i.e. many less than one and therefore process of the oscillations is represented by a large number of periods. This process has been well studied experimentally, as illustrated in Fig. 28 (b).

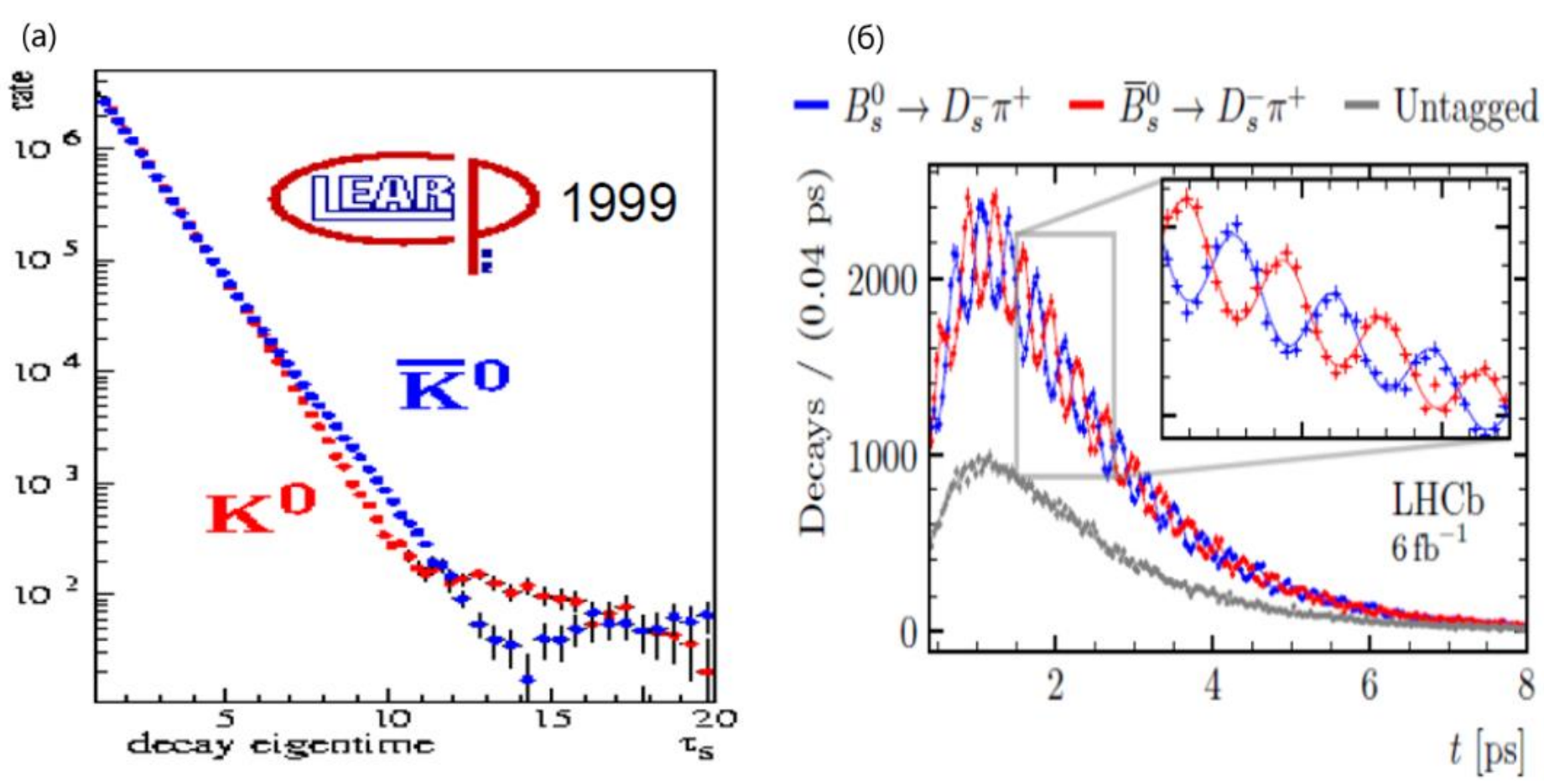

Fig. 28. Experimental results of flavor oscillations: a) for oscillations of K $^0$ mesons, b) for oscillations $B_S^0\bar{B}_S^0$

The experiment measures the effects of CP violation as a function of the decay time, so calculations of these effects as a function of time must be performed. Figure 29 presents the results of these calculations: for $K^0$ meson, for $D^0$ meson, for $B^0$ meson, and for $B_s^0$ meson. All asymmetries except for $B_s^0$ are multiplied by a decaying exponential to account for the statistical contribution in a real experiment.
Although, as can be seen in Table 2, the integral effects for $K^0$ meson, $B^0$ meson and $D^0$ meson are practically determined by the value: $A_T^{LR} = -2\delta\zeta = 5.5 \times 10^{-3}$, in fact, these CP-violating asymmetries, presented in differential form, differ significantly.

For example, for $K^0$ meson, interference between the short-lived and long-lived states is observed at short times. This dependence can apparently explain the presence of two experimental results: for mixing $A_T^{exp} = (6.6 \pm 1.3 \pm 1.0) \times 10^{-3}$ and for decay in the final state: $A_T^{exp} = (3.32 \pm 0.06) \times 10^{-3}$. Thus, it can be said that at least a qualitative description of the experimental results is observed.

For $D^0$ meson, the maximum differential asymmetry of $\approx 2 \times 10^{-3}$ is observed at a time equal to the meson lifetime. Experimental asymmetry [13] $a_{cp}(\pi^-\pi^+) = (23.2 \pm 6.1) \times 10^{-4}$ fits well into our calculations, as follows from the description of the time dependence of the CP violation process in our model in Fig. 29 (b).

For $B^0$ meson, a smooth maximum of the asymmetry is observed at the meson lifetime, and already at the tail of the decay exponent, an alternating-sign feature associated with the oscillation frequency is observed. The oscillation period is most clearly manifested for $B_s^0$ meson, where such alternating features are clearly traced over three meson lifetimes. They clearly coincide with the experimental oscillation periodicity shown in Fig. 29d. This completely compensates for the CP violation effect. Experimental observations with more statistics in 2021 confirmed the absence of asymmetry [67]. Thus, the result of our calculations for $B_s^0$ meson is confirmed by the experimental results.

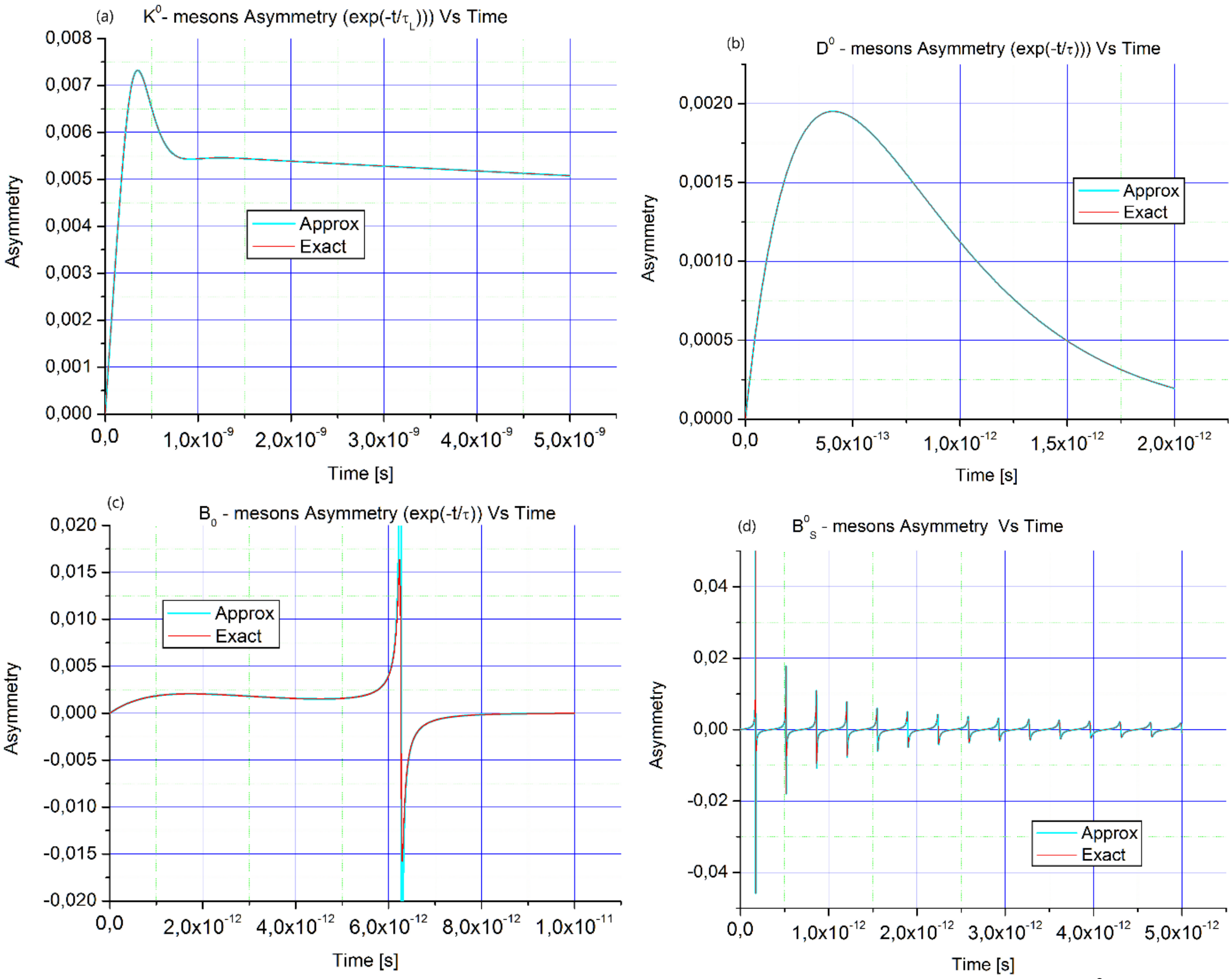

Fig. 29. CP-violating asymmetries for neutral mesons as a function of decay time. a) calculation results for $K^0$ meson, b) for $D^0$ meson, c) for $B^0$ meson, d) for $B_s^0$ meson. It will be useful to indicate the lifetimes for different mesons here: $\tau(K_s^0) = 0.89 \times 10^{-10} s$, $\tau(K_L^0) = 5.2 \times 10^{-8} s$, $\tau(K_L^0) = 5.2 \times 10^{-8} s$, $\tau(D^0) = 4.1 \times 10^{-13} s$, $\tau(B^0) = 1.530 \times 10^{-12} s$, $\tau(B_s^0) = 1.470 \times 10^{-12} s$.

In conclusion, it should be emphasized that our left-right CP-violating model included a sign reversal of mixing with the right-handed vector boson upon transition from $W^-$(particle) to $W^+$(antiparticle). The CP-violating phase change upon transition from particles to antiparticles occurs by 180 degrees, so CP invariance is violated 100%.

Finally, it should be noted that this rule, laid down for W bosons living $\approx 10^{-25}$ s, is preserved for mesons living $\approx 10^{-12}$ s. The reason that the effect is reproduced on such a time interval is the conservation of the vector (axial) current, which is preserved due to the conservation of charge (due to the conservation of spin).

### 12. Evolution of the Universe in a left-right model with CP violation

In this section, we will examine the role of the weak interaction potential in the evolution of the Universe. To do this, we will use the following matrix:

$$\begin{pmatrix} M-\Delta U - i\dfrac{\Gamma-\delta\Gamma}{2} & \Delta m - i\dfrac{\Delta\Gamma}{2} \\ \Delta m - i\dfrac{\Delta\Gamma}{2} & M+\Delta U - i\dfrac{\Gamma+\delta\Gamma}{2} \end{pmatrix} \tag{5.1}$$

Indeed, in the early Universe, there was no asymmetry in the number of particles and antiparticles, and the cause of this asymmetry is unknown. In the left-right CP-violated model, an asymmetry arises in the interaction potential for particles and antiparticles, which was previously estimated using the model parameters $\delta$ and $\zeta$ by the following relation $A_T^{LR} = -2\delta\zeta = (5.5 \pm 2.1) \times 10^{-3}$ $(2.6\sigma)$.
The weak interaction potential is discussed in detail in [68] and was also used in [69] and can be represented by the following equation:

$$U = \eta\frac{11\zeta(3)}{\pi^2\sqrt{2}}G_F T^3 - \frac{14}{45}\frac{\pi\left(3-\sin^2\theta_W\right)\sin^2\theta_W}{\alpha}G_F^2 T^4 E \tag{5.2}$$

or in numerical expression:

$$U = \eta \times 1.1\times 10^{-23}\left[\frac{1}{\mathrm{eV}^2}\right]T^3 - 1.1\times 10^{-44}\left[\frac{1}{\mathrm{eV}^4}\right]T^4 E \tag{5.3}$$

where $G_F$ is the Fermi constant, $T$ is the plasma temperature, $E$ is the particle energy, $\eta = \frac{N_f - N_{\bar{f}}}{N_f + N_{\bar{f}}}$ is the lepton asymmetry, $\frac{11\zeta(3)}{\pi^2\sqrt{2}}T^3$ is the density of cosmic plasma.
We will assume that in the early Universe, lepton asymmetry $\eta$ was zero. However, neutral mesons, being in the particle-antiparticle state, will have different interaction potentials with the medium even if there is no asymmetry in the number of particles and antiparticles in the medium, i.e. $\eta = 0$ . Then the potential difference for neutral mesons and antimesons will be: $U \approx \pm A^{LR}\frac{11\zeta(3)}{\pi^2\sqrt{2}}G_F T^3$ or in numerical expression - $\Delta U \approx \pm 6 \times 10^{-23}[eV^2] \times T^3[eV^3]$. Using this dependence of the potential difference from the temperature of the cosmic plasma, one can calculate the influence of the potential difference on the process of oscillation suppression, which is shown in Fig. 30. As can be seen, with an increase in the distance between the levels for particles and antiparticles, the oscillation process is suppressed.

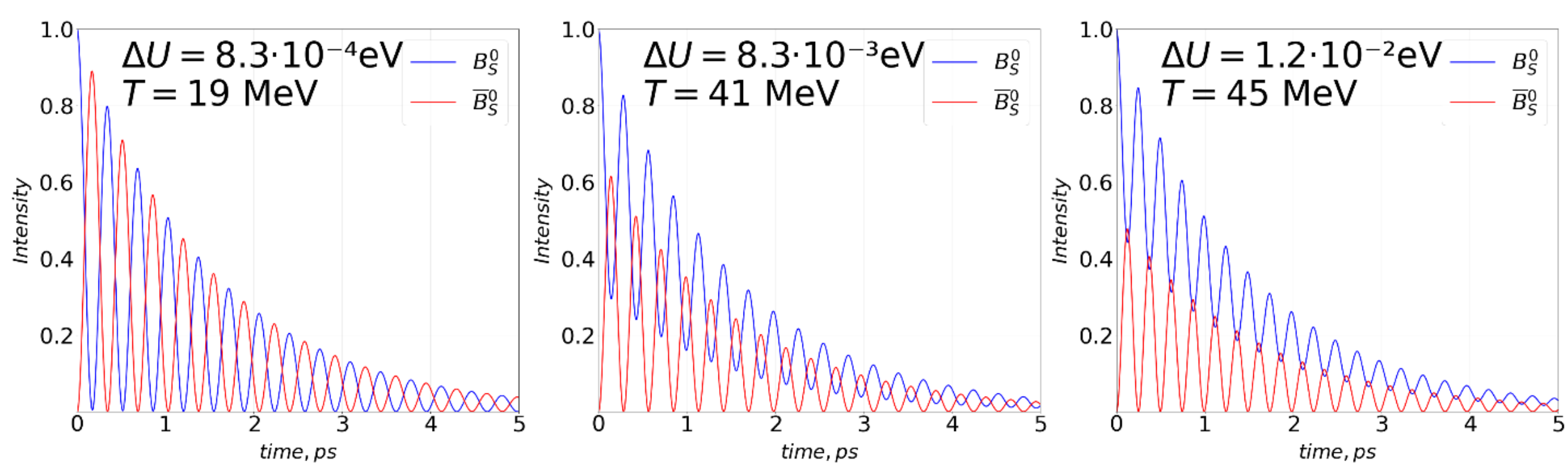

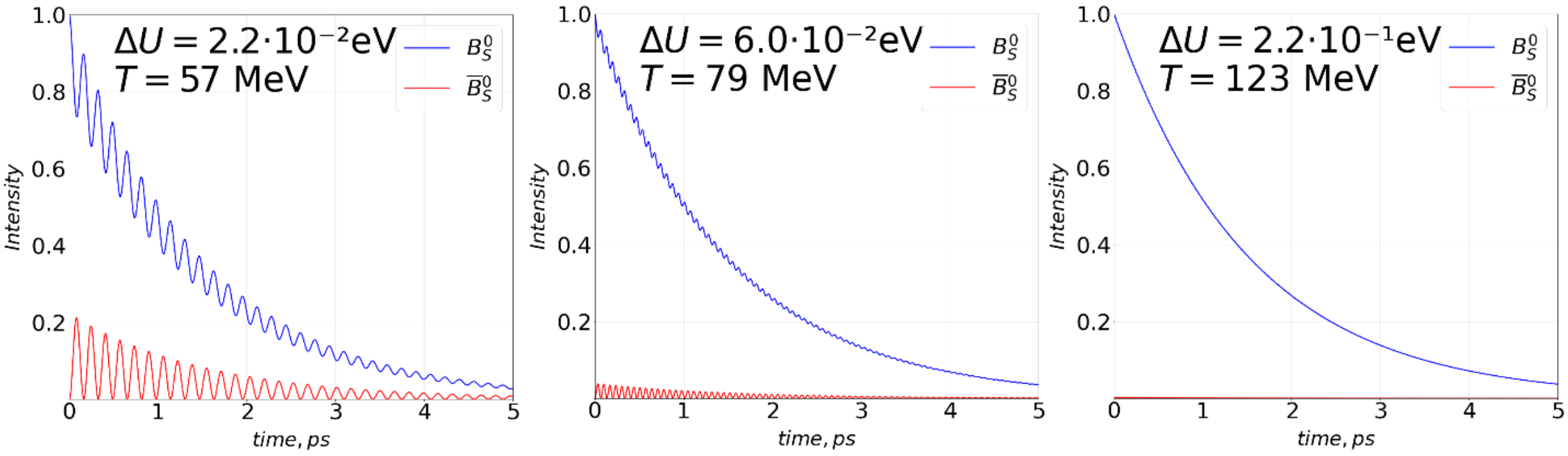

Fig. 30. The process of suppression of oscillations with an increase in the distance between levels for particles and antiparticles *ΔU*, which depends on the plasma density, which depends on the plasma temperature.

$B_s^0$ meson oscillation suppression factor due to the difference in the weak interaction potentials for particles and antiparticles is shown in Fig. 31. It reaches $10^9$ at a plasma temperature of 1 GeV.

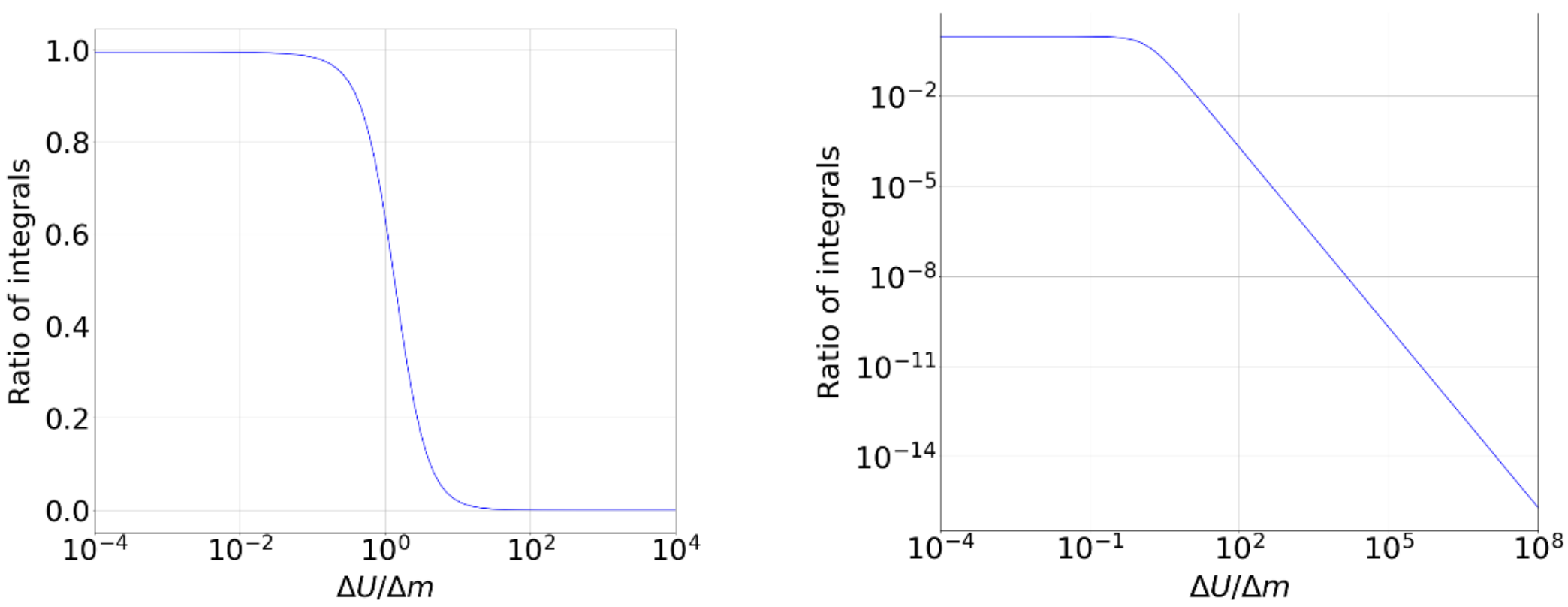

Fig. 31. $B_s^0$ meson oscillation suppression factor due to the difference in weak interaction potentials for particles and antiparticles on a linear and logarithmic scale.

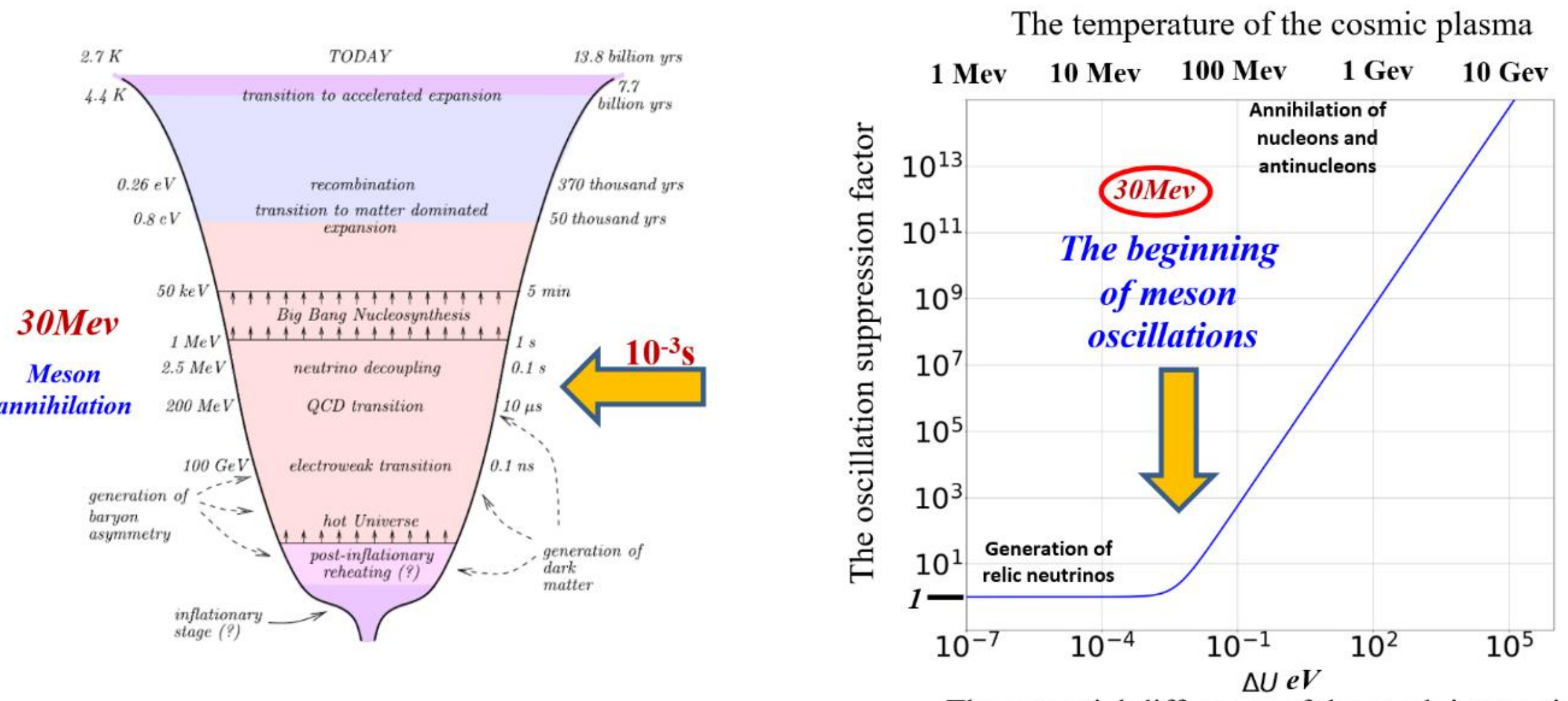

Fig. 32. Left: diagram of the stages of evolution of the Universe [70]. Right: dependence of $B_s^0$ meson oscillation suppression factor due to the difference in the weak interaction potentials for particles and antiparticles and the temperature of the cosmic plasma. The onset of oscillations occurs at $10^{-3}$ s and at a temperature of 30 MeV, which coincides with the onset of the meson annihilation process.

Table 7. Stages of evolution of the Universe

| Eras | Time | Stage of evolution | Temperature , K | Density, g/cm$^3$ | Energy, GeV |
|---|---|---|---|---|---|
| **Planck** | | Unknown laws of physics, quantum properties of space-time | | | |
| **Hadron** | $10^{-43}$ s | The limit of applicability of the relativistic theory of gravitation | $10^{32}$ | $10^{94}$ | $8.6 \cdot 10^{19}$ |
| | $10^{-35}$ s | The emergence of charge asymmetry | $10^{28}$ | $10^{78}$ | $8.6 \cdot 10^{15}$ |
| | **$10^{-5}$ s** | **Annihilation of nucleons and antinucleons** | **$3 \cdot 10^{12}$** | **$10^{16}$** | **$2.58 \cdot 10^{-1}$** |
| **Lepton** | **$10^{-4}$ s** | **The limit of applicability of experimentally verified laws of physics** | **$10^{12}$** | **$10^{14}$** | **$8.6 \cdot 10^{-1}$** |
| | **$10^{-3}$ s** | **Meson annihilation** | **$3 \cdot 10^{11}$** | **$10^{12}$** | **$2.58 \cdot 10^{-2}$** |
| | **0.2 s** | **Formation of relic neutrinos** | **$2 \cdot 10^{10}$** | **$10^{7}$** | **$1.72 \cdot 10^{-3}$** |
| **Radiation** | **10 s** | **Annihilation of electrons and positrons** | **$10^{10}$** | **$10^{4}$** | **$8.6 \cdot 10^{-3}$** |
| | **100 s** | **Formation of primary helium** | **$10^{8}$** | **$10^{2}$** | **$8.6 \cdot 10^{-5}$** |
| **Substances** | $10^{6}$ years | Separation of relic radiation from matter | $4 \cdot 10^{3}$ | $10^{-20}$ | $3.44 \cdot 10^{-10}$ |
| | $10^{9}$ years | The beginning of the emergence of stars and galaxies | 30 | $10^{-26}$ | $2.58 \cdot 10^{-12}$ |
| | $1 - 2 \cdot 10^{10}$ years | Modern era | 2.7 | $10^{-29} - 10^{-30}$ | $2.32 \cdot 10^{-13}$ |

Figure 33 shows the oscillation suppression factor for $K^0$ meson, $D^0$ meson, $B^0$ meson, and $B_s^0$ meson. $D^0$ meson decay occurs before the oscillation process begins. The oscillation process for $B_s^0$ meson begins at a temperature of 30 MeV and is accompanied by annihilation.

Above, an analysis of the stages of evolution of the Universe was presented within the framework of the left-right model with CP violation. The lepton asymmetry was used as the initial asymmetry $A_T^{LR} = -2\delta\zeta = 5.5 \times 10^{-3}$. It is important to note that the process of active oscillation development coincides with the process of active meson annihilation at a temperature of 30 MeV, which also coincides with the stages of the evolution of the Universe (Fig. 32, Table 7). As for the process of baryon annihilation, it can be assumed that this process occurs during neutron- antineutron oscillations during the hadronization of quarks at temperatures below $\approx 170$ MeV.

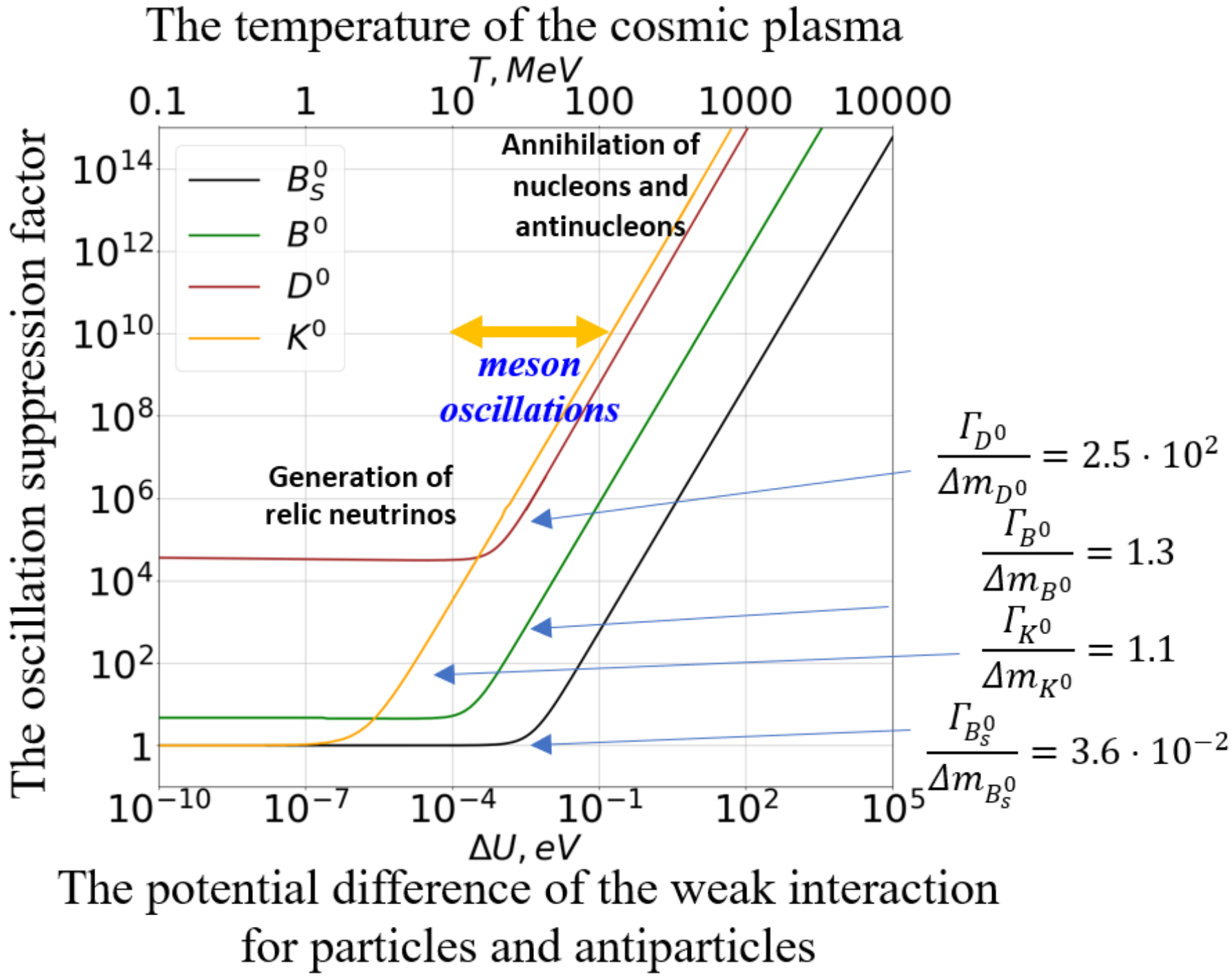

Fig. 33. Oscillation suppression factor for $K^0$ meson, for $D^0$ meson, $B^0$ meson and $B_s^0$ meson. $D^0$ meson decays before the oscillation process begins.

As calculations in Fig. 33 show, at cosmic plasma temperatures above 100 MeV, the process of neutral meson oscillations is suppressed by the potential difference of the weak interaction. At the same time, the process of their mutual annihilation is also suppressed; it is at this stage that lepton asymmetry arises, since the process of CP violation leads to the preferential decay of antimesons. In the next stage, at temperatures below 50 MeV, annihilation of the remaining mesons and antimesons occurs, increasing the degree of lepton asymmetry. For $B_s^0$ mesons, the annihilation process occurs at a temperature of approximately 30 MeV, for $D^0$ mesons the annihilation process occurs at a temperature of approximately 20 MeV, for $B^0$ mesons, the annihilation process occurs at a temperature of approximately 10 MeV and for $K^0$ mesons, the annihilation process occurs at a temperature of approximately 2 MeV.

In conclusion, I would like to present a general picture of the stages of the Universe's development, using A.D. Sakharov's fundamental principles, supplemented by our understanding of the violation of left-right symmetry, which determines the nature of its origin. At temperatures above 300 GeV, i.e., at temperatures greater than the mass of the right-handed $W_R$ boson, there was symmetry of the weak interaction with respect to right and left-handed interactions. At temperatures below 300 GeV, a bifurcation occurs in the choice of the interaction type, determined by the mass of the left-handed vector boson $W_L$. A compromise between the left-handed and right-handed bosons was found thanks to the admixture of heavy $W_R$ particles in the squared mass ratio with a small mixing angle. The process of CP violation, which is fundamentally important according to A.D. Sakharov's theorem, is ensured in our model by the different signs of the mixing angle for particles and antiparticles, i.e., for $W^-$and $W^+$. The latter circumstance is key to the left-right model with CP violation under consideration.

Finally, it is important to remember that the existence of a right-handed vector boson implies the existence of right-handed neutrinos, which are responsible for the formation of dark matter at the stages of meson annihilation at the temperatures indicated above.

## 13. PROSPECTS FOR IMPROVMENT THE PRECISION NEUTRON DECAY MEASUREMENTS

The left-right model of the weak interaction with CP violation presented here requires a significant increase in the experimental accuracy. The possibility of further increasing the accuracy of measurements in neutron decay exists. This is the goal of, for example, the PNPI NRC KI project "Neutron Beta Decay" for the PIK reactor [71-73], in which it is planned to use a superconducting solenoid with a long flight base for neutron decay in order to increase the statistics of decay events and with a magnetic mirror collimator to isolate the electron emission direction. The magnetic field in the region of the uniform magnetic field is 0.34 T, and in the region of the mirror 0.86 T. It is a development of the PNPI RAS experiment of 1998 [20], it is planned to achieve a relative measurement accuracy of $10^{-3}$ for neutrino and electron decay asymmetries, and most importantly, the asymmetries a-small and B, for which the greatest contradiction of 3.7σ is observed, will be measured with three times better accuracy.

General scheme of the experiment

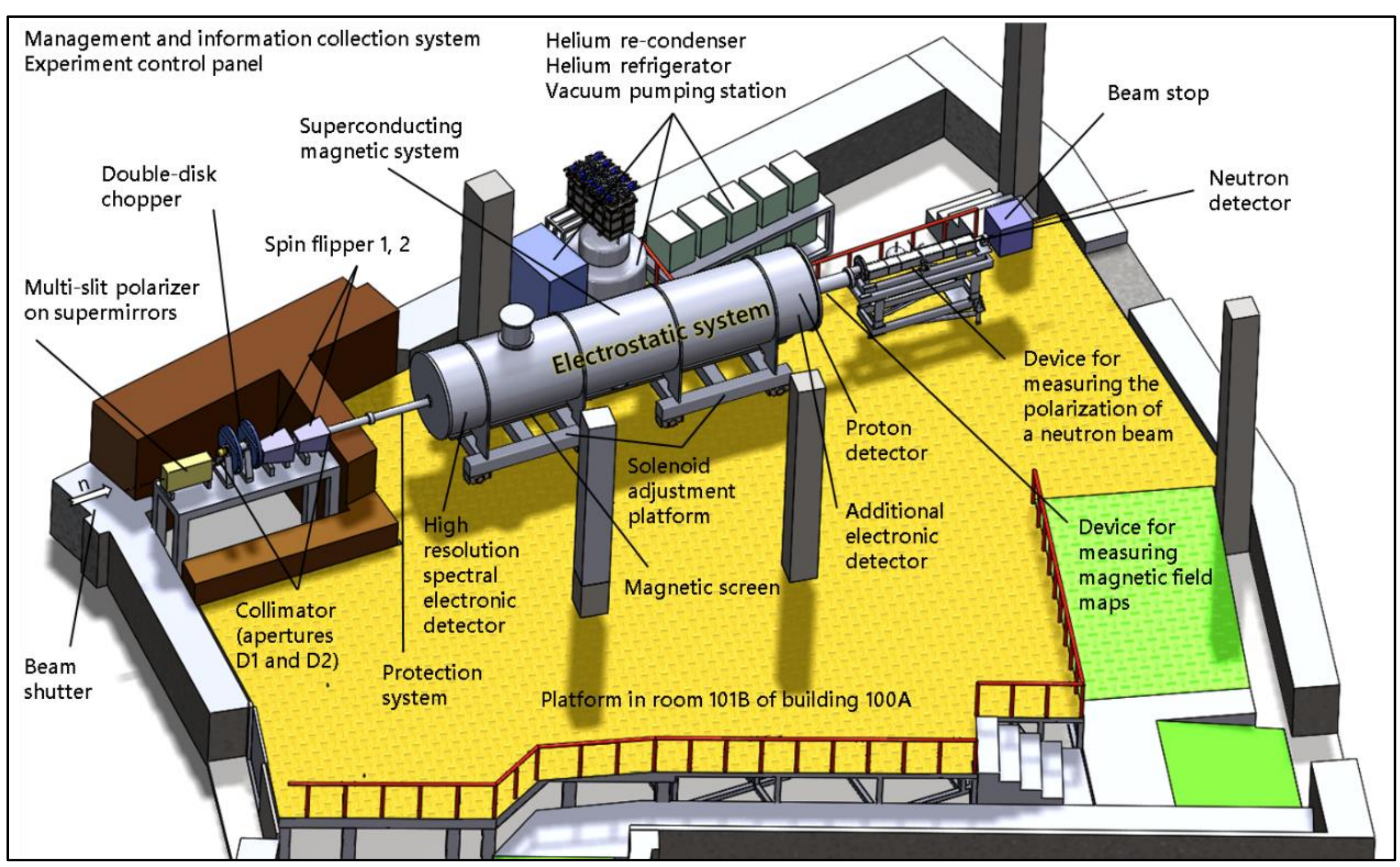

Superconducting solenoid.

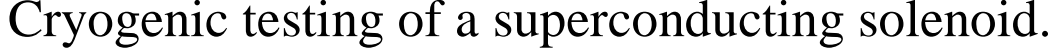
Cryogenic testing of a superconducting solenoid.

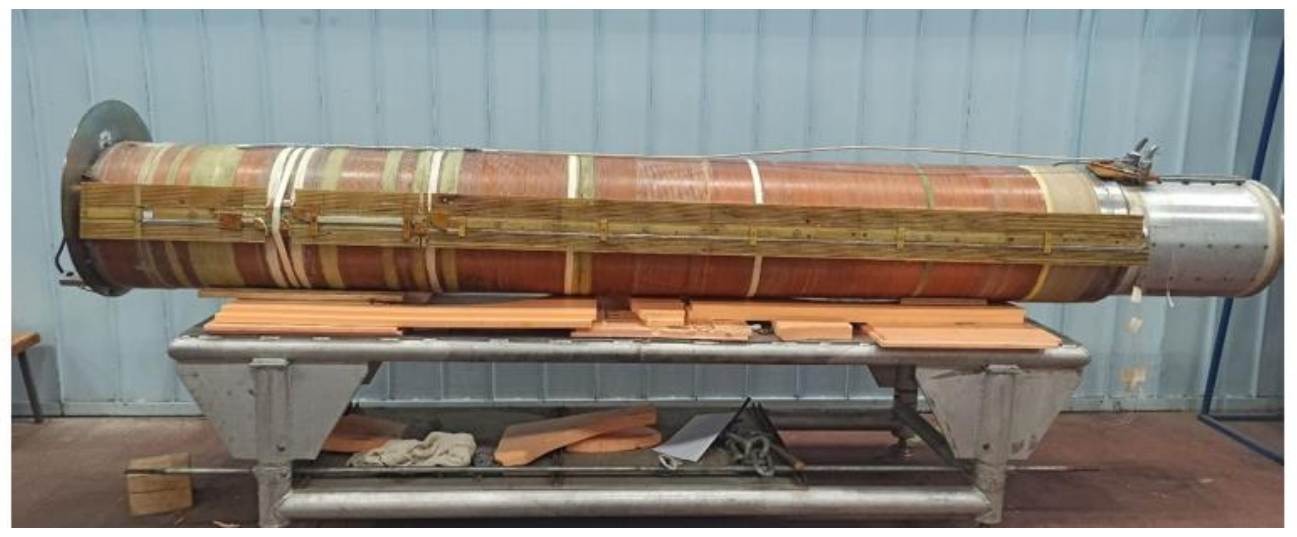

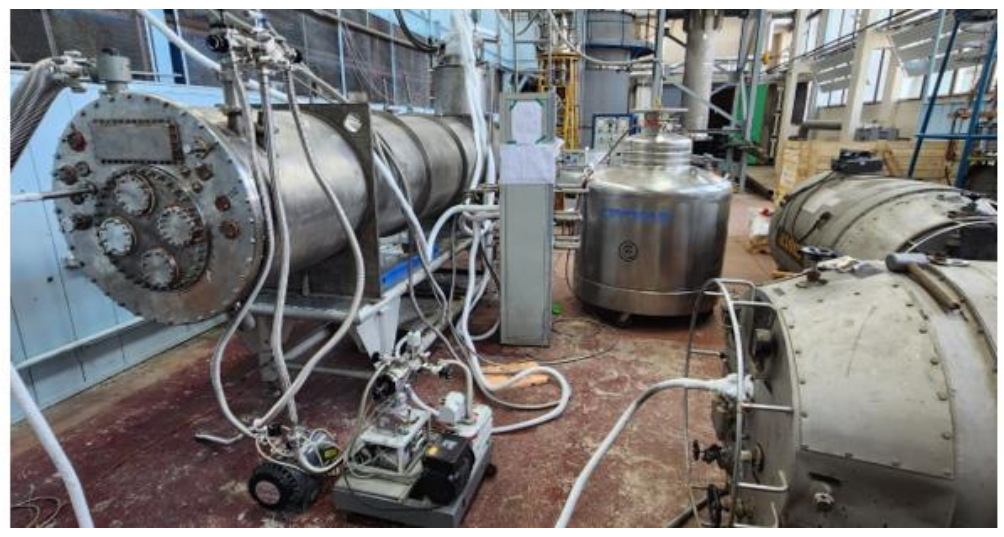

Scheme of the setup for measuring neutron decay asymmetries.

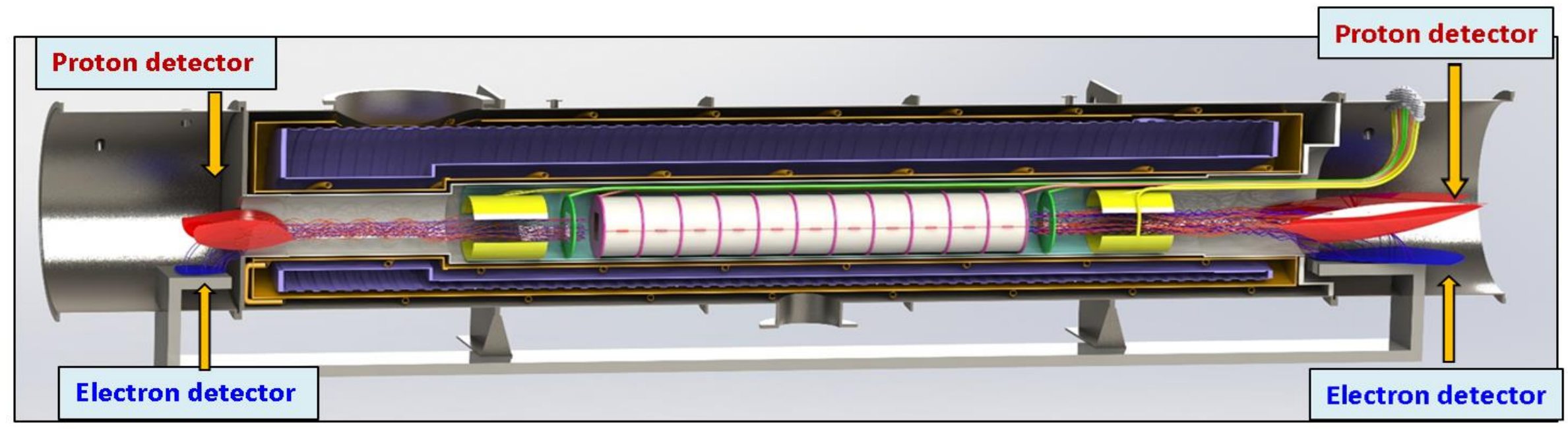

Fig. 34. 1) General experimental setup, 2) elements of the superconducting system, and 3) detailed diagram of the setup for measuring neutron decay asymmetries. Electron trajectories are shown in blue, proton trajectories are shown in red. The white cylinder is at a potential of +30 kV. The yellow half-cylinders are the plates of high-voltage capacitors with potentials of 20 kV. The green diaphragms are at zero potential, so protons from the decay region inside the white cylinder are accelerated at a potential of +30 kV and can be detected by the proton detector.

Another project of the Technical University of Munich "PERC" for the FRM2 reactor [74] also uses a long superconducting solenoid with a magnetic mirror to measure the neutrino and electron asymmetries of neutron decay with a relative accuracy of $10^{-3} - 10^{-4}$ [75]. Thus, there is reason to believe that the question of the existence of W_R mixing, with the above parameters, will be clarified.

## 14. CONCLUSION

1. The results of the latest most accurate experimental data on neutron decay are presented. It has been shown that the accuracy of measurements has increased by more than an order of magnitude over the past 30 years, which allows for an analysis of inconsistencies of the Standard Model.

2. An analysis was carried out on the possibility of the existence of the right vector boson $W_R$. As a result of the analysis within the framework of the extended left-right symmetric model, it was found that there is an indication of the existence of the right vector boson $W_R$ with a mass $M_{W_R} = 304^{+24}_{-20}\, GeV$, and a mixing angle with $W_L$: $\zeta = -0.039 \pm 0.014$.

3. It is shown that this result does not contradict the experiments at colliders to search for a hypothetical vector boson.

4. CP violation was found in the baryon sector at the $2\sigma$ confidence level.

5. It is shown that it is possible to describe the effects of CP violation in decays of neutral $K$-mesons, $D$-mesons and $B^0-$mesons using the parameters of the extended left-right model, which are obtained from neutron decay.

Thus, the extended left-right symmetric model of weak interaction within the limits of the available accuracy, allows to describe the effects of CP violation in baryons and mesons using relation $A^{LR} = -2\delta\zeta$.

6. It is shown that in the left-right model with CP violation, a splitting of the interaction potential for particles and antiparticles with cosmic plasma occurs, which is the cause of the emergence of lepton asymmetry of the Universe.

7. The opposite signs of the baryon and lepton CP-violating asymmetry are associated with the different signs of the baryon and lepton asymmetry of the Universe, and B-L is conserved [62,63].

8 Finally, we can consider possible consequences, assuming the results presented in this work will be confirmed. First, an extension of SM by introducing right vector bosons $W_R^{\pm}, Z_R$ and righthanded neutrinos is required. Second, righthanded neutrinos can be considered as candidates for dark matter.

## ACKNOWLEDGMENTS

The authors express their gratitude to the staff of the High Energy Physics Department of PNPI for assistance in analyzing the ATLAS experiment data.

## FINANCING

This work was supported by the Russian Science Foundation (Project No. 24-12-00091 https://rscf.ru/p r oject/24-12-00091/ ).

## CONFLICT OF INTEREST

The authors declare no conflict of interest.